\documentclass[reprint,amsmath,amssymb,aps]{revtex4-2}
\usepackage{graphicx}
\usepackage{bm}
\usepackage{tabstackengine}
\usepackage{eucal}
\usepackage{bm}
\usepackage{booktabs}
\usepackage{xcolor}
\usepackage{amsmath, graphics, epsfig, color, verbatim, amsfonts, tensor}
\usepackage{hhline}
\usepackage{multirow}
\usepackage{siunitx}
\usepackage{tabularx}
\makeatletter
\renewcommand*\env@matrix[1][c]{\hskip -\arraycolsep
  \let\@ifnextchar\new@ifnextchar
  \array{*\c@MaxMatrixCols #1}}
\makeatother

\begin{document}
\title{Parameterizing empirical interatomic potentials for predicting thermophysical properties via an irreducible derivative approach: the case of ThO$_2$ and UO$_2$}
\author{Shuxiang Zhou$^1$}
\email{shuxiang.zhou@inl.gov}
\author{Chao Jiang$^1$, Enda Xiao$^2$, Sasaank Bandi$^3$, Michael W.D. Cooper$^4$, Miaomiao Jin$^5$, David H Hurley$^1$, Marat Khafizov$^6$}
\author{Chris A. Marianetti$^3$}
\affiliation{$^1$Idaho National Laboratory, Idaho Falls, ID 83415, USA\\$^2$Department of Chemistry, Columbia University, New York, NY 10027, USA\\$^3$Department of Applied Physics and Applied Mathematics, Columbia University, New York, NY 10027, USA\\$^4$Los Alamos National Laboratory, Los Alamos, NM 87545, USA\\$^5$Department of Nuclear Engineering, The Pennsylvania State University, University Park, PA 16802, USA\\$^6$Department of Mechanical and Aerospace Engineering, The Ohio State University, Columbus, OH 43210, USA}

\begin{abstract}
The accuracy of classical physical property predictions using molecular
dynamics simulations is determined by the quality of the interatomic
potentials. Here we introduce a training approach for empirical interatomic
potentials (EIPs) which is well suited for capturing phonons
and phonon-related properties.
Our approach is based on direct comparisons of the second- and third-order
irreducible derivatives between an EIP and the Born-Oppenheimer potential
within density functional theory (DFT) calculations.  Irreducible derivatives
fully exploit space group symmetry and allow for training without redundant
information.  We demonstrate the fidelity of our approach in
the context of  ThO$_2$ and UO$_2$, where we optimize parameters of an
embedded-atom method potential in addition to core-shell interactions.  Our
EIPs provide thermophysical properties in good agreement with DFT
and outperform widely utilized EIPs for phonon dispersion and thermal
conductivity predictions.  Reasonable estimates of thermal expansion and
formation energies of Frenkel pairs are also obtained. 

\end{abstract}

\maketitle

\section{\label{sec:intro}Introduction}

Interatomic potentials are mathematical models used to encode the Born-Oppenheimer potential of some collection of atoms.
Traditionally, empirical interatomic potentials (EIPs) are derived using analytical functional forms with a few parameters: examples include the Lennard-Jones potential \cite{lennard1931cohesion}, which has only two parameters in simple systems, and the Tersoff potential \cite{tersoff_empirical_1988} and embedded atom method (EAM) potential \cite{daw_embedded-atom_1984}, which both have tens of parameters. 
The analytical functional forms and the limited number of parameters of EIPs offer simplicity for modeling, but also restrict accuracy and limit the potential for capturing complex phenomena. 
Recently, machine-learning interatomic potentials (MLIPs) have emerged as a valuable complement to traditional EIPs by leveraging machine-learning algorithms like neural networks and Gaussian process regression \cite{behler2007generalized, bartok2010gaussian}, exhibiting higher accuracy and transferability. However, MLIPs are computationally more demanding, often using thousands to millions of parameters.  
Despite the success of MLIPs, EIPs are still useful given their computational efficiency and physical interpretability.

In the last few decades, extensive studies have been performed to advance the training procedure for interatomic potentials. EIPs were previously trained by fitting to experimental measurements, including the lattice parameters, elastic constants, thermal expansion, and specific heat \cite{daw_embedded-atom_1984,tersoff_empirical_1988,allen_computer_2017,cooper_many-body_2014}. As training data, experimental measurements are limited by both quantity and quality.  In terms of quantity, experiments are often limited by high-cost and time-consuming processes, especially for complex systems such as defect structures. 
Additionally, experiments are limited in terms of directly probing the details of the Born-Oppenheimer potential, and instead only measure averaged quantities.
This lack of data can limit the accuracy of the EIP and can yield substantial errors for selected observables.
For example, traditional parametrization of EIPs to elastic constants and thermal expansion is often inadequate for accurately predicting phonon dispersions, not to mention third-order phonon interactions \cite{Chernatynskiy2012critical}. Both the challenge of quantity and quality of training data can be resolved by utilizing first-principles calculations, where approaches such as density functional theory (DFT) can produce a multitude of atomistic data at both equilibrium and non-equilibrium. 

Given that the goal of parameterizing an interatomic potential is to faithfully encode the Born-Oppenheimer potential, a natural source of training data
would be a direct sampling of the Born-Oppenheimer potential over some relevant domain,
which can be achieved when using data from sufficiently reliable first-principles calculations.
There are various approaches for sampling the Born-Oppenheimer potential. One approach would
be to compute the value of the Born-Oppenheimer potential (i.e. energy) and the first derivatives (i.e. forces) at a collection of configurations, which is practical for approaches such as density functional theory where the forces can be computed
at a small computational overhead. The collection of configurations might be generated using a
molecular dynamics trajectory. 
EIPs trained on DFT-calculated energies and forces have generated robust results, including reasonable predictions of phonons \cite{ercolessi_interatomic_1994,mishin1999interatomic,fellinger2010force,lee2012force,brommer_classical_2015,schopf_effective_2014,fan_minimal_2019,qiu_molecular_2009,roy_chowdhury_development_2019,fan2019minimal,tanaka2023interatomic} and defect energies \cite{cooper_development_2016}, and it is standard to train MLIP on DFT-calculated forces and energies \cite{novikov2020mlip}.
An alternative approach would be to use a single minimum energy configuration and construct the second and third derivatives of the Born-Oppenheimer potential of this configuration.
The use of second- and higher-order derivatives in potential training is relatively uncommon compared to the utilization of energies and forces, both due to the increased computational cost of generating the data and training the model. However, if the targeted observables to be generated using the EIP directly probe the derivatives (e.g. scattering function), using the second and higher
order derivatives as training data may be worthwhile. 
For example, interatomic potentials trained using DFT-calculated second energy derivatives can significantly improve the predicted phonon dispersion, for both EIP \cite{lindsay2010optimized, murakami_importance_2013, han_interatomic_2011, rohskopf_empirical_2017,muraleedharan2017phonon,rohskopf2020fast,shi2021molecular} and MLIP \cite{loew2024training,fang_phonon_2024}.
When training to second- and higher-order energy derivatives, it is important not to have any redundancy in the training data, which can be achieved using space group irreducible derivatives (ID) \cite{fu_group_2019}. The IDs are generated using the group theoretical selection rules which dictate
which irreducible representations are allowed to couple, providing a minimal set of derivatives that characterize a given discretization of the Brillouin zone.

\begin{table*}[t]
\caption{\label{tab:pairParameter}%
Parameters for the short-range pairwise potentials described by Eqs. (\ref{eq:phiB})--(\ref{eq:phiM}).  
}
\begin{ruledtabular}
\begin{tabular}{@{}c
S [table-format=5.2]
S [table-format=1.4]
S [table-format=2.4]
S [table-format=1.4]
S [table-format=1.4]
S [table-format=1.4]}
\textrm{Interaction}&
\multicolumn{3}{c}{$\phi_B(r_{ij})$}&
\multicolumn{3}{c}{$\phi_M(r_{ij})$}\\ \cmidrule(lr){1-1}\cmidrule(lr){2-4}\cmidrule(lr){5-7} 
{$\alpha$-$\beta$} &
{$A_{\alpha\beta}(eV)$} &
{$\rho_{\alpha\beta}(\AA)$} &
{$C_{\alpha\beta}(eV\AA^6)$} &
{$D_{\alpha\beta}(eV)$} & 
{$\gamma_{\alpha\beta}(\AA^{-1})$}&
{$r_0(\AA)$}\\[0.2em]
\colrule\\[-0.8em]
Th-Th & 16078.23 & 0.2985 & 1.6499 & 5.0866 & 5.7739 & 2.4611 \\ [0.2em]
U-U   & 18447.13 & 0.1452 & 12.9937 & 6.2571 & 4.2394 & 2.1289  \\ [0.2em]
Th-O  & 859.98 & 0.3820 & 9.7412 & 0.3241 & 2.1530 & 2.4440 \\ [0.2em]
U-O   & 807.24 & 0.3895 & 6.0011 & 0.3088 & 1.9770 &  2.5195 \\ [0.2em]
O-O   & 958.50 & 0.3202 & 7.0602 & {-} & {-} & {-}  
\end{tabular}
\end{ruledtabular}
\end{table*}
\normalsize

\begin{table}[t]
\caption{\label{tab:pairManybody}%
Parameters for the many-body term, the Coulomb potential, and core-shell models described by Eqs. (\ref{eq:potential}), (\ref{eq:phiC}), and (\ref{eq:sigmaBeta}).  
}
\begin{ruledtabular}
\begin{tabular}{@{}c
S [table-format=1.4]
S [table-format=4.2]
S [table-format=4.4]
S [table-format=3.4]
S [table-format=4.2]
}
\textrm{Species}&
{$G_\alpha(eV\AA^{1.5})$}&
{$n_\alpha(\AA^5)$}&
{$q_\alpha^c(|e|)$}&
{$q_\alpha^s(|e|)$}&
{$k_\alpha(eV\AA^{-2})$}\\[0.2em]
\colrule\\[-0.8em]
\textrm{Th} & 1.1122 & 899.84 & -9.7268 & 12.6813 & 1204.99\\[0.2em]
\textrm{U}  & 2.0053 & 991.90 & -7.2900 & 10.2445 &  551.98\\[0.2em]
\textrm{O}  & 0.4772 & 986.88 & 1.7688 & -3.2461 &  436.86
\end{tabular}
\end{ruledtabular}
\end{table}
\normalsize

In this work, we develop an ID-based potential training approach, by including the second- and third-order displacement IDs and the second-order strain IDs in the training data. Our approach is used to parameterize an EIP for the nuclear fuel materials ThO$_2$ and UO$_2$,  where
there is a critical need to understand defect formation, microstructure evolution, and thermal transport degradation \cite{hurley2022thermal,jin2020systematic,dennett_integrated_2021,He2022phase, khafizov2019impact,khafizov2020combining,deskins2022combined,jin2023unfaulting,yu2024establishing,neilson2024irradiation,neilson2024oxygen,malakkal2024first,jiang2024machine}.
A previously developed EIP for the actinide oxides \cite{cooper_many-body_2014},  referred to as the CRG potential, 
has found widespread use in classical molecular dynamics (MD) 
studies of actinide oxides \cite{liu2016molecular,park_sensitivity_2018,cooper_modelling_2015,park_thermal_2018,balboa2017assessment,malakkal2019thermal,zhu2020effect,jiang2022unraveling,jin2024extended,fossati2024superionic,cooper2025role}. The CRG potential was parameterized in a traditional fashion, based
on the experimental elastic constant and thermal expansion.
Recently, the accuracy of the CRG potential for predicting phonons and thermal transport
within the  Boltzmann transport equation (BTE) framework was assessed, demonstrating that the CRG potential has nontrivial differences with the experimental phonon dispersion of optical branches as well as with the thermal conductivity \cite{jin_assessment_2021}. 
These differences are not unexpected, given that neither the phonons nor the phonon interactions 
were included in the CRG training data.
In this work, we demonstrate that our ID-based training procedure yields an EIP 
that reliably characterizes the first-principles data. Given that DFT and DFT+$U$ can reliably
describe ThO$_2$ and UO$_2$ \cite{dennett_integrated_2021,mathis2022generalized,xiao2022validating, zhouCapturingGroundState2022a, zhou_phonon_2024}, respectively, our resulting EIP 
yields a substantial improvement over the CRG potential for the predicted phonon related properties relative to experiment.

\section{\label{sec:methods}Methods}

\subsection{\label{sec:methodPotential}Potential form}

In this work, the analytical expression of the EIP is based on the CRG potential \cite{cooper_many-body_2014}, which uses a pair-potential for each atom pair in addition to the many-body EAM potential \cite{daw_embedded-atom_1984}. The pair-potential contains three contributions: the Buckingham potential \cite{buckingham_classical_1938}, the Morse potential \cite{morse_diatomic_1929}, and the long-range electrostatic Coulomb interaction. Additionally, a core-shell spring model \cite{mitchell_shell_1993} is added, as it substantially improves the prediction of selected optical phonons. Within the core-shell model, each ion splits into two particles, a core and a shell, where the core and shell are attached by a spring force with spring constant $k$ (see the last term of Eq.~\ref{eq:potential}). The charge and mass of the ion also split into a core and shell contribution, and here we apply the massless shell model, as implemented in the General Utility Lattice Program (GULP) package \cite{gale_gulp_1997}. The non-Coulombic interactions (i.e., Buckingham potential, Morse potential, and EAM) are only defined between the shells, while Coulombic interactions are applied between all cores and shells, except the core and shell of the same ion (see Eq.~\ref{eq:phiC}). 

We denote the distances between the core-shell ion $i$ and $j$ as a vector $\mathbf{r_{ij}}=(r_{ij}^{cc}, r_{ij}^{cs}, r_{ij}^{sc}, r_{ij}^{ss})$, where $r_{ij}^{cc}$, $r_{ij}^{cs}$, $r_{ij}^{sc}$, $r_{ij}^{ss}$ represent the distance between $i$'s core and $j$'s core, $i$'s core and $j$'s shell, $i$'s shell and $j$'s core, and $i$'s shell and $j$'s shell, respectively. 
The potential energy $E_i$ of an ion $i$ concerning all other ions is given by:
\begin{equation}
E_i = \frac{1}{2}\sum_{j\ne i}\phi_{\alpha\beta}(\mathbf{r_{ij}})-G_\alpha\sqrt{\sum_{j\ne i}\sigma_\beta(r_{ij}^{ss})}+\frac{1}{2}{k_i}{r_{ii}^{cs}}^2,\label{eq:potential}\\    
\end{equation}
which is a sum of pairwise components, many-body components, and harmonic spring component terms. The pairwise potential between ions $i$ and $j$, $\phi_{\alpha\beta}(\mathbf{r_{ij}})$, is given by the sum of the Coulomb potential $\phi_{C}(\mathbf{r_{ij}})$, the Buckingham potential $\phi_{B}(r_{ij}^{ss})$, and the Morse potential $\phi_{M}(r_{ij}^{ss})$: 

\begin{align}
    &\phi_{\alpha\beta}(\mathbf{r_{ij}}) = \phi_C(\mathbf{r_{ij}}) + \phi_B(r_{ij}^{ss}) + \phi_M(r_{ij}^{ss}), \label{eq:pair}\\
    &\phi_C(\mathbf{r_{ij}}) = \frac{1}{4\pi\epsilon_0}(\frac{q_\alpha^c q_\beta^c}{ r_{ij}^{cc}}+\frac{q_\alpha^c q_\beta^s}{r_{ij}^{cs}}+\frac{q_\alpha^s q_\beta^c}{r_{ij}^{sc}}+\frac{q_\alpha^s q_\beta^s}{r_{ij}^{ss}}), \label{eq:phiC}\\
    &\phi_B(r_{ij}^{ss}) = A_{\alpha\beta}  \mathrm{exp} \left( \frac{-r_{ij}^{ss}}{\rho_{\alpha\beta}} \right) -\frac{C_{\alpha\beta}}{(r_{ij}^{ss})^6}, \label{eq:phiB}\\
    &\phi_M(r_{ij}^{ss})=D_{\alpha\beta} [ \mathrm{e}^{-2\gamma_{\alpha\beta}(r_{ij}^{ss}-r_0)} - 2\mathrm{e}^{-\gamma_{\alpha\beta}(r_{ij}^{ss}-r_0)} ], \label{eq:phiM}
\end{align}
where $\alpha$ and $\beta$ are the labels of species for ions $i$ and $j$, respectively. 
The many-body term in Eq. (\ref{eq:potential}) is given  by the square root of the
sum of the pairwise function $\sigma_\beta(r_{ij}^{ss})$, given by:
\begin{equation}
    \sigma_\beta(r_{ij}^{ss}) = \frac{n_\beta}{{r_{ij}^{ss}}^8}. \label{eq:sigmaBeta}
\end{equation}

\subsection{\label{sec:methodTraining}Training procedure and model assessment }

\begin{table*}[t]
\caption{\label{tab:agreementFitting}%
A portion of training data, including the second-order displacement IDs, the second-order strain IDs, lattice parameter at $T=0$ (denoted $a(0)$), normalized Born effective charges $Z^*_{Th}$ or $Z^*_{U}$, and  the defect formation energy $E_F$ for a Th or U FP in the conventional cubic supercell, computed by DFT, and compared with the CRG  \cite{cooper_many-body_2014} and the PW.
}
\begin{ruledtabular}
\begin{tabular}{@{}c
S [table-format=4.3]
S [table-format=4.3]
S [table-format=4.3]
S [table-format=4.3]
S [table-format=4.3]
S [table-format=4.3]
}
\multirow{ 2}{*}{Property} &
\multicolumn{3}{c}{ThO$_2$}&
\multicolumn{3}{c}{UO$_2$}\\
\cmidrule(lr){2-4}\cmidrule(lr){5-7}
& {SCAN} & {CRG} & {PW} & {PBE+$U$} & {CRG} & {PW}\\ [0.2em]
\hline\hline \\[-0.8em]
\multicolumn{7}{c}{Second-order irreducible derivatives within the $2\times2\times2$ supercell (eV/\AA$^2$)}\\[0.2em]
\colrule\\[-0.8em]
$ d^{T_{2g}^{}T_{2g}^{}}_{\Gamma \Gamma } $ 
& 12.65 & 16.02 & 12.82 &  11.64  &  17.35  &  12.90  \\[0.2em]
$ d^{T_{1u}^{}T_{1u}^{}}_{\Gamma \Gamma } $ 
& 12.75 & 16.60 & 11.62 &  10.91  &  15.63  &  10.06  \\[0.2em]

$ d^{A_{1g}^{}A_{1g}^{}}_{L L } $ 
&   17.47 &  16.28 &   16.87 &  16.11  &  17.34  &  17.67  \\[0.2em]
$ d^{E_{g}^{}E_{g}^{}}_{L L } $   
&   9.57  &  11.10 &   9.14  &  8.69   & 11.64   &   9.01 \\[0.2em]
$ d^{A_{2u}^{}A_{2u}^{}}_{L L } $ 
&   35.76 &  32.75 &   34.01 &  30.50  &  32.04  &   28.83  \\[0.2em]
$ d^{A_{2u}^{}\tensor*[^{1}]{\hspace{-0.2em}A}{}_{2u}^{}}_{L L } $ 
&   7.78 &    6.69 &   5.73  &   6.63  &  7.72   &  4.14  \\[0.2em]
$ d^{\tensor*[^{1}]{\hspace{-0.2em}A}{}_{2u}^{}\tensor*[^{1}]{\hspace{-0.2em}A}{}_{2u}^{}}_{L L } $ 
&   10.07 &  16.10 &   9.97 &    8.72 &  17.14  &   8.36  \\[0.2em]
$ d^{E_{u}^{}E_{u}^{}}_{L L } $   
&   9.59  &  12.22 &   9.62  &   7.93  &  11.25  &  8.15 \\[0.2em] 
$ d^{E_{u}^{}\tensor*[^{1}]{\hspace{-0.2em}E}{}_{u}^{}}_{L L } $ 
&  -3.09 &   -6.57 &   -3.19 &  -2.62  &  -5.85  &   -2.49  \\[0.2em]
$ d^{\tensor*[^{1}]{\hspace{-0.2em}E}{}_{u}^{}\tensor*[^{1}]{\hspace{-0.2em}E}{}_{u}^{}}_{L L } $
&   7.55  &  11.10 &    7.47 &  6.89   &  11.64  &  6.75  \\[0.2em] 

$ d^{A_{1g}^{}A_{1g}^{}}_{X X } $ 
& 21.02 & 27.18 & 25.23 &  19.74   &  29.87  &  26.53 \\[0.2em]
$ d^{E_{g}^{}E_{g}^{}}_{X X } $ 
& 4.61 & 4.11   &  3.03 &  3.70    &   3.55  &  2.57 \\[0.2em]
$ d^{\tensor*[^{2}]{\hspace{-0.2em}A}{}_{2u}^{}\tensor*[^{2}]{\hspace{-0.2em}A}{}_{2u}^{}}_{X X } $ 
& 41.53 & 28.64 & 31.60 &   37.25  &  28.23  &  26.95 \\[0.2em]
$ d^{B_{1u}^{}B_{1u}^{}}_{X X } $
& 2.53 & 2.86   &  1.91 &    1.55  &   2.10  &  1.39 \\[0.2em]
$ d^{E_{u}^{}E_{u}^{}}_{X X } $
& 12.09 & 19.62 & 12.79 &   11.38  &  21.45  &  11.68 \\[0.2em]
$ d^{E_{u}^{}\tensor*[^{1}]{\hspace{-0.2em}E}{}_{u}^{}}_{X X } $
& -0.59 & -4.12 & -0.63 &   -0.24  &  -3.13  &  -0.23 \\[0.2em]
$ d^{\tensor*[^{1}]{\hspace{-0.2em}E}{}_{u}^{}\tensor*[^{1}]{\hspace{-0.2em}E}{}_{u}^{}}_{X X } $
& 11.79 & 14.76 & 13.67 &   9.40   &  13.33  &  11.69 \\[0.8em]


\multicolumn{7}{c}{Elastic energy irreducible strain derivatives (eV)}\\[0.2em]
\colrule\\[-0.8em]
$d\indices*{*_{A_{1g}^{}}^{}_{A_{1g}^{}}^{}}$ & 166.73 & 157.02 & 166.52 & 164.91 & 165.92 & 165.88 \\ [0.2em]
$d\indices*{*_{E_{g}^{}}^{}_{E_{g}^{}}^{}}$   &  70.81 &  64.78 &  68.89 &  69.48 &  71.25 &  66.55 \\ [0.2em]
$d\indices*{*_{T_{2g}^{}}^{}_{T_{2g}^{}}^{}}$ &  22.13 &  19.43 &  20.56 &  16.66 &  16.17 &  16.33  \\[0.8em]

\multicolumn{7}{c}{Other properties}\\[0.2em]
\colrule\\[-0.8em]
$a(0)$ (\AA) & 5.594 & 5.580 & 5.594 & 5.546 & 5.453 & 5.546 \\ [0.2em]
$Z^*_{Th}$ or $Z^*_{U}$ $(|e|)$   &  2.448 &  2.221 &  2.448 &  2.325 &  2.221 &  2.325 \\ [0.2em]
$E_F$(Th or U FP) (eV) &  18.000 &  17.529 &  17.999 &  14.023 &  11.165 &  14.033  

\end{tabular}
\end{ruledtabular}

\end{table*}
\normalsize

Due to the specific interests in phonon predictions, our dataset was separated into two categories, phonon-related properties and other properties. Phonon-related properties contain the second-order displacement IDs, the third-order displacement IDs, the second-order strain IDs,  and phonon thermal conductivity. Besides these phonon-related properties, we consider the classical lattice parameter $a(T)$ at $T=0$ K, the normalized Born effective charges $Z^*_\alpha$, and the defect formation energy ($E_F$) of Frenkel pairs (FPs). Here $Z^*_\alpha$ is obtained by the Born effective charge $Z_\alpha$ normalized by the dielectric constant $\epsilon$: $Z^*_\alpha=Z_\alpha/\sqrt{\epsilon}$. Among the different types of point defects, only the FP is considered, so a direct comparison can be made between DFT and the EIP, avoiding approximations needed to treat unbalanced charges. The training dataset contained the second-order displacement IDs within the $2\times2\times2$ supercell, the third-order displacement IDs within the conventional cubic supercell $\mathbf{\hat{S}}_\mathrm{C}$, 
the second-order strain IDs, $a(0)$, $Z^*_\alpha$, and $E_F$ for a Th or U FP in the conventional cubic supercell, while the remaining data were used to assess the accuracy of the model. 
The details of properties used in the training procedure and assessment are tabulated in Table S1 of the Supplemental Materials (SM) \cite{Supplemental_Material}. 
Additionally, we also predicted thermal expansion using
the EIP and directly compared with experimental results.
See Section I of SM \cite{Supplemental_Material} for the mathematical representation of the supercells used in calculations, following Ref.\cite{fu_group_2019}.

All DFT calculations were carried out via a projector augmented-wave (PAW) method \cite{blochl_projector_1994, kresse_ultrasoft_1999}, as implemented in the Vienna ab initio simulation package \cite{kresse_ab_1993, kresse_efficient_1996}. 
For the exchange-correlation functionals, the strongly constrained and appropriately normed (SCAN) \cite{sunStronglyConstrainedAppropriately2015} functional 
was used for ThO$_2$, while the Perdew, Burke, Ernzerhof (PBE) generalized gradient approximation (GGA)  \cite{perdewGeneralizedGradientApproximation1996a} was used for UO$_2$, following previous DFT studies  \cite{maldonado2016crystal,dennett_integrated_2021,mathis2022generalized,xiao2022validating, zhouCapturingGroundState2022a, zhou_phonon_2024} which yield phonons in good agreement with experiments. Additionally, the UO$_2$ calculations applied spin-orbit coupling, DFT+$U$ \cite{anisimov_band_1991,dudarev_effect_1997} with $U = 4$ eV, and occupation matrix control with the initial values of the occupation matrices from the 3\textbf{k} AFM state reported in Ref. \cite{zhouCapturingGroundState2022a}.  
The plane-wave cutoff energy was set to 550 eV, and the energy convergence criterion was set to $10^{-6}$ eV. For the primitive cell, a $\Gamma$-centered 13$\times$13$\times$13  \emph{k}-point mesh \cite{monkhorst_special_1976} was applied; for supercells, the \emph{k}-point densities were kept approximately the same. 
The second-order displacement IDs and the second-order strain IDs were calculated using the lone irreducible derivative approach, and the third-order displacement IDs were calculated using the bundled irreducible derivative approach \cite{fu_group_2019}. All second-order displacement IDs (phonon dispersion) commensurate with the $4\times4\times4$ supercell were computed, while all third-order displacement IDs commensurate with the non-diagonal supercell  $\mathbf{\hat{S}}_\mathrm{O}$ were computed. The thermal conductivity is computed within the relaxation time approximation (RTA) using a $12\times12\times12$ $\textbf{q}$-mesh.
The formation energy $E_F$ of FPs is computed with the volume fixed, while the interstitial is located in the octahedral site. 

The GULP package \cite{gale_gulp_1997} was applied to calculate the static properties of EIPs, including lattice parameter, force, energy, dielectric constant, and Born effective charge. 
The Coulomb interactions were implemented using the Ewald summation  \cite{ewald_berechnung_1921}, where an 11.0 \AA\ cut-off was applied for both the pairwise and many-body interactions. 
For thermal expansion, MD simulations were performed by GULP in an NPT ensemble at $T$ up to $3000$ K using the non-diagonal supercell $4\mathbf{\hat{S}}_\mathrm{C}$. Each MD simulation runs 5 ps with a time step of 2 fs, while $a(T)$ is obtained by averaging the last 4 ps.  

The parameter optimization for the EIPs was carried out using the Potential Pro-Fit package \cite{potential-pro-fit}, which minimizes the total error $e$, given by:
\begin{equation}
    e=\sqrt{\sum(w_i e_i)^2},
\end{equation}
where $w_i$ and $e_i$ are the weight and error, respectively, for the $i$-th training data. The parameter optimization method was provided as a flowchart in Section II of SM \cite{Supplemental_Material}, and the weight of training data is tabulated in Table S1 of SM \cite{Supplemental_Material}. Throughout this work, we denote our trained EIP as the present work (PW), and its optimized parameter values are presented in Tables~\ref{tab:pairParameter} and \ref{tab:pairManybody}. To demonstrate the robustness of our training method, two alternative potential forms without the core-shell model have also been trained, and improvements of thermophysical predictions relative to the CRG potential are also observed (see in Section VII of SM \cite{Supplemental_Material}).

\section{\label{sec:results}Results and discussions}

\begin{figure*}[t]
    \centering
    \includegraphics[width=0.65\columnwidth]{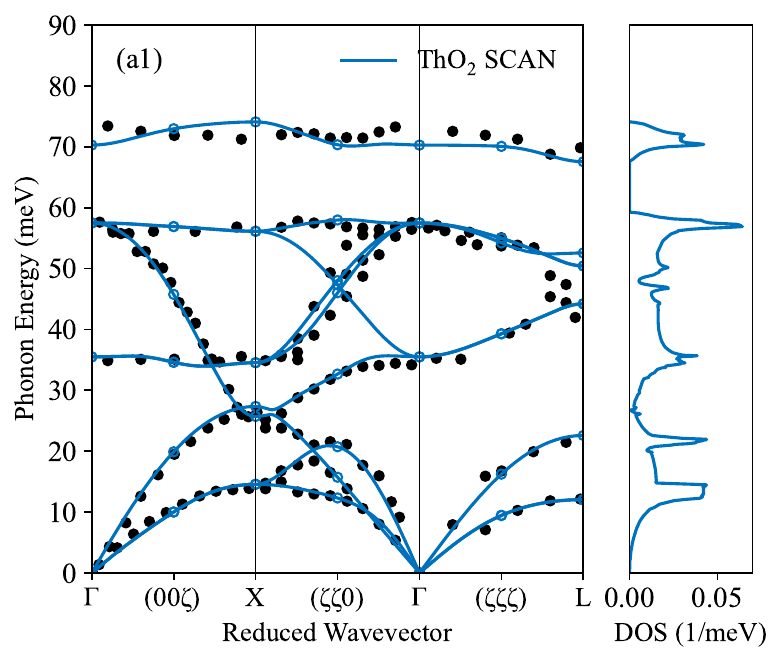}
    \includegraphics[width=0.65\columnwidth]{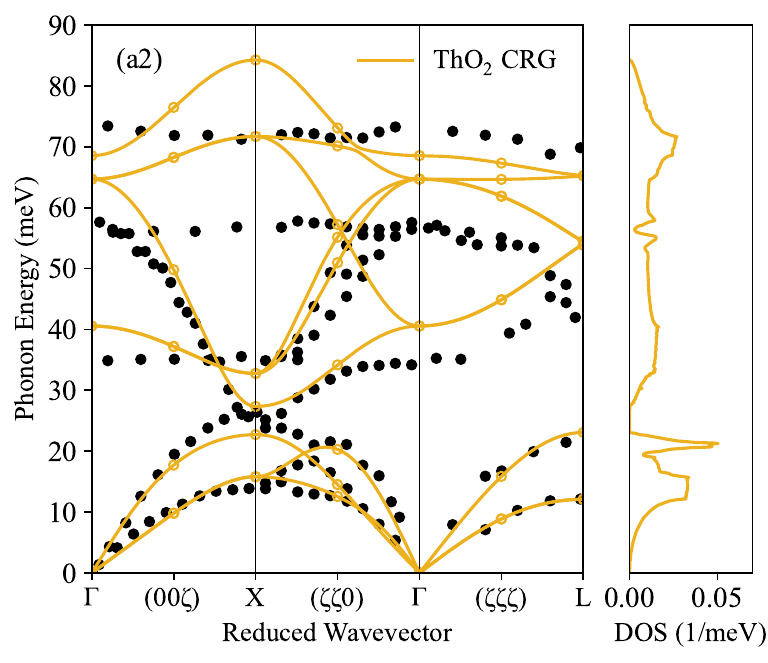}
    \includegraphics[width=0.65\columnwidth]{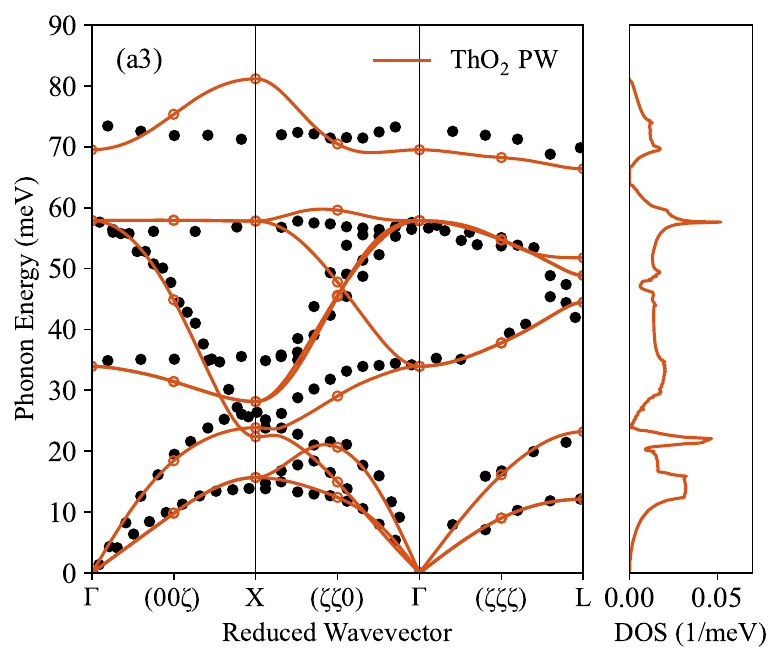}

    \includegraphics[width=0.65\columnwidth]{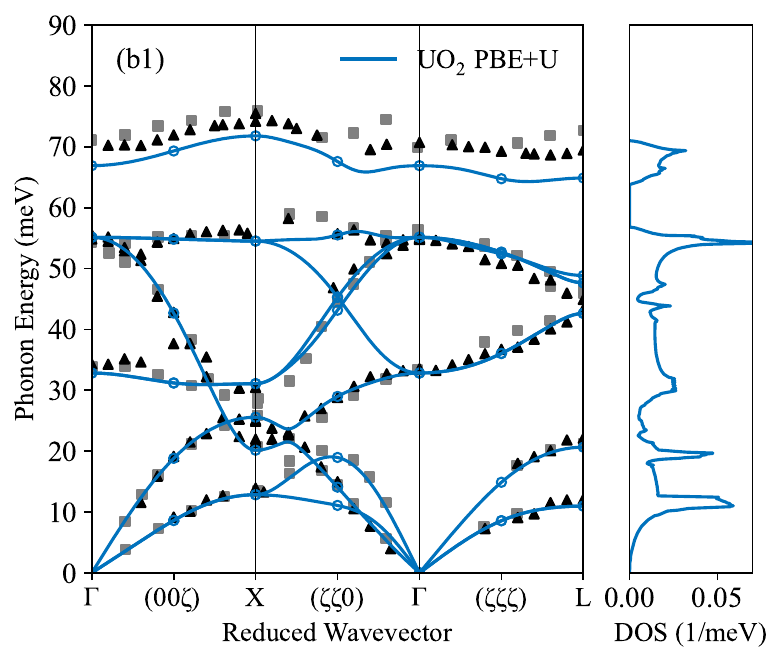}
    \includegraphics[width=0.65\columnwidth]{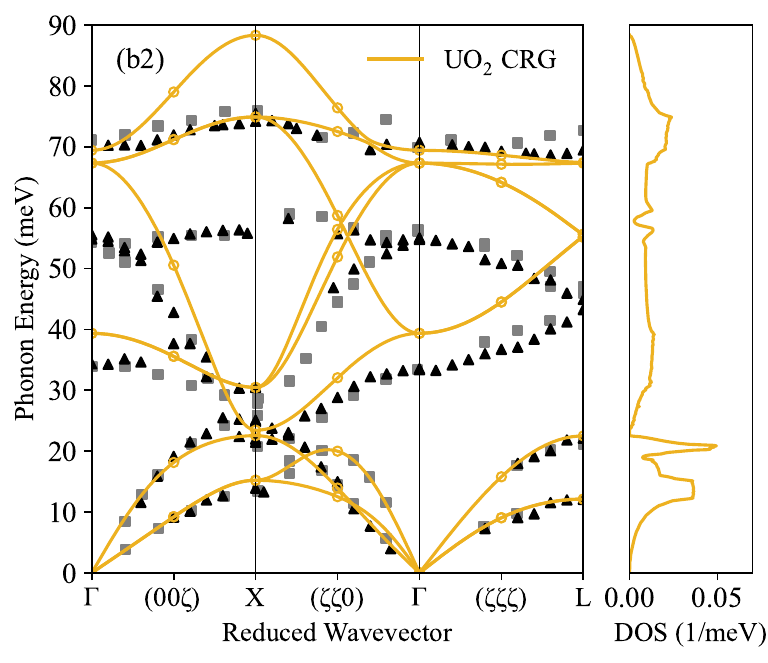}
    \includegraphics[width=0.65\columnwidth]{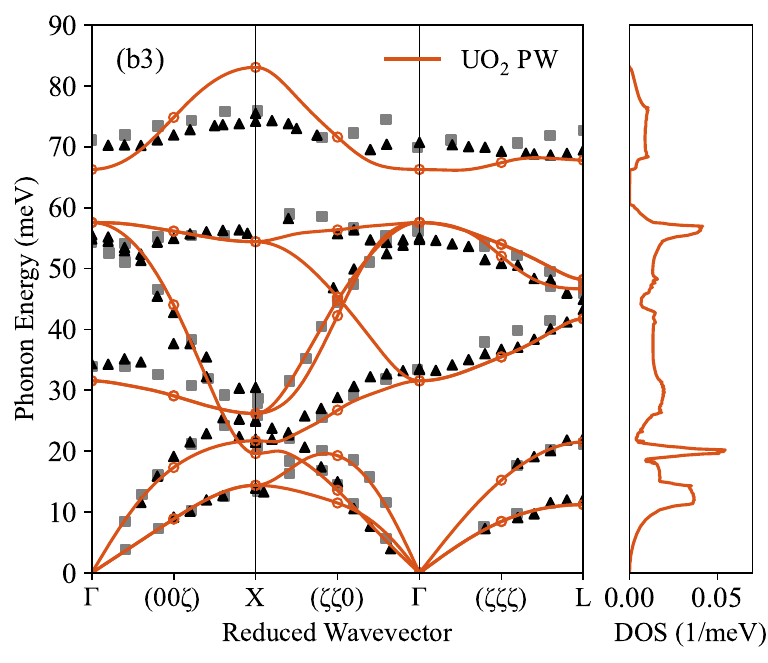}
    \caption{Calculated phonon dispersion and density of states for (a) ThO$_2$ and (b) UO$_2$, by using (1) DFT, (2) the CRG, and (3) the PW, compared to experimental data of ThO$_2$ (black circles) \cite{bryan_impact_2019} and UO$_2$ (grey squares \cite{pang_phonon_2013} and black triangles \cite{zhou_phonon_2024}). The lines are a Fourier interpolation of the computed data points, which are presented as open circles.}
    \label{fig:phononSpectra}
\end{figure*}


We begin by comparing the PW with the DFT training and assessment data to ensure that a good fit was achieved. For training data, the comparison is tabulated in Table~\ref{tab:agreementFitting}, except the third-order displacement IDs which are tabulated in Table S2 of SM \cite{Supplemental_Material}. For the assessment data, the comparison for each property can be found by using Table S1 of SM \cite{Supplemental_Material}, including the second- and third-order displacement IDs in larger supercells, the formation energy for FPs in larger supercells, and thermal conductivity computed by using the BTE within the RTA. Overall, our parameter optimization process yields an EIP with relatively small errors relative to the DFT data. Interestingly, while the PW only has two training data points for the defect formation energy (i.e., FP of Th or U in a conventional cell of ThO$_2$ or UO$_2$, respectively), the PW can predict the formation energies for FPs of O, Th, and U in a larger supercell ($2\times2\times2$ of the conventional cell) reasonably well, all within 9\% error compared with DFT (see Table~\ref{tab:elasticConstant}).
For the sake of comparison, all of the aforementioned properties are also computed using the CRG.


\begin{table*}[t]
\caption{\label{tab:elasticConstant}%
Elastic constants and defect formation energy $E_F$ for FPs calculated using DFT, the CRG, and the PW in comparison to experimental data \cite{macedo_elastic_1964,khanolkar2023temperature,brandt_temperature_1967}. See Section IV of SM for definitions of FP1 and FP2 \cite{Supplemental_Material}.
}
\begin{ruledtabular}
\begin{tabular}{@{}c
cccccccc
}
\multirow{ 2}{*}{Property} &
\multicolumn{4}{c}{ThO$_2$}&
\multicolumn{4}{c}{UO$_2$}\\
\cmidrule(lr){2-5}\cmidrule(lr){6-9}
& {SCAN} & {CRG} & {PW} & {exp} & {PBE+$U$} & {CRG} & {PW} & {exp}\\ [0.2em]
\hline\hline \\[-0.8em]
\multicolumn{9}{c}{Elastic constants (GPa)}\\[0.2em]
\colrule\\[-0.8em]
$ C_{11} $ 
& 376 & 352 & 371 &  367\textsuperscript{a}, 366\textsuperscript{b}  &  380  &  406  &  374 & 400\textsuperscript{c} \\[0.2em]
$ C_{12} $ 
& 117 & 113 & 119 &  106\textsuperscript{a}, 114\textsuperscript{b}  &  120  &  125 & 124 & 125\textsuperscript{c}  \\[0.2em]
$ C_{44} $ 
&  81 &  72 &  75 &  80\textsuperscript{a}, 81\textsuperscript{b}  &  63  & 64 & 61 & 59\textsuperscript{c}  \\[0.8em]

\multicolumn{9}{c}{Defect formation energy $E_F$ (eV)}\\[0.2em]
\colrule\\[-0.8em]
O FP1 & 4.55 & 5.25 & 4.78 &  & 4.02 & 5.36 & 4.30 & \\ [0.2em]
O FP2 & 4.63 & 5.61 & 4.95 &  & 4.06 & 5.77 & 4.41 & \\ [0.2em]
Th or U FP & 13.15 & 13.63 & 13.93 &  & 10.26 & 11.07 & 10.62 & 

\end{tabular}

\end{ruledtabular}
{\raggedright
\textsuperscript{a} Ref.~\cite{macedo_elastic_1964}, \textsuperscript{b} Ref.~\cite{khanolkar2023temperature}, \textsuperscript{c} Ref.~\cite{brandt_temperature_1967}.
\par}

\end{table*}

\normalsize

\begin{figure}[t]
    \centering
    \includegraphics[width=0.8\columnwidth]{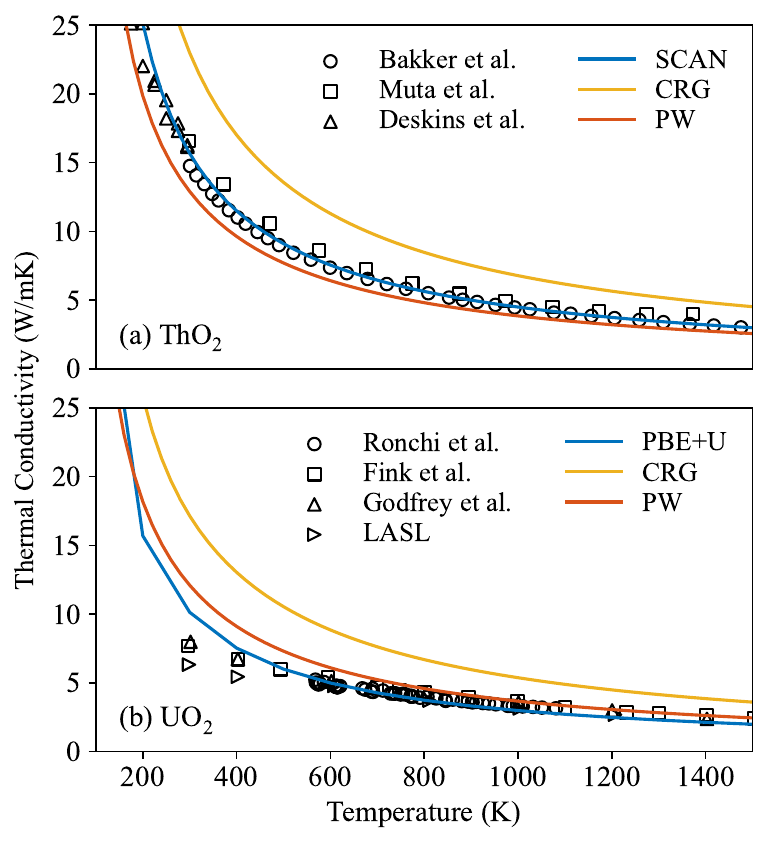}
    \caption{Calculated phonon thermal conductivity obtained from the BTE within the RTA for (a) ThO$_2$ and (b) UO$_2$, using DFT, the CRG, and the PW at $T=100-1500$ K, compared with the experimental data of ThO$_2$ \cite{deskins2022combined,muta_thermophysical_2013,bakker_critical_1997} and UO$_2$ \cite{finkThermophysicalPropertiesUranium2000, batesVisibleInfraredAbsorption1965, godfreyThermalConductivityUranium1965, ronchiEffectBurnupThermal2004}. }
    \label{fig:thermalConductivity}
\end{figure}

\begin{figure}[t]
    \centering
    \includegraphics[width=0.80\columnwidth]{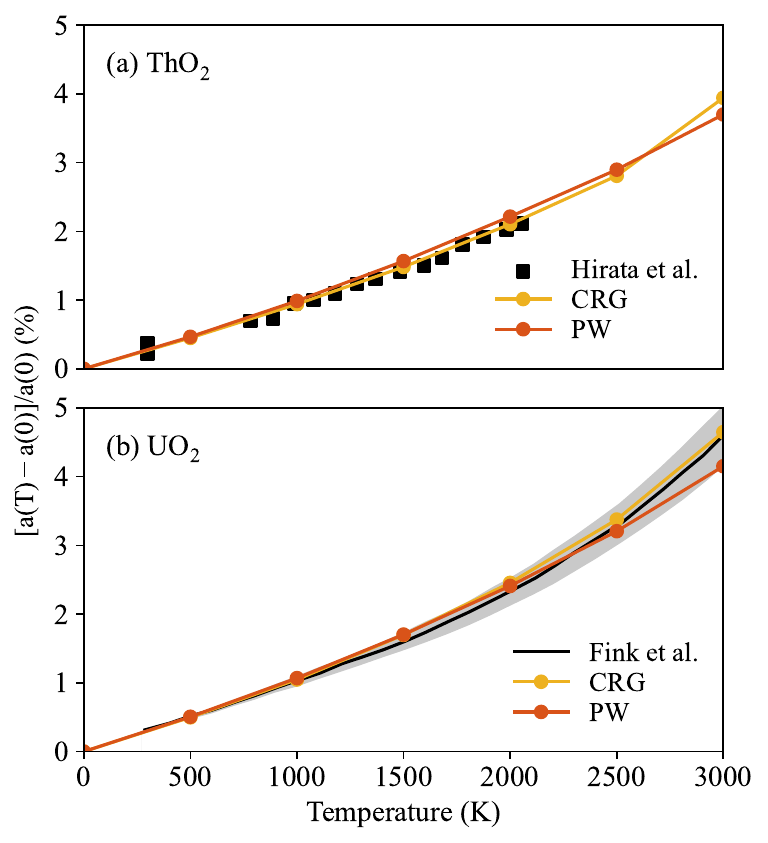}
    \caption{Calculated percentage change of the lattice parameter as a function of temperature at $T=0-3000$ K obtained from molecular dynamics simulations for (a) ThO$_2$ and (b) UO$_2$ using the CRG and the PW, compared with the experimental data of ThO$_2$ \cite{hirata_high_1977} and UO$_2$ \cite{finkThermophysicalPropertiesUranium2000}.}
    \label{fig:thermalExpansion}
\end{figure}

We proceed to assess the PW via comparison with experiments and the CRG potential, which is fitted to experimental elastic constants and thermal expansion. 
For the phonon dispersion, Figure~\ref{fig:phononSpectra} presents the comparison between DFT, EIPs, and experiments \cite{bryan_impact_2019, pang_phonon_2013, zhouCapturingGroundState2022a}.
The CRG potential exhibits two deficiencies: the overprediction of all optical phonon branches and the inability to capture the phonon frequency gap between the highest two optical branches. 
As DFT can accurately capture the phonon dispersion in comparison with experiments, the PW, which is fitted to DFT, significantly improves the agreement of the optical phonon branches. Furthermore, a gap between the highest two optical branches is successfully predicted, which can be attributed to the core-shell model (see Section VII of SM \cite{Supplemental_Material} for more details). The only major deficiency of the PW for the phonon dispersion is the overprediction of the highest optical branch near the $X$ point. However, as the highest phonon branch has a negligible contribution to the thermal conductivity in ThO$_2$ and UO$_2$ \cite{jin_assessment_2021, xiao2022validating, zhou_phonon_2024}, this error should not significantly affect the thermal transport (see Section VI of SM \cite{Supplemental_Material} for more details).
Table~\ref{tab:elasticConstant} tabulates the calculated elastic constants. Compared with experiments \cite{macedo_elastic_1964,khanolkar2023temperature, brandt_temperature_1967}, the PW has better predictions than the CRG for $C_{11}$ and $C_{44}$ in ThO$_2$, and $C_{44}$ in UO$_2$, though the CRG was directly fitted to experimental elastic constants. For the predictions of $C_{12}$ in ThO$_2$ and $C_{11}$ in UO$_2$, the PW is not as good as the CRG, but this is mainly due to the discrepancy between DFT and experiments. 
The thermal conductivity is computed and compared with experiments \cite{deskins2022combined,muta_thermophysical_2013,bakker_critical_1997, finkThermophysicalPropertiesUranium2000, batesVisibleInfraredAbsorption1965, godfreyThermalConductivityUranium1965, ronchiEffectBurnupThermal2004} in Figure~\ref{fig:thermalConductivity}. 
The PW significantly improves the thermal conductivity prediction as compared to the CRG,
which is expected given that DFT has good agreement with experiments. For example, 
at $T=1000$ K, the errors of predicted thermal conductivity by the CRG for ThO$_2$ and UO$_2$ are 54\% and 63\%, respectively, which are reduced to 17\% and 9\% by the PW, respectively, compared with experiments (see Section VI of SM \cite{Supplemental_Material} for more details).  
Finally, we compare the thermal expansion predictions with experiments \cite{hirata_high_1977, finkThermophysicalPropertiesUranium2000}. As the CRG and the PW predict different $a(0)$, rather than comparing $a$ directly, we compare the percentage change in lattice parameter, $\frac{a(T)-a(0)}{a(0)}$ (see in Figure~\ref{fig:thermalExpansion}), with experiments. 
The PW gives good agreement with experiment up to $T=2500$ K for both ThO$_2$ and UO$_2$, and slightly underpredicts the lattice parameter when $T>2500$ K. 
In summary, the PW yields good agreement with the DFT data of irreducible derivatives, $a(0)$, $Z^*_\alpha$, the formation energy of FPs, and thermal conductivity. Compared with experiments, the PW also outperforms the CRG for phonon dispersions, elastic constants, and thermal conductivity, except for certain elastic constants where DFT has discrepancies with experiments, and thermal expansion at high temperatures where there is likely insufficient training data.

\section{\label{sec:conclusion}Conclusion}

In this work, we developed an interatomic potential training approach by utilizing irreducible derivatives (IDs), including second- and third-order displacement IDs and second-order strain IDs, calculated from DFT. This ID-based potential training approach was used to construct an empirical interatomic potential (EIP) for ThO$_2$ and UO$_2$ crystals, yielding an EIP with relatively small errors relative to the DFT data, including the aforementioned IDs, phonon dispersion, thermal conductivity, and the formation energy of Frenkel pairs. Compared with experiments, the PW outperforms the widely-used CRG potential \cite{cooper_many-body_2014} for phonon dispersions and thermal conductivity. In addition to using a far more expansive training data set for our EIP, we also enhanced the analytical functional form for our EIP to include a core-shell model, which was essential for capturing selected optical phonon modes. Training EIPs based on IDs is clearly a promising direction for developing accurate interatomic potentials focussed on predicting phonon-related thermophysical properties, and this approach will likely be useful in the broader context of machine learning-based potentials.

\section{Acknowledgements}
This work was supported by the Center for Thermal Energy Transport under Irradiation, an Energy Frontier Research Center funded by the U.S. Department of Energy (DOE) Office of Basic Energy Sciences. This research made use of the resources of the High Performance Computing Center at Idaho National Laboratory, which is supported by the Office of Nuclear Energy of the U.S. Department of Energy and the Nuclear Science User Facilities under Contract No. DE-AC07-05ID14517. M.W.D.C. acknowledges support from the Nuclear Energy Advanced Modeling and Simulation (NEAMS) Program funded by the U.S. DOE Office of Nuclear Energy.
S.B., E.X., and C.A.M. acknowledge resources of the National Energy Research Scientific Computing Center, a DOE Office of Science User Facility supported by the Office of Science of the U.S. Department of Energy under Contract No. DE-AC02-05CH11231. Grant DE-SC0016507 supported S.B. and C.A.M. for integrating empirical potentials in the irreducible derivative framework.


\normalsize

\bibliographystyle{apsrev4-1} 
\bibliography{tho2,extra}

\begin{thebibliography}{90}%
\makeatletter
\providecommand \@ifxundefined [1]{%
 \@ifx{#1\undefined}
}%
\providecommand \@ifnum [1]{%
 \ifnum #1\expandafter \@firstoftwo
 \else \expandafter \@secondoftwo
 \fi
}%
\providecommand \@ifx [1]{%
 \ifx #1\expandafter \@firstoftwo
 \else \expandafter \@secondoftwo
 \fi
}%
\providecommand \natexlab [1]{#1}%
\providecommand \enquote  [1]{``#1''}%
\providecommand \bibnamefont  [1]{#1}%
\providecommand \bibfnamefont [1]{#1}%
\providecommand \citenamefont [1]{#1}%
\providecommand \href@noop [0]{\@secondoftwo}%
\providecommand \href [0]{\begingroup \@sanitize@url \@href}%
\providecommand \@href[1]{\@@startlink{#1}\@@href}%
\providecommand \@@href[1]{\endgroup#1\@@endlink}%
\providecommand \@sanitize@url [0]{\catcode `\\12\catcode `\$12\catcode
  `\&12\catcode `\#12\catcode `\^12\catcode `\_12\catcode `\%12\relax}%
\providecommand \@@startlink[1]{}%
\providecommand \@@endlink[0]{}%
\providecommand \url  [0]{\begingroup\@sanitize@url \@url }%
\providecommand \@url [1]{\endgroup\@href {#1}{\urlprefix }}%
\providecommand \urlprefix  [0]{URL }%
\providecommand \Eprint [0]{\href }%
\providecommand \doibase [0]{http://dx.doi.org/}%
\providecommand \selectlanguage [0]{\@gobble}%
\providecommand \bibinfo  [0]{\@secondoftwo}%
\providecommand \bibfield  [0]{\@secondoftwo}%
\providecommand \translation [1]{[#1]}%
\providecommand \BibitemOpen [0]{}%
\providecommand \bibitemStop [0]{}%
\providecommand \bibitemNoStop [0]{.\EOS\space}%
\providecommand \EOS [0]{\spacefactor3000\relax}%
\providecommand \BibitemShut  [1]{\csname bibitem#1\endcsname}%
\let\auto@bib@innerbib\@empty
\bibitem [{\citenamefont {Lennard-Jones}(1931)}]{lennard1931cohesion}%
  \BibitemOpen
  \bibfield  {author} {\bibinfo {author} {\bibfnamefont {J.~E.}\ \bibnamefont
  {Lennard-Jones}},\ }\href@noop {} {\bibfield  {journal} {\bibinfo  {journal}
  {Proceedings of the Physical Society}\ }\textbf {\bibinfo {volume} {43}},\
  \bibinfo {pages} {461} (\bibinfo {year} {1931})}\BibitemShut {NoStop}%
\bibitem [{\citenamefont {Tersoff}(1988)}]{tersoff_empirical_1988}%
  \BibitemOpen
  \bibfield  {author} {\bibinfo {author} {\bibfnamefont {J.}~\bibnamefont
  {Tersoff}},\ }\href {\doibase 10.1103/PhysRevB.38.9902} {\bibfield  {journal}
  {\bibinfo  {journal} {Physical Review B}\ }\textbf {\bibinfo {volume} {38}},\
  \bibinfo {pages} {9902} (\bibinfo {year} {1988})}\BibitemShut {NoStop}%
\bibitem [{\citenamefont {Daw}\ and\ \citenamefont
  {Baskes}(1984)}]{daw_embedded-atom_1984}%
  \BibitemOpen
  \bibfield  {author} {\bibinfo {author} {\bibfnamefont {M.~S.}\ \bibnamefont
  {Daw}}\ and\ \bibinfo {author} {\bibfnamefont {M.~I.}\ \bibnamefont
  {Baskes}},\ }\href {\doibase 10.1103/PhysRevB.29.6443} {\bibfield  {journal}
  {\bibinfo  {journal} {Physical Review B}\ }\textbf {\bibinfo {volume} {29}},\
  \bibinfo {pages} {6443} (\bibinfo {year} {1984})}\BibitemShut {NoStop}%
\bibitem [{\citenamefont {Behler}\ and\ \citenamefont
  {Parrinello}(2007)}]{behler2007generalized}%
  \BibitemOpen
  \bibfield  {author} {\bibinfo {author} {\bibfnamefont {J.}~\bibnamefont
  {Behler}}\ and\ \bibinfo {author} {\bibfnamefont {M.}~\bibnamefont
  {Parrinello}},\ }\href {\doibase 10.1103/PhysRevLett.98.146401} {\bibfield
  {journal} {\bibinfo  {journal} {Physical Review Letters}\ }\textbf {\bibinfo
  {volume} {98}},\ \bibinfo {pages} {146401} (\bibinfo {year}
  {2007})}\BibitemShut {NoStop}%
\bibitem [{\citenamefont {Bart{\'o}k}\ \emph {et~al.}(2010)\citenamefont
  {Bart{\'o}k}, \citenamefont {Payne}, \citenamefont {Kondor},\ and\
  \citenamefont {Cs{\'a}nyi}}]{bartok2010gaussian}%
  \BibitemOpen
  \bibfield  {author} {\bibinfo {author} {\bibfnamefont {A.~P.}\ \bibnamefont
  {Bart{\'o}k}}, \bibinfo {author} {\bibfnamefont {M.~C.}\ \bibnamefont
  {Payne}}, \bibinfo {author} {\bibfnamefont {R.}~\bibnamefont {Kondor}}, \
  and\ \bibinfo {author} {\bibfnamefont {G.}~\bibnamefont {Cs{\'a}nyi}},\
  }\href {\doibase 10.1103/PhysRevLett.104.136403} {\bibfield  {journal}
  {\bibinfo  {journal} {Physical Review Letters}\ }\textbf {\bibinfo {volume}
  {104}},\ \bibinfo {pages} {136403} (\bibinfo {year} {2010})}\BibitemShut
  {NoStop}%
\bibitem [{\citenamefont {Allen}\ and\ \citenamefont
  {Tildesley}(2017)}]{allen_computer_2017}%
  \BibitemOpen
  \bibfield  {author} {\bibinfo {author} {\bibfnamefont {M.~P.}\ \bibnamefont
  {Allen}}\ and\ \bibinfo {author} {\bibfnamefont {D.~J.}\ \bibnamefont
  {Tildesley}},\ }\href
  {https://oxford.universitypressscholarship.com/10.1093/oso/9780198803195.001.0001/oso-9780198803195}
  {\emph {\bibinfo {title} {Computer {Simulation} of {Liquids}: {Second}
  {Edition}}}}\ (\bibinfo {address} {Oxford},\ \bibinfo {year}
  {2017})\BibitemShut {NoStop}%
\bibitem [{\citenamefont {Cooper}\ \emph {et~al.}(2014)\citenamefont {Cooper},
  \citenamefont {Rushton},\ and\ \citenamefont
  {Grimes}}]{cooper_many-body_2014}%
  \BibitemOpen
  \bibfield  {author} {\bibinfo {author} {\bibfnamefont {M.~W.~D.}\
  \bibnamefont {Cooper}}, \bibinfo {author} {\bibfnamefont {M.~J.~D.}\
  \bibnamefont {Rushton}}, \ and\ \bibinfo {author} {\bibfnamefont {R.~W.}\
  \bibnamefont {Grimes}},\ }\href {\doibase 10.1088/0953-8984/26/10/105401}
  {\bibfield  {journal} {\bibinfo  {journal} {Journal of Physics: Condensed
  Matter}\ }\textbf {\bibinfo {volume} {26}},\ \bibinfo {pages} {105401}
  (\bibinfo {year} {2014})}\BibitemShut {NoStop}%
\bibitem [{\citenamefont {Chernatynskiy}\ \emph {et~al.}(2012)\citenamefont
  {Chernatynskiy}, \citenamefont {Flint}, \citenamefont {Sinnott},\ and\
  \citenamefont {Phillpot}}]{Chernatynskiy2012critical}%
  \BibitemOpen
  \bibfield  {author} {\bibinfo {author} {\bibfnamefont {A.}~\bibnamefont
  {Chernatynskiy}}, \bibinfo {author} {\bibfnamefont {C.}~\bibnamefont
  {Flint}}, \bibinfo {author} {\bibfnamefont {S.}~\bibnamefont {Sinnott}}, \
  and\ \bibinfo {author} {\bibfnamefont {S.}~\bibnamefont {Phillpot}},\ }\href
  {\doibase 10.1007/s10853-011-6230-0} {\bibfield  {journal} {\bibinfo
  {journal} {Journal of Materials Science}\ }\textbf {\bibinfo {volume} {47}},\
  \bibinfo {pages} {7693} (\bibinfo {year} {2012})}\BibitemShut {NoStop}%
\bibitem [{\citenamefont {Ercolessi}\ and\ \citenamefont
  {Adams}(1994)}]{ercolessi_interatomic_1994}%
  \BibitemOpen
  \bibfield  {author} {\bibinfo {author} {\bibfnamefont {F.}~\bibnamefont
  {Ercolessi}}\ and\ \bibinfo {author} {\bibfnamefont {J.~B.}\ \bibnamefont
  {Adams}},\ }\href {\doibase 10.1209/0295-5075/26/8/005} {\bibfield  {journal}
  {\bibinfo  {journal} {Europhysics Letters}\ }\textbf {\bibinfo {volume}
  {26}},\ \bibinfo {pages} {583} (\bibinfo {year} {1994})}\BibitemShut
  {NoStop}%
\bibitem [{\citenamefont {Mishin}\ \emph {et~al.}(1999)\citenamefont {Mishin},
  \citenamefont {Farkas}, \citenamefont {Mehl},\ and\ \citenamefont
  {Papaconstantopoulos}}]{mishin1999interatomic}%
  \BibitemOpen
  \bibfield  {author} {\bibinfo {author} {\bibfnamefont {Y.}~\bibnamefont
  {Mishin}}, \bibinfo {author} {\bibfnamefont {D.}~\bibnamefont {Farkas}},
  \bibinfo {author} {\bibfnamefont {M.}~\bibnamefont {Mehl}}, \ and\ \bibinfo
  {author} {\bibfnamefont {D.}~\bibnamefont {Papaconstantopoulos}},\
  }\href@noop {} {\bibfield  {journal} {\bibinfo  {journal} {Physical Review
  B}\ }\textbf {\bibinfo {volume} {59}},\ \bibinfo {pages} {3393} (\bibinfo
  {year} {1999})}\BibitemShut {NoStop}%
\bibitem [{\citenamefont {Fellinger}\ \emph {et~al.}(2010)\citenamefont
  {Fellinger}, \citenamefont {Park},\ and\ \citenamefont
  {Wilkins}}]{fellinger2010force}%
  \BibitemOpen
  \bibfield  {author} {\bibinfo {author} {\bibfnamefont {M.~R.}\ \bibnamefont
  {Fellinger}}, \bibinfo {author} {\bibfnamefont {H.}~\bibnamefont {Park}}, \
  and\ \bibinfo {author} {\bibfnamefont {J.~W.}\ \bibnamefont {Wilkins}},\
  }\href@noop {} {\bibfield  {journal} {\bibinfo  {journal} {Physical Review
  B—Condensed Matter and Materials Physics}\ }\textbf {\bibinfo {volume}
  {81}},\ \bibinfo {pages} {144119} (\bibinfo {year} {2010})}\BibitemShut
  {NoStop}%
\bibitem [{\citenamefont {Lee}\ and\ \citenamefont
  {Hwang}(2012)}]{lee2012force}%
  \BibitemOpen
  \bibfield  {author} {\bibinfo {author} {\bibfnamefont {Y.}~\bibnamefont
  {Lee}}\ and\ \bibinfo {author} {\bibfnamefont {G.~S.}\ \bibnamefont
  {Hwang}},\ }\href@noop {} {\bibfield  {journal} {\bibinfo  {journal}
  {Physical Review B—Condensed Matter and Materials Physics}\ }\textbf
  {\bibinfo {volume} {85}},\ \bibinfo {pages} {125204} (\bibinfo {year}
  {2012})}\BibitemShut {NoStop}%
\bibitem [{\citenamefont {Brommer}\ \emph {et~al.}(2015)\citenamefont
  {Brommer}, \citenamefont {Kiselev}, \citenamefont {Schopf}, \citenamefont
  {Beck}, \citenamefont {Roth},\ and\ \citenamefont
  {Trebin}}]{brommer_classical_2015}%
  \BibitemOpen
  \bibfield  {author} {\bibinfo {author} {\bibfnamefont {P.}~\bibnamefont
  {Brommer}}, \bibinfo {author} {\bibfnamefont {A.}~\bibnamefont {Kiselev}},
  \bibinfo {author} {\bibfnamefont {D.}~\bibnamefont {Schopf}}, \bibinfo
  {author} {\bibfnamefont {P.}~\bibnamefont {Beck}}, \bibinfo {author}
  {\bibfnamefont {J.}~\bibnamefont {Roth}}, \ and\ \bibinfo {author}
  {\bibfnamefont {H.-R.}\ \bibnamefont {Trebin}},\ }\href {\doibase
  10.1088/0965-0393/23/7/074002} {\bibfield  {journal} {\bibinfo  {journal}
  {Modelling and Simulation in Materials Science and Engineering}\ }\textbf
  {\bibinfo {volume} {23}},\ \bibinfo {pages} {074002} (\bibinfo {year}
  {2015})}\BibitemShut {NoStop}%
\bibitem [{\citenamefont {Schopf}\ \emph {et~al.}(2014)\citenamefont {Schopf},
  \citenamefont {Euchner},\ and\ \citenamefont
  {Trebin}}]{schopf_effective_2014}%
  \BibitemOpen
  \bibfield  {author} {\bibinfo {author} {\bibfnamefont {D.}~\bibnamefont
  {Schopf}}, \bibinfo {author} {\bibfnamefont {H.}~\bibnamefont {Euchner}}, \
  and\ \bibinfo {author} {\bibfnamefont {H.-R.}\ \bibnamefont {Trebin}},\
  }\href {\doibase 10.1103/PhysRevB.89.214306} {\bibfield  {journal} {\bibinfo
  {journal} {Physical Review B}\ }\textbf {\bibinfo {volume} {89}},\ \bibinfo
  {pages} {214306} (\bibinfo {year} {2014})}\BibitemShut {NoStop}%
\bibitem [{\citenamefont {Fan}\ \emph {et~al.}(2019{\natexlab{a}})\citenamefont
  {Fan}, \citenamefont {Wang}, \citenamefont {Gu}, \citenamefont {Qian},
  \citenamefont {Su},\ and\ \citenamefont {Ala-Nissila}}]{fan_minimal_2019}%
  \BibitemOpen
  \bibfield  {author} {\bibinfo {author} {\bibfnamefont {Z.}~\bibnamefont
  {Fan}}, \bibinfo {author} {\bibfnamefont {Y.}~\bibnamefont {Wang}}, \bibinfo
  {author} {\bibfnamefont {X.}~\bibnamefont {Gu}}, \bibinfo {author}
  {\bibfnamefont {P.}~\bibnamefont {Qian}}, \bibinfo {author} {\bibfnamefont
  {Y.}~\bibnamefont {Su}}, \ and\ \bibinfo {author} {\bibfnamefont
  {T.}~\bibnamefont {Ala-Nissila}},\ }\href {\doibase 10.1088/1361-648X/ab5c5f}
  {\bibfield  {journal} {\bibinfo  {journal} {Journal of Physics: Condensed
  Matter}\ }\textbf {\bibinfo {volume} {32}},\ \bibinfo {pages} {135901}
  (\bibinfo {year} {2019}{\natexlab{a}})}\BibitemShut {NoStop}%
\bibitem [{\citenamefont {Qiu}\ and\ \citenamefont
  {Ruan}(2009)}]{qiu_molecular_2009}%
  \BibitemOpen
  \bibfield  {author} {\bibinfo {author} {\bibfnamefont {B.}~\bibnamefont
  {Qiu}}\ and\ \bibinfo {author} {\bibfnamefont {X.}~\bibnamefont {Ruan}},\
  }\href {\doibase 10.1103/PhysRevB.80.165203} {\bibfield  {journal} {\bibinfo
  {journal} {Physical Review B}\ }\textbf {\bibinfo {volume} {80}},\ \bibinfo
  {pages} {165203} (\bibinfo {year} {2009})}\BibitemShut {NoStop}%
\bibitem [{\citenamefont {Roy~Chowdhury}\ \emph {et~al.}(2019)\citenamefont
  {Roy~Chowdhury}, \citenamefont {Feng},\ and\ \citenamefont
  {Ruan}}]{roy_chowdhury_development_2019}%
  \BibitemOpen
  \bibfield  {author} {\bibinfo {author} {\bibfnamefont {P.}~\bibnamefont
  {Roy~Chowdhury}}, \bibinfo {author} {\bibfnamefont {T.}~\bibnamefont {Feng}},
  \ and\ \bibinfo {author} {\bibfnamefont {X.}~\bibnamefont {Ruan}},\ }\href
  {\doibase 10.1103/PhysRevB.99.155202} {\bibfield  {journal} {\bibinfo
  {journal} {Physical Review B}\ }\textbf {\bibinfo {volume} {99}},\ \bibinfo
  {pages} {155202} (\bibinfo {year} {2019})}\BibitemShut {NoStop}%
\bibitem [{\citenamefont {Fan}\ \emph {et~al.}(2019{\natexlab{b}})\citenamefont
  {Fan}, \citenamefont {Wang}, \citenamefont {Gu}, \citenamefont {Qian},
  \citenamefont {Su},\ and\ \citenamefont {Ala-Nissila}}]{fan2019minimal}%
  \BibitemOpen
  \bibfield  {author} {\bibinfo {author} {\bibfnamefont {Z.}~\bibnamefont
  {Fan}}, \bibinfo {author} {\bibfnamefont {Y.}~\bibnamefont {Wang}}, \bibinfo
  {author} {\bibfnamefont {X.}~\bibnamefont {Gu}}, \bibinfo {author}
  {\bibfnamefont {P.}~\bibnamefont {Qian}}, \bibinfo {author} {\bibfnamefont
  {Y.}~\bibnamefont {Su}}, \ and\ \bibinfo {author} {\bibfnamefont
  {T.}~\bibnamefont {Ala-Nissila}},\ }\href@noop {} {\bibfield  {journal}
  {\bibinfo  {journal} {Journal of Physics: Condensed Matter}\ }\textbf
  {\bibinfo {volume} {32}},\ \bibinfo {pages} {135901} (\bibinfo {year}
  {2019}{\natexlab{b}})}\BibitemShut {NoStop}%
\bibitem [{\citenamefont {Tanaka}\ \emph {et~al.}(2023)\citenamefont {Tanaka},
  \citenamefont {Sakai}, \citenamefont {Taniguchi}, \citenamefont {Shimomai},\
  and\ \citenamefont {Iwazaki}}]{tanaka2023interatomic}%
  \BibitemOpen
  \bibfield  {author} {\bibinfo {author} {\bibfnamefont {K.}~\bibnamefont
  {Tanaka}}, \bibinfo {author} {\bibfnamefont {Y.}~\bibnamefont {Sakai}},
  \bibinfo {author} {\bibfnamefont {S.}~\bibnamefont {Taniguchi}}, \bibinfo
  {author} {\bibfnamefont {K.}~\bibnamefont {Shimomai}}, \ and\ \bibinfo
  {author} {\bibfnamefont {Y.}~\bibnamefont {Iwazaki}},\ }\href@noop {}
  {\bibfield  {journal} {\bibinfo  {journal} {Journal of the Ceramic Society of
  Japan}\ }\textbf {\bibinfo {volume} {131}},\ \bibinfo {pages} {252} (\bibinfo
  {year} {2023})}\BibitemShut {NoStop}%
\bibitem [{\citenamefont {Cooper}\ \emph {et~al.}(2016)\citenamefont {Cooper},
  \citenamefont {Kuganathan}, \citenamefont {Burr}, \citenamefont {Rushton},
  \citenamefont {Grimes}, \citenamefont {Stanek},\ and\ \citenamefont
  {Andersson}}]{cooper_development_2016}%
  \BibitemOpen
  \bibfield  {author} {\bibinfo {author} {\bibfnamefont {M.~W.~D.}\
  \bibnamefont {Cooper}}, \bibinfo {author} {\bibfnamefont {N.}~\bibnamefont
  {Kuganathan}}, \bibinfo {author} {\bibfnamefont {P.~A.}\ \bibnamefont
  {Burr}}, \bibinfo {author} {\bibfnamefont {M.~J.~D.}\ \bibnamefont
  {Rushton}}, \bibinfo {author} {\bibfnamefont {R.~W.}\ \bibnamefont {Grimes}},
  \bibinfo {author} {\bibfnamefont {C.~R.}\ \bibnamefont {Stanek}}, \ and\
  \bibinfo {author} {\bibfnamefont {D.~A.}\ \bibnamefont {Andersson}},\ }\href
  {\doibase 10.1088/0953-8984/28/40/405401} {\bibfield  {journal} {\bibinfo
  {journal} {Journal of Physics: Condensed Matter}\ }\textbf {\bibinfo {volume}
  {28}},\ \bibinfo {pages} {405401} (\bibinfo {year} {2016})}\BibitemShut
  {NoStop}%
\bibitem [{\citenamefont {Novikov}\ \emph {et~al.}(2020)\citenamefont
  {Novikov}, \citenamefont {Gubaev}, \citenamefont {Podryabinkin},\ and\
  \citenamefont {Shapeev}}]{novikov2020mlip}%
  \BibitemOpen
  \bibfield  {author} {\bibinfo {author} {\bibfnamefont {I.~S.}\ \bibnamefont
  {Novikov}}, \bibinfo {author} {\bibfnamefont {K.}~\bibnamefont {Gubaev}},
  \bibinfo {author} {\bibfnamefont {E.~V.}\ \bibnamefont {Podryabinkin}}, \
  and\ \bibinfo {author} {\bibfnamefont {A.~V.}\ \bibnamefont {Shapeev}},\
  }\href@noop {} {\bibfield  {journal} {\bibinfo  {journal} {Machine Learning:
  Science and Technology}\ }\textbf {\bibinfo {volume} {2}},\ \bibinfo {pages}
  {025002} (\bibinfo {year} {2020})}\BibitemShut {NoStop}%
\bibitem [{\citenamefont {Lindsay}\ and\ \citenamefont
  {Broido}(2010)}]{lindsay2010optimized}%
  \BibitemOpen
  \bibfield  {author} {\bibinfo {author} {\bibfnamefont {L.}~\bibnamefont
  {Lindsay}}\ and\ \bibinfo {author} {\bibfnamefont {D.}~\bibnamefont
  {Broido}},\ }\href@noop {} {\bibfield  {journal} {\bibinfo  {journal}
  {Physical Review B—Condensed Matter and Materials Physics}\ }\textbf
  {\bibinfo {volume} {81}},\ \bibinfo {pages} {205441} (\bibinfo {year}
  {2010})}\BibitemShut {NoStop}%
\bibitem [{\citenamefont {Murakami}\ \emph {et~al.}(2013)\citenamefont
  {Murakami}, \citenamefont {Shiga}, \citenamefont {Hori}, \citenamefont
  {Esfarjani},\ and\ \citenamefont {Shiomi}}]{murakami_importance_2013}%
  \BibitemOpen
  \bibfield  {author} {\bibinfo {author} {\bibfnamefont {T.}~\bibnamefont
  {Murakami}}, \bibinfo {author} {\bibfnamefont {T.}~\bibnamefont {Shiga}},
  \bibinfo {author} {\bibfnamefont {T.}~\bibnamefont {Hori}}, \bibinfo {author}
  {\bibfnamefont {K.}~\bibnamefont {Esfarjani}}, \ and\ \bibinfo {author}
  {\bibfnamefont {J.}~\bibnamefont {Shiomi}},\ }\href {\doibase
  10.1209/0295-5075/102/46002} {\bibfield  {journal} {\bibinfo  {journal} {EPL
  (Europhysics Letters)}\ }\textbf {\bibinfo {volume} {102}},\ \bibinfo {pages}
  {46002} (\bibinfo {year} {2013})}\BibitemShut {NoStop}%
\bibitem [{\citenamefont {Han}\ and\ \citenamefont
  {Bester}(2011)}]{han_interatomic_2011}%
  \BibitemOpen
  \bibfield  {author} {\bibinfo {author} {\bibfnamefont {P.}~\bibnamefont
  {Han}}\ and\ \bibinfo {author} {\bibfnamefont {G.}~\bibnamefont {Bester}},\
  }\href {\doibase 10.1103/PhysRevB.83.174304} {\bibfield  {journal} {\bibinfo
  {journal} {Physical Review B}\ }\textbf {\bibinfo {volume} {83}},\ \bibinfo
  {pages} {174304} (\bibinfo {year} {2011})}\BibitemShut {NoStop}%
\bibitem [{\citenamefont {Rohskopf}\ \emph {et~al.}(2017)\citenamefont
  {Rohskopf}, \citenamefont {Seyf}, \citenamefont {Gordiz}, \citenamefont
  {Tadano},\ and\ \citenamefont {Henry}}]{rohskopf_empirical_2017}%
  \BibitemOpen
  \bibfield  {author} {\bibinfo {author} {\bibfnamefont {A.}~\bibnamefont
  {Rohskopf}}, \bibinfo {author} {\bibfnamefont {H.~R.}\ \bibnamefont {Seyf}},
  \bibinfo {author} {\bibfnamefont {K.}~\bibnamefont {Gordiz}}, \bibinfo
  {author} {\bibfnamefont {T.}~\bibnamefont {Tadano}}, \ and\ \bibinfo {author}
  {\bibfnamefont {A.}~\bibnamefont {Henry}},\ }\href {\doibase
  10.1038/s41524-017-0026-y} {\bibfield  {journal} {\bibinfo  {journal} {npj
  Computational Materials}\ }\textbf {\bibinfo {volume} {3}},\ \bibinfo {pages}
  {1} (\bibinfo {year} {2017})}\BibitemShut {NoStop}%
\bibitem [{\citenamefont {Muraleedharan}\ \emph {et~al.}(2017)\citenamefont
  {Muraleedharan}, \citenamefont {Rohskopf}, \citenamefont {Yang},\ and\
  \citenamefont {Henry}}]{muraleedharan2017phonon}%
  \BibitemOpen
  \bibfield  {author} {\bibinfo {author} {\bibfnamefont {M.~G.}\ \bibnamefont
  {Muraleedharan}}, \bibinfo {author} {\bibfnamefont {A.}~\bibnamefont
  {Rohskopf}}, \bibinfo {author} {\bibfnamefont {V.}~\bibnamefont {Yang}}, \
  and\ \bibinfo {author} {\bibfnamefont {A.}~\bibnamefont {Henry}},\
  }\href@noop {} {\bibfield  {journal} {\bibinfo  {journal} {AIP Advances}\
  }\textbf {\bibinfo {volume} {7}} (\bibinfo {year} {2017})}\BibitemShut
  {NoStop}%
\bibitem [{\citenamefont {Rohskopf}\ \emph {et~al.}(2020)\citenamefont
  {Rohskopf}, \citenamefont {Wyant}, \citenamefont {Gordiz}, \citenamefont
  {Seyf}, \citenamefont {Muraleedharan},\ and\ \citenamefont
  {Henry}}]{rohskopf2020fast}%
  \BibitemOpen
  \bibfield  {author} {\bibinfo {author} {\bibfnamefont {A.}~\bibnamefont
  {Rohskopf}}, \bibinfo {author} {\bibfnamefont {S.}~\bibnamefont {Wyant}},
  \bibinfo {author} {\bibfnamefont {K.}~\bibnamefont {Gordiz}}, \bibinfo
  {author} {\bibfnamefont {H.~R.}\ \bibnamefont {Seyf}}, \bibinfo {author}
  {\bibfnamefont {M.~G.}\ \bibnamefont {Muraleedharan}}, \ and\ \bibinfo
  {author} {\bibfnamefont {A.}~\bibnamefont {Henry}},\ }\href@noop {}
  {\bibfield  {journal} {\bibinfo  {journal} {Computational Materials Science}\
  }\textbf {\bibinfo {volume} {184}},\ \bibinfo {pages} {109884} (\bibinfo
  {year} {2020})}\BibitemShut {NoStop}%
\bibitem [{\citenamefont {Shi}\ \emph {et~al.}(2021)\citenamefont {Shi},
  \citenamefont {Ma}, \citenamefont {Li}, \citenamefont {Zhong}, \citenamefont
  {Yang}, \citenamefont {Yin},\ and\ \citenamefont {He}}]{shi2021molecular}%
  \BibitemOpen
  \bibfield  {author} {\bibinfo {author} {\bibfnamefont {L.}~\bibnamefont
  {Shi}}, \bibinfo {author} {\bibfnamefont {X.}~\bibnamefont {Ma}}, \bibinfo
  {author} {\bibfnamefont {M.}~\bibnamefont {Li}}, \bibinfo {author}
  {\bibfnamefont {Y.}~\bibnamefont {Zhong}}, \bibinfo {author} {\bibfnamefont
  {L.}~\bibnamefont {Yang}}, \bibinfo {author} {\bibfnamefont {W.}~\bibnamefont
  {Yin}}, \ and\ \bibinfo {author} {\bibfnamefont {X.}~\bibnamefont {He}},\
  }\href@noop {} {\bibfield  {journal} {\bibinfo  {journal} {Physical Chemistry
  Chemical Physics}\ }\textbf {\bibinfo {volume} {23}},\ \bibinfo {pages}
  {8336} (\bibinfo {year} {2021})}\BibitemShut {NoStop}%
\bibitem [{\citenamefont {Loew}\ \emph {et~al.}(2024)\citenamefont {Loew},
  \citenamefont {Wang}, \citenamefont {Cerqueira},\ and\ \citenamefont
  {Marques}}]{loew2024training}%
  \BibitemOpen
  \bibfield  {author} {\bibinfo {author} {\bibfnamefont {A.}~\bibnamefont
  {Loew}}, \bibinfo {author} {\bibfnamefont {H.-C.}\ \bibnamefont {Wang}},
  \bibinfo {author} {\bibfnamefont {T.~F.}\ \bibnamefont {Cerqueira}}, \ and\
  \bibinfo {author} {\bibfnamefont {M.~A.}\ \bibnamefont {Marques}},\
  }\href@noop {} {\bibfield  {journal} {\bibinfo  {journal} {Machine Learning:
  Science and Technology}\ }\textbf {\bibinfo {volume} {5}},\ \bibinfo {pages}
  {045019} (\bibinfo {year} {2024})}\BibitemShut {NoStop}%
\bibitem [{\citenamefont {Fang}\ \emph {et~al.}(2024)\citenamefont {Fang},
  \citenamefont {Geiger}, \citenamefont {Checkelsky},\ and\ \citenamefont
  {Smidt}}]{fang_phonon_2024}%
  \BibitemOpen
  \bibfield  {author} {\bibinfo {author} {\bibfnamefont {S.}~\bibnamefont
  {Fang}}, \bibinfo {author} {\bibfnamefont {M.}~\bibnamefont {Geiger}},
  \bibinfo {author} {\bibfnamefont {J.~G.}\ \bibnamefont {Checkelsky}}, \ and\
  \bibinfo {author} {\bibfnamefont {T.}~\bibnamefont {Smidt}},\ }\href
  {\doibase 10.48550/arXiv.2403.11347} {\enquote {\bibinfo {title} {Phonon
  predictions with {E}(3)-equivariant graph neural networks},}\ } (\bibinfo
  {year} {2024}),\ \bibinfo {note} {arXiv:2403.11347 [cond-mat]}\BibitemShut
  {NoStop}%
\bibitem [{\citenamefont {Fu}\ \emph {et~al.}(2019)\citenamefont {Fu},
  \citenamefont {Kornbluth}, \citenamefont {Cheng},\ and\ \citenamefont
  {Marianetti}}]{fu_group_2019}%
  \BibitemOpen
  \bibfield  {author} {\bibinfo {author} {\bibfnamefont {L.}~\bibnamefont
  {Fu}}, \bibinfo {author} {\bibfnamefont {M.}~\bibnamefont {Kornbluth}},
  \bibinfo {author} {\bibfnamefont {Z.}~\bibnamefont {Cheng}}, \ and\ \bibinfo
  {author} {\bibfnamefont {C.~A.}\ \bibnamefont {Marianetti}},\ }\href
  {\doibase 10.1103/PhysRevB.100.014303} {\bibfield  {journal} {\bibinfo
  {journal} {Physical Review B}\ }\textbf {\bibinfo {volume} {100}},\ \bibinfo
  {pages} {014303} (\bibinfo {year} {2019})}\BibitemShut {NoStop}%
\bibitem [{\citenamefont {Hurley}\ \emph {et~al.}(2022)\citenamefont {Hurley},
  \citenamefont {El-Azab}, \citenamefont {Bryan}, \citenamefont {Cooper},
  \citenamefont {Dennett}, \citenamefont {Gofryk}, \citenamefont {He},
  \citenamefont {Khafizov}, \citenamefont {Lander}, \citenamefont {Manley},
  \citenamefont {Mann}, \citenamefont {Marianetti}, \citenamefont {Rickert},
  \citenamefont {Selim}, \citenamefont {Tonks},\ and\ \citenamefont
  {Wharry}}]{hurley2022thermal}%
  \BibitemOpen
  \bibfield  {author} {\bibinfo {author} {\bibfnamefont {D.~H.}\ \bibnamefont
  {Hurley}}, \bibinfo {author} {\bibfnamefont {A.}~\bibnamefont {El-Azab}},
  \bibinfo {author} {\bibfnamefont {M.~S.}\ \bibnamefont {Bryan}}, \bibinfo
  {author} {\bibfnamefont {M.~W.~D.}\ \bibnamefont {Cooper}}, \bibinfo {author}
  {\bibfnamefont {C.~A.}\ \bibnamefont {Dennett}}, \bibinfo {author}
  {\bibfnamefont {K.}~\bibnamefont {Gofryk}}, \bibinfo {author} {\bibfnamefont
  {L.}~\bibnamefont {He}}, \bibinfo {author} {\bibfnamefont {M.}~\bibnamefont
  {Khafizov}}, \bibinfo {author} {\bibfnamefont {G.~H.}\ \bibnamefont
  {Lander}}, \bibinfo {author} {\bibfnamefont {M.~E.}\ \bibnamefont {Manley}},
  \bibinfo {author} {\bibfnamefont {J.~M.}\ \bibnamefont {Mann}}, \bibinfo
  {author} {\bibfnamefont {C.~A.}\ \bibnamefont {Marianetti}}, \bibinfo
  {author} {\bibfnamefont {K.}~\bibnamefont {Rickert}}, \bibinfo {author}
  {\bibfnamefont {F.~A.}\ \bibnamefont {Selim}}, \bibinfo {author}
  {\bibfnamefont {M.~R.}\ \bibnamefont {Tonks}}, \ and\ \bibinfo {author}
  {\bibfnamefont {J.~P.}\ \bibnamefont {Wharry}},\ }\href {\doibase
  10.1021/acs.chemrev.1c00262} {\bibfield  {journal} {\bibinfo  {journal}
  {Chemical Reviews}\ }\textbf {\bibinfo {volume} {122}},\ \bibinfo {pages}
  {3711} (\bibinfo {year} {2022})},\ \bibinfo {note} {pMID:
  34919381}\BibitemShut {NoStop}%
\bibitem [{\citenamefont {Jin}\ \emph {et~al.}(2020)\citenamefont {Jin},
  \citenamefont {Jiang}, \citenamefont {Gan},\ and\ \citenamefont
  {Hurley}}]{jin2020systematic}%
  \BibitemOpen
  \bibfield  {author} {\bibinfo {author} {\bibfnamefont {M.}~\bibnamefont
  {Jin}}, \bibinfo {author} {\bibfnamefont {C.}~\bibnamefont {Jiang}}, \bibinfo
  {author} {\bibfnamefont {J.}~\bibnamefont {Gan}}, \ and\ \bibinfo {author}
  {\bibfnamefont {D.~H.}\ \bibnamefont {Hurley}},\ }\href@noop {} {\bibfield
  {journal} {\bibinfo  {journal} {Journal of Nuclear Materials}\ }\textbf
  {\bibinfo {volume} {536}},\ \bibinfo {pages} {152144} (\bibinfo {year}
  {2020})}\BibitemShut {NoStop}%
\bibitem [{\citenamefont {Dennett}\ \emph {et~al.}(2021)\citenamefont
  {Dennett}, \citenamefont {Deskins}, \citenamefont {Khafizov}, \citenamefont
  {Hua}, \citenamefont {Khanolkar}, \citenamefont {Bawane}, \citenamefont {Fu},
  \citenamefont {Mann}, \citenamefont {Marianetti}, \citenamefont {He},
  \citenamefont {Hurley},\ and\ \citenamefont
  {El-Azab}}]{dennett_integrated_2021}%
  \BibitemOpen
  \bibfield  {author} {\bibinfo {author} {\bibfnamefont {C.~A.}\ \bibnamefont
  {Dennett}}, \bibinfo {author} {\bibfnamefont {W.~R.}\ \bibnamefont
  {Deskins}}, \bibinfo {author} {\bibfnamefont {M.}~\bibnamefont {Khafizov}},
  \bibinfo {author} {\bibfnamefont {Z.}~\bibnamefont {Hua}}, \bibinfo {author}
  {\bibfnamefont {A.}~\bibnamefont {Khanolkar}}, \bibinfo {author}
  {\bibfnamefont {K.}~\bibnamefont {Bawane}}, \bibinfo {author} {\bibfnamefont
  {L.}~\bibnamefont {Fu}}, \bibinfo {author} {\bibfnamefont {J.~M.}\
  \bibnamefont {Mann}}, \bibinfo {author} {\bibfnamefont {C.~A.}\ \bibnamefont
  {Marianetti}}, \bibinfo {author} {\bibfnamefont {L.}~\bibnamefont {He}},
  \bibinfo {author} {\bibfnamefont {D.~H.}\ \bibnamefont {Hurley}}, \ and\
  \bibinfo {author} {\bibfnamefont {A.}~\bibnamefont {El-Azab}},\ }\href
  {\doibase 10.1016/j.actamat.2021.116934} {\bibfield  {journal} {\bibinfo
  {journal} {Acta Materialia}\ }\textbf {\bibinfo {volume} {213}},\ \bibinfo
  {pages} {116934} (\bibinfo {year} {2021})}\BibitemShut {NoStop}%
\bibitem [{\citenamefont {He}\ \emph {et~al.}(2021)\citenamefont {He},
  \citenamefont {Khafizov}, \citenamefont {Jiang}, \citenamefont
  {Tyburska-Püschel}, \citenamefont {Jaques}, \citenamefont {Xiu},
  \citenamefont {Xu}, \citenamefont {Meyer}, \citenamefont {Sridharan},
  \citenamefont {Butt},\ and\ \citenamefont {Gan}}]{He2022phase}%
  \BibitemOpen
  \bibfield  {author} {\bibinfo {author} {\bibfnamefont {L.}~\bibnamefont
  {He}}, \bibinfo {author} {\bibfnamefont {M.}~\bibnamefont {Khafizov}},
  \bibinfo {author} {\bibfnamefont {C.}~\bibnamefont {Jiang}}, \bibinfo
  {author} {\bibfnamefont {B.}~\bibnamefont {Tyburska-Püschel}}, \bibinfo
  {author} {\bibfnamefont {B.~J.}\ \bibnamefont {Jaques}}, \bibinfo {author}
  {\bibfnamefont {P.}~\bibnamefont {Xiu}}, \bibinfo {author} {\bibfnamefont
  {P.}~\bibnamefont {Xu}}, \bibinfo {author} {\bibfnamefont {M.~K.}\
  \bibnamefont {Meyer}}, \bibinfo {author} {\bibfnamefont {K.}~\bibnamefont
  {Sridharan}}, \bibinfo {author} {\bibfnamefont {D.~P.}\ \bibnamefont {Butt}},
  \ and\ \bibinfo {author} {\bibfnamefont {J.}~\bibnamefont {Gan}},\ }\href
  {\doibase https://doi.org/10.1016/j.actamat.2021.116778} {\bibfield
  {journal} {\bibinfo  {journal} {Acta Materialia}\ }\textbf {\bibinfo {volume}
  {208}},\ \bibinfo {pages} {116778} (\bibinfo {year} {2021})}\BibitemShut
  {NoStop}%
\bibitem [{\citenamefont {Khafizov}\ \emph {et~al.}(2019)\citenamefont
  {Khafizov}, \citenamefont {Pakarinen}, \citenamefont {He},\ and\
  \citenamefont {Hurley}}]{khafizov2019impact}%
  \BibitemOpen
  \bibfield  {author} {\bibinfo {author} {\bibfnamefont {M.}~\bibnamefont
  {Khafizov}}, \bibinfo {author} {\bibfnamefont {J.}~\bibnamefont {Pakarinen}},
  \bibinfo {author} {\bibfnamefont {L.}~\bibnamefont {He}}, \ and\ \bibinfo
  {author} {\bibfnamefont {D.~H.}\ \bibnamefont {Hurley}},\ }\href@noop {}
  {\bibfield  {journal} {\bibinfo  {journal} {Journal of the American Ceramic
  Society}\ }\textbf {\bibinfo {volume} {102}},\ \bibinfo {pages} {7533}
  (\bibinfo {year} {2019})}\BibitemShut {NoStop}%
\bibitem [{\citenamefont {Khafizov}\ \emph {et~al.}(2020)\citenamefont
  {Khafizov}, \citenamefont {Riyad}, \citenamefont {Wang}, \citenamefont
  {Pakarinen}, \citenamefont {He}, \citenamefont {Yao}, \citenamefont
  {El-Azab},\ and\ \citenamefont {Hurley}}]{khafizov2020combining}%
  \BibitemOpen
  \bibfield  {author} {\bibinfo {author} {\bibfnamefont {M.}~\bibnamefont
  {Khafizov}}, \bibinfo {author} {\bibfnamefont {M.~F.}\ \bibnamefont {Riyad}},
  \bibinfo {author} {\bibfnamefont {Y.}~\bibnamefont {Wang}}, \bibinfo {author}
  {\bibfnamefont {J.}~\bibnamefont {Pakarinen}}, \bibinfo {author}
  {\bibfnamefont {L.}~\bibnamefont {He}}, \bibinfo {author} {\bibfnamefont
  {T.}~\bibnamefont {Yao}}, \bibinfo {author} {\bibfnamefont {A.}~\bibnamefont
  {El-Azab}}, \ and\ \bibinfo {author} {\bibfnamefont {D.}~\bibnamefont
  {Hurley}},\ }\href@noop {} {\bibfield  {journal} {\bibinfo  {journal} {Acta
  Materialia}\ }\textbf {\bibinfo {volume} {193}},\ \bibinfo {pages} {61}
  (\bibinfo {year} {2020})}\BibitemShut {NoStop}%
\bibitem [{\citenamefont {Deskins}\ \emph {et~al.}(2022)\citenamefont
  {Deskins}, \citenamefont {Khanolkar}, \citenamefont {Mazumder}, \citenamefont
  {Dennett}, \citenamefont {Bawane}, \citenamefont {Hua}, \citenamefont
  {Ferrigno}, \citenamefont {He}, \citenamefont {Mann}, \citenamefont
  {Khafizov} \emph {et~al.}}]{deskins2022combined}%
  \BibitemOpen
  \bibfield  {author} {\bibinfo {author} {\bibfnamefont {W.~R.}\ \bibnamefont
  {Deskins}}, \bibinfo {author} {\bibfnamefont {A.}~\bibnamefont {Khanolkar}},
  \bibinfo {author} {\bibfnamefont {S.}~\bibnamefont {Mazumder}}, \bibinfo
  {author} {\bibfnamefont {C.~A.}\ \bibnamefont {Dennett}}, \bibinfo {author}
  {\bibfnamefont {K.}~\bibnamefont {Bawane}}, \bibinfo {author} {\bibfnamefont
  {Z.}~\bibnamefont {Hua}}, \bibinfo {author} {\bibfnamefont {J.}~\bibnamefont
  {Ferrigno}}, \bibinfo {author} {\bibfnamefont {L.}~\bibnamefont {He}},
  \bibinfo {author} {\bibfnamefont {J.~M.}\ \bibnamefont {Mann}}, \bibinfo
  {author} {\bibfnamefont {M.}~\bibnamefont {Khafizov}},  \emph {et~al.},\
  }\href@noop {} {\bibfield  {journal} {\bibinfo  {journal} {Acta Materialia}\
  }\textbf {\bibinfo {volume} {241}},\ \bibinfo {pages} {118379} (\bibinfo
  {year} {2022})}\BibitemShut {NoStop}%
\bibitem [{\citenamefont {Jin}\ \emph {et~al.}(2023)\citenamefont {Jin},
  \citenamefont {Miao}, \citenamefont {Zhang}, \citenamefont {Khafizov},
  \citenamefont {Bawane}, \citenamefont {Kombaiah}, \citenamefont {Zhang},\
  and\ \citenamefont {Hurley}}]{jin2023unfaulting}%
  \BibitemOpen
  \bibfield  {author} {\bibinfo {author} {\bibfnamefont {M.}~\bibnamefont
  {Jin}}, \bibinfo {author} {\bibfnamefont {J.}~\bibnamefont {Miao}}, \bibinfo
  {author} {\bibfnamefont {Y.}~\bibnamefont {Zhang}}, \bibinfo {author}
  {\bibfnamefont {M.}~\bibnamefont {Khafizov}}, \bibinfo {author}
  {\bibfnamefont {K.~K.}\ \bibnamefont {Bawane}}, \bibinfo {author}
  {\bibfnamefont {B.}~\bibnamefont {Kombaiah}}, \bibinfo {author}
  {\bibfnamefont {Y.}~\bibnamefont {Zhang}}, \ and\ \bibinfo {author}
  {\bibfnamefont {D.~H.}\ \bibnamefont {Hurley}},\ }\href@noop {} {\bibfield
  {journal} {\bibinfo  {journal} {Scripta Materialia}\ }\textbf {\bibinfo
  {volume} {237}},\ \bibinfo {pages} {115706} (\bibinfo {year}
  {2023})}\BibitemShut {NoStop}%
\bibitem [{\citenamefont {Yu}\ \emph {et~al.}(2024)\citenamefont {Yu},
  \citenamefont {Zhou}, \citenamefont {Jin}, \citenamefont {Khafizov},
  \citenamefont {Hurley},\ and\ \citenamefont {Zhang}}]{yu2024establishing}%
  \BibitemOpen
  \bibfield  {author} {\bibinfo {author} {\bibfnamefont {L.-C.}\ \bibnamefont
  {Yu}}, \bibinfo {author} {\bibfnamefont {S.}~\bibnamefont {Zhou}}, \bibinfo
  {author} {\bibfnamefont {M.}~\bibnamefont {Jin}}, \bibinfo {author}
  {\bibfnamefont {M.}~\bibnamefont {Khafizov}}, \bibinfo {author}
  {\bibfnamefont {D.}~\bibnamefont {Hurley}}, \ and\ \bibinfo {author}
  {\bibfnamefont {Y.}~\bibnamefont {Zhang}},\ }\href@noop {} {\bibfield
  {journal} {\bibinfo  {journal} {Nuclear Materials and Energy}\ }\textbf
  {\bibinfo {volume} {41}},\ \bibinfo {pages} {101774} (\bibinfo {year}
  {2024})}\BibitemShut {NoStop}%
\bibitem [{\citenamefont {Neilson}\ \emph
  {et~al.}(2024{\natexlab{a}})\citenamefont {Neilson}, \citenamefont {Galvin},
  \citenamefont {Dillon}, \citenamefont {Cooper},\ and\ \citenamefont
  {Andersson}}]{neilson2024irradiation}%
  \BibitemOpen
  \bibfield  {author} {\bibinfo {author} {\bibfnamefont {W.~D.}\ \bibnamefont
  {Neilson}}, \bibinfo {author} {\bibfnamefont {C.~O.}\ \bibnamefont {Galvin}},
  \bibinfo {author} {\bibfnamefont {S.~J.}\ \bibnamefont {Dillon}}, \bibinfo
  {author} {\bibfnamefont {M.~W.}\ \bibnamefont {Cooper}}, \ and\ \bibinfo
  {author} {\bibfnamefont {D.~A.}\ \bibnamefont {Andersson}},\ }\href@noop {}
  {\bibfield  {journal} {\bibinfo  {journal} {Physical Review Materials}\
  }\textbf {\bibinfo {volume} {8}},\ \bibinfo {pages} {103602} (\bibinfo {year}
  {2024}{\natexlab{a}})}\BibitemShut {NoStop}%
\bibitem [{\citenamefont {Neilson}\ \emph
  {et~al.}(2024{\natexlab{b}})\citenamefont {Neilson}, \citenamefont {Rizk},
  \citenamefont {Cooper},\ and\ \citenamefont {Andersson}}]{neilson2024oxygen}%
  \BibitemOpen
  \bibfield  {author} {\bibinfo {author} {\bibfnamefont {W.~D.}\ \bibnamefont
  {Neilson}}, \bibinfo {author} {\bibfnamefont {J.}~\bibnamefont {Rizk}},
  \bibinfo {author} {\bibfnamefont {M.~W.}\ \bibnamefont {Cooper}}, \ and\
  \bibinfo {author} {\bibfnamefont {D.~A.}\ \bibnamefont {Andersson}},\
  }\href@noop {} {\bibfield  {journal} {\bibinfo  {journal} {The Journal of
  Physical Chemistry C}\ }\textbf {\bibinfo {volume} {128}},\ \bibinfo {pages}
  {21559} (\bibinfo {year} {2024}{\natexlab{b}})}\BibitemShut {NoStop}%
\bibitem [{\citenamefont {Malakkal}\ \emph {et~al.}(2024)\citenamefont
  {Malakkal}, \citenamefont {Katre}, \citenamefont {Zhou}, \citenamefont
  {Jiang}, \citenamefont {Hurley}, \citenamefont {Marianetti},\ and\
  \citenamefont {Khafizov}}]{malakkal2024first}%
  \BibitemOpen
  \bibfield  {author} {\bibinfo {author} {\bibfnamefont {L.}~\bibnamefont
  {Malakkal}}, \bibinfo {author} {\bibfnamefont {A.}~\bibnamefont {Katre}},
  \bibinfo {author} {\bibfnamefont {S.}~\bibnamefont {Zhou}}, \bibinfo {author}
  {\bibfnamefont {C.}~\bibnamefont {Jiang}}, \bibinfo {author} {\bibfnamefont
  {D.~H.}\ \bibnamefont {Hurley}}, \bibinfo {author} {\bibfnamefont {C.~A.}\
  \bibnamefont {Marianetti}}, \ and\ \bibinfo {author} {\bibfnamefont
  {M.}~\bibnamefont {Khafizov}},\ }\href@noop {} {\bibfield  {journal}
  {\bibinfo  {journal} {Physical Review Materials}\ }\textbf {\bibinfo {volume}
  {8}},\ \bibinfo {pages} {025401} (\bibinfo {year} {2024})}\BibitemShut
  {NoStop}%
\bibitem [{\citenamefont {Jiang}\ \emph {et~al.}(2024)\citenamefont {Jiang},
  \citenamefont {Marianetti}, \citenamefont {Khafizov},\ and\ \citenamefont
  {Hurley}}]{jiang2024machine}%
  \BibitemOpen
  \bibfield  {author} {\bibinfo {author} {\bibfnamefont {C.}~\bibnamefont
  {Jiang}}, \bibinfo {author} {\bibfnamefont {C.~A.}\ \bibnamefont
  {Marianetti}}, \bibinfo {author} {\bibfnamefont {M.}~\bibnamefont
  {Khafizov}}, \ and\ \bibinfo {author} {\bibfnamefont {D.~H.}\ \bibnamefont
  {Hurley}},\ }\href@noop {} {\bibfield  {journal} {\bibinfo  {journal} {npj
  Computational Materials}\ }\textbf {\bibinfo {volume} {10}},\ \bibinfo
  {pages} {21} (\bibinfo {year} {2024})}\BibitemShut {NoStop}%
\bibitem [{\citenamefont {Liu}\ \emph {et~al.}(2016)\citenamefont {Liu},
  \citenamefont {Cooper}, \citenamefont {McClellan}, \citenamefont {Lashley},
  \citenamefont {Byler}, \citenamefont {Bell}, \citenamefont {Grimes},
  \citenamefont {Stanek},\ and\ \citenamefont {Andersson}}]{liu2016molecular}%
  \BibitemOpen
  \bibfield  {author} {\bibinfo {author} {\bibfnamefont {X.-Y.}\ \bibnamefont
  {Liu}}, \bibinfo {author} {\bibfnamefont {M.~W.~D.}\ \bibnamefont {Cooper}},
  \bibinfo {author} {\bibfnamefont {K.~J.}\ \bibnamefont {McClellan}}, \bibinfo
  {author} {\bibfnamefont {J.~C.}\ \bibnamefont {Lashley}}, \bibinfo {author}
  {\bibfnamefont {D.~D.}\ \bibnamefont {Byler}}, \bibinfo {author}
  {\bibfnamefont {B.}~\bibnamefont {Bell}}, \bibinfo {author} {\bibfnamefont
  {R.}~\bibnamefont {Grimes}}, \bibinfo {author} {\bibfnamefont {C.~R.}\
  \bibnamefont {Stanek}}, \ and\ \bibinfo {author} {\bibfnamefont {D.~A.}\
  \bibnamefont {Andersson}},\ }\href@noop {} {\bibfield  {journal} {\bibinfo
  {journal} {Physical Review Applied}\ }\textbf {\bibinfo {volume} {6}},\
  \bibinfo {pages} {044015} (\bibinfo {year} {2016})}\BibitemShut {NoStop}%
\bibitem [{\citenamefont {Park}\ \emph
  {et~al.}(2018{\natexlab{a}})\citenamefont {Park}, \citenamefont {Farfán},
  \citenamefont {Mitchell}, \citenamefont {Resnick}, \citenamefont {Enriquez},\
  and\ \citenamefont {Yee}}]{park_sensitivity_2018}%
  \BibitemOpen
  \bibfield  {author} {\bibinfo {author} {\bibfnamefont {J.}~\bibnamefont
  {Park}}, \bibinfo {author} {\bibfnamefont {E.~B.}\ \bibnamefont {Farfán}},
  \bibinfo {author} {\bibfnamefont {K.}~\bibnamefont {Mitchell}}, \bibinfo
  {author} {\bibfnamefont {A.}~\bibnamefont {Resnick}}, \bibinfo {author}
  {\bibfnamefont {C.}~\bibnamefont {Enriquez}}, \ and\ \bibinfo {author}
  {\bibfnamefont {T.}~\bibnamefont {Yee}},\ }\href {\doibase
  10.1016/j.jnucmat.2018.03.043} {\bibfield  {journal} {\bibinfo  {journal}
  {Journal of Nuclear Materials}\ }\textbf {\bibinfo {volume} {504}},\ \bibinfo
  {pages} {198} (\bibinfo {year} {2018}{\natexlab{a}})}\BibitemShut {NoStop}%
\bibitem [{\citenamefont {Cooper}\ \emph {et~al.}(2015)\citenamefont {Cooper},
  \citenamefont {Middleburgh},\ and\ \citenamefont
  {Grimes}}]{cooper_modelling_2015}%
  \BibitemOpen
  \bibfield  {author} {\bibinfo {author} {\bibfnamefont {M.~W.~D.}\
  \bibnamefont {Cooper}}, \bibinfo {author} {\bibfnamefont {S.~C.}\
  \bibnamefont {Middleburgh}}, \ and\ \bibinfo {author} {\bibfnamefont {R.~W.}\
  \bibnamefont {Grimes}},\ }\href {\doibase 10.1016/j.jnucmat.2015.07.022}
  {\bibfield  {journal} {\bibinfo  {journal} {Journal of Nuclear Materials}\
  }\textbf {\bibinfo {volume} {466}},\ \bibinfo {pages} {29} (\bibinfo {year}
  {2015})}\BibitemShut {NoStop}%
\bibitem [{\citenamefont {Park}\ \emph
  {et~al.}(2018{\natexlab{b}})\citenamefont {Park}, \citenamefont {Farfán},\
  and\ \citenamefont {Enriquez}}]{park_thermal_2018}%
  \BibitemOpen
  \bibfield  {author} {\bibinfo {author} {\bibfnamefont {J.}~\bibnamefont
  {Park}}, \bibinfo {author} {\bibfnamefont {E.~B.}\ \bibnamefont {Farfán}}, \
  and\ \bibinfo {author} {\bibfnamefont {C.}~\bibnamefont {Enriquez}},\ }\href
  {\doibase 10.1016/j.net.2018.02.002} {\bibfield  {journal} {\bibinfo
  {journal} {Nuclear Engineering and Technology}\ }\textbf {\bibinfo {volume}
  {50}},\ \bibinfo {pages} {731} (\bibinfo {year}
  {2018}{\natexlab{b}})}\BibitemShut {NoStop}%
\bibitem [{\citenamefont {Balboa}\ \emph {et~al.}(2017)\citenamefont {Balboa},
  \citenamefont {Van~Brutzel}, \citenamefont {Chartier},\ and\ \citenamefont
  {Le~Bouar}}]{balboa2017assessment}%
  \BibitemOpen
  \bibfield  {author} {\bibinfo {author} {\bibfnamefont {H.}~\bibnamefont
  {Balboa}}, \bibinfo {author} {\bibfnamefont {L.}~\bibnamefont {Van~Brutzel}},
  \bibinfo {author} {\bibfnamefont {A.}~\bibnamefont {Chartier}}, \ and\
  \bibinfo {author} {\bibfnamefont {Y.}~\bibnamefont {Le~Bouar}},\ }\href@noop
  {} {\bibfield  {journal} {\bibinfo  {journal} {Journal of Nuclear Materials}\
  }\textbf {\bibinfo {volume} {495}},\ \bibinfo {pages} {67} (\bibinfo {year}
  {2017})}\BibitemShut {NoStop}%
\bibitem [{\citenamefont {Malakkal}\ \emph {et~al.}(2019)\citenamefont
  {Malakkal}, \citenamefont {Prasad}, \citenamefont {Jossou}, \citenamefont
  {Ranasinghe}, \citenamefont {Szpunar}, \citenamefont {Bichler},\ and\
  \citenamefont {Szpunar}}]{malakkal2019thermal}%
  \BibitemOpen
  \bibfield  {author} {\bibinfo {author} {\bibfnamefont {L.}~\bibnamefont
  {Malakkal}}, \bibinfo {author} {\bibfnamefont {A.}~\bibnamefont {Prasad}},
  \bibinfo {author} {\bibfnamefont {E.}~\bibnamefont {Jossou}}, \bibinfo
  {author} {\bibfnamefont {J.}~\bibnamefont {Ranasinghe}}, \bibinfo {author}
  {\bibfnamefont {B.}~\bibnamefont {Szpunar}}, \bibinfo {author} {\bibfnamefont
  {L.}~\bibnamefont {Bichler}}, \ and\ \bibinfo {author} {\bibfnamefont
  {J.}~\bibnamefont {Szpunar}},\ }\href@noop {} {\bibfield  {journal} {\bibinfo
   {journal} {Journal of Alloys and Compounds}\ }\textbf {\bibinfo {volume}
  {798}},\ \bibinfo {pages} {507} (\bibinfo {year} {2019})}\BibitemShut
  {NoStop}%
\bibitem [{\citenamefont {Zhu}\ \emph {et~al.}(2020)\citenamefont {Zhu},
  \citenamefont {Gong}, \citenamefont {Zhao}, \citenamefont {Lin},
  \citenamefont {Han}, \citenamefont {Liu},\ and\ \citenamefont
  {Song}}]{zhu2020effect}%
  \BibitemOpen
  \bibfield  {author} {\bibinfo {author} {\bibfnamefont {X.}~\bibnamefont
  {Zhu}}, \bibinfo {author} {\bibfnamefont {H.}~\bibnamefont {Gong}}, \bibinfo
  {author} {\bibfnamefont {Y.-F.}\ \bibnamefont {Zhao}}, \bibinfo {author}
  {\bibfnamefont {D.-Y.}\ \bibnamefont {Lin}}, \bibinfo {author} {\bibfnamefont
  {G.}~\bibnamefont {Han}}, \bibinfo {author} {\bibfnamefont {T.}~\bibnamefont
  {Liu}}, \ and\ \bibinfo {author} {\bibfnamefont {H.}~\bibnamefont {Song}},\
  }\href@noop {} {\bibfield  {journal} {\bibinfo  {journal} {Journal of Nuclear
  Materials}\ }\textbf {\bibinfo {volume} {533}},\ \bibinfo {pages} {152080}
  (\bibinfo {year} {2020})}\BibitemShut {NoStop}%
\bibitem [{\citenamefont {Jiang}\ \emph {et~al.}(2022)\citenamefont {Jiang},
  \citenamefont {He}, \citenamefont {Dennett}, \citenamefont {Khafizov},
  \citenamefont {Mann},\ and\ \citenamefont {Hurley}}]{jiang2022unraveling}%
  \BibitemOpen
  \bibfield  {author} {\bibinfo {author} {\bibfnamefont {C.}~\bibnamefont
  {Jiang}}, \bibinfo {author} {\bibfnamefont {L.}~\bibnamefont {He}}, \bibinfo
  {author} {\bibfnamefont {C.~A.}\ \bibnamefont {Dennett}}, \bibinfo {author}
  {\bibfnamefont {M.}~\bibnamefont {Khafizov}}, \bibinfo {author}
  {\bibfnamefont {J.~M.}\ \bibnamefont {Mann}}, \ and\ \bibinfo {author}
  {\bibfnamefont {D.~H.}\ \bibnamefont {Hurley}},\ }\href@noop {} {\bibfield
  {journal} {\bibinfo  {journal} {Scripta Materialia}\ }\textbf {\bibinfo
  {volume} {214}},\ \bibinfo {pages} {114684} (\bibinfo {year}
  {2022})}\BibitemShut {NoStop}%
\bibitem [{\citenamefont {Jin}\ \emph {et~al.}(2024)\citenamefont {Jin},
  \citenamefont {Miao}, \citenamefont {Chen}, \citenamefont {Khafizov},
  \citenamefont {Zhang},\ and\ \citenamefont {Hurley}}]{jin2024extended}%
  \BibitemOpen
  \bibfield  {author} {\bibinfo {author} {\bibfnamefont {M.}~\bibnamefont
  {Jin}}, \bibinfo {author} {\bibfnamefont {J.}~\bibnamefont {Miao}}, \bibinfo
  {author} {\bibfnamefont {B.}~\bibnamefont {Chen}}, \bibinfo {author}
  {\bibfnamefont {M.}~\bibnamefont {Khafizov}}, \bibinfo {author}
  {\bibfnamefont {Y.}~\bibnamefont {Zhang}}, \ and\ \bibinfo {author}
  {\bibfnamefont {D.~H.}\ \bibnamefont {Hurley}},\ }\href@noop {} {\bibfield
  {journal} {\bibinfo  {journal} {Computational Materials Science}\ }\textbf
  {\bibinfo {volume} {235}},\ \bibinfo {pages} {112842} (\bibinfo {year}
  {2024})}\BibitemShut {NoStop}%
\bibitem [{\citenamefont {Fossati}\ \emph {et~al.}(2024)\citenamefont
  {Fossati}, \citenamefont {Burr}, \citenamefont {Cooper}, \citenamefont
  {Galvin},\ and\ \citenamefont {Grimes}}]{fossati2024superionic}%
  \BibitemOpen
  \bibfield  {author} {\bibinfo {author} {\bibfnamefont {P.~C.}\ \bibnamefont
  {Fossati}}, \bibinfo {author} {\bibfnamefont {P.~A.}\ \bibnamefont {Burr}},
  \bibinfo {author} {\bibfnamefont {M.~W.}\ \bibnamefont {Cooper}}, \bibinfo
  {author} {\bibfnamefont {C.~O.}\ \bibnamefont {Galvin}}, \ and\ \bibinfo
  {author} {\bibfnamefont {R.~W.}\ \bibnamefont {Grimes}},\ }\href@noop {}
  {\bibfield  {journal} {\bibinfo  {journal} {Physical Review Materials}\
  }\textbf {\bibinfo {volume} {8}},\ \bibinfo {pages} {115404} (\bibinfo {year}
  {2024})}\BibitemShut {NoStop}%
\bibitem [{\citenamefont {Cooper}\ \emph {et~al.}(2025)\citenamefont {Cooper},
  \citenamefont {Matthews},\ and\ \citenamefont {Andersson}}]{cooper2025role}%
  \BibitemOpen
  \bibfield  {author} {\bibinfo {author} {\bibfnamefont {M.}~\bibnamefont
  {Cooper}}, \bibinfo {author} {\bibfnamefont {C.}~\bibnamefont {Matthews}}, \
  and\ \bibinfo {author} {\bibfnamefont {D.}~\bibnamefont {Andersson}},\ }\href
  {\doibase https://doi.org/10.1016/j.jnucmat.2024.155452} {\bibfield
  {journal} {\bibinfo  {journal} {Journal of Nuclear Materials}\ }\textbf
  {\bibinfo {volume} {604}},\ \bibinfo {pages} {155452} (\bibinfo {year}
  {2025})}\BibitemShut {NoStop}%
\bibitem [{\citenamefont {Jin}\ \emph {et~al.}(2021)\citenamefont {Jin},
  \citenamefont {Khafizov}, \citenamefont {Jiang}, \citenamefont {Zhou},
  \citenamefont {Marianetti}, \citenamefont {Bryan}, \citenamefont {Manley},\
  and\ \citenamefont {Hurley}}]{jin_assessment_2021}%
  \BibitemOpen
  \bibfield  {author} {\bibinfo {author} {\bibfnamefont {M.}~\bibnamefont
  {Jin}}, \bibinfo {author} {\bibfnamefont {M.}~\bibnamefont {Khafizov}},
  \bibinfo {author} {\bibfnamefont {C.}~\bibnamefont {Jiang}}, \bibinfo
  {author} {\bibfnamefont {S.}~\bibnamefont {Zhou}}, \bibinfo {author}
  {\bibfnamefont {C.~A.}\ \bibnamefont {Marianetti}}, \bibinfo {author}
  {\bibfnamefont {M.~S.}\ \bibnamefont {Bryan}}, \bibinfo {author}
  {\bibfnamefont {M.~E.}\ \bibnamefont {Manley}}, \ and\ \bibinfo {author}
  {\bibfnamefont {D.~H.}\ \bibnamefont {Hurley}},\ }\href {\doibase
  10.1088/1361-648X/abdc8f} {\bibfield  {journal} {\bibinfo  {journal} {Journal
  of Physics: Condensed Matter}\ }\textbf {\bibinfo {volume} {33}},\ \bibinfo
  {pages} {275402} (\bibinfo {year} {2021})}\BibitemShut {NoStop}%
\bibitem [{\citenamefont {Mathis}\ \emph {et~al.}(2022)\citenamefont {Mathis},
  \citenamefont {Khanolkar}, \citenamefont {Fu}, \citenamefont {Bryan},
  \citenamefont {Dennett}, \citenamefont {Rickert}, \citenamefont {Mann},
  \citenamefont {Winn}, \citenamefont {Abernathy}, \citenamefont {Manley},
  \citenamefont {Hurley},\ and\ \citenamefont
  {Marianetti}}]{mathis2022generalized}%
  \BibitemOpen
  \bibfield  {author} {\bibinfo {author} {\bibfnamefont {M.~A.}\ \bibnamefont
  {Mathis}}, \bibinfo {author} {\bibfnamefont {A.}~\bibnamefont {Khanolkar}},
  \bibinfo {author} {\bibfnamefont {L.}~\bibnamefont {Fu}}, \bibinfo {author}
  {\bibfnamefont {M.~S.}\ \bibnamefont {Bryan}}, \bibinfo {author}
  {\bibfnamefont {C.~A.}\ \bibnamefont {Dennett}}, \bibinfo {author}
  {\bibfnamefont {K.}~\bibnamefont {Rickert}}, \bibinfo {author} {\bibfnamefont
  {J.~M.}\ \bibnamefont {Mann}}, \bibinfo {author} {\bibfnamefont
  {B.}~\bibnamefont {Winn}}, \bibinfo {author} {\bibfnamefont {D.~L.}\
  \bibnamefont {Abernathy}}, \bibinfo {author} {\bibfnamefont {M.~E.}\
  \bibnamefont {Manley}}, \bibinfo {author} {\bibfnamefont {D.~H.}\
  \bibnamefont {Hurley}}, \ and\ \bibinfo {author} {\bibfnamefont {C.~A.}\
  \bibnamefont {Marianetti}},\ }\href {\doibase 10.1103/PhysRevB.106.014314}
  {\bibfield  {journal} {\bibinfo  {journal} {Physical Review B}\ }\textbf
  {\bibinfo {volume} {106}},\ \bibinfo {pages} {014314} (\bibinfo {year}
  {2022})}\BibitemShut {NoStop}%
\bibitem [{\citenamefont {Xiao}\ \emph {et~al.}(2022)\citenamefont {Xiao},
  \citenamefont {Ma}, \citenamefont {Bryan}, \citenamefont {Fu}, \citenamefont
  {Mann}, \citenamefont {Winn}, \citenamefont {Abernathy}, \citenamefont
  {Hermann}, \citenamefont {Khanolkar}, \citenamefont {Dennett}, \citenamefont
  {Hurley}, \citenamefont {Manley},\ and\ \citenamefont
  {Marianetti}}]{xiao2022validating}%
  \BibitemOpen
  \bibfield  {author} {\bibinfo {author} {\bibfnamefont {E.}~\bibnamefont
  {Xiao}}, \bibinfo {author} {\bibfnamefont {H.}~\bibnamefont {Ma}}, \bibinfo
  {author} {\bibfnamefont {M.~S.}\ \bibnamefont {Bryan}}, \bibinfo {author}
  {\bibfnamefont {L.}~\bibnamefont {Fu}}, \bibinfo {author} {\bibfnamefont
  {J.~M.}\ \bibnamefont {Mann}}, \bibinfo {author} {\bibfnamefont
  {B.}~\bibnamefont {Winn}}, \bibinfo {author} {\bibfnamefont {D.~L.}\
  \bibnamefont {Abernathy}}, \bibinfo {author} {\bibfnamefont {R.~P.}\
  \bibnamefont {Hermann}}, \bibinfo {author} {\bibfnamefont {A.~R.}\
  \bibnamefont {Khanolkar}}, \bibinfo {author} {\bibfnamefont {C.~A.}\
  \bibnamefont {Dennett}}, \bibinfo {author} {\bibfnamefont {D.~H.}\
  \bibnamefont {Hurley}}, \bibinfo {author} {\bibfnamefont {M.~E.}\
  \bibnamefont {Manley}}, \ and\ \bibinfo {author} {\bibfnamefont {C.~A.}\
  \bibnamefont {Marianetti}},\ }\href {\doibase 10.1103/PhysRevB.106.144310}
  {\bibfield  {journal} {\bibinfo  {journal} {Physical Review B}\ }\textbf
  {\bibinfo {volume} {106}},\ \bibinfo {pages} {144310} (\bibinfo {year}
  {2022})}\BibitemShut {NoStop}%
\bibitem [{\citenamefont {Zhou}\ \emph {et~al.}(2022)\citenamefont {Zhou},
  \citenamefont {Ma}, \citenamefont {Xiao}, \citenamefont {Gofryk},
  \citenamefont {Jiang}, \citenamefont {Manley}, \citenamefont {Hurley},\ and\
  \citenamefont {Marianetti}}]{zhouCapturingGroundState2022a}%
  \BibitemOpen
  \bibfield  {author} {\bibinfo {author} {\bibfnamefont {S.}~\bibnamefont
  {Zhou}}, \bibinfo {author} {\bibfnamefont {H.}~\bibnamefont {Ma}}, \bibinfo
  {author} {\bibfnamefont {E.}~\bibnamefont {Xiao}}, \bibinfo {author}
  {\bibfnamefont {K.}~\bibnamefont {Gofryk}}, \bibinfo {author} {\bibfnamefont
  {C.}~\bibnamefont {Jiang}}, \bibinfo {author} {\bibfnamefont {M.~E.}\
  \bibnamefont {Manley}}, \bibinfo {author} {\bibfnamefont {D.~H.}\
  \bibnamefont {Hurley}}, \ and\ \bibinfo {author} {\bibfnamefont {C.~A.}\
  \bibnamefont {Marianetti}},\ }\href {\doibase 10.1103/PhysRevB.106.125134}
  {\bibfield  {journal} {\bibinfo  {journal} {Physical Review B}\ }\textbf
  {\bibinfo {volume} {106}},\ \bibinfo {pages} {125134} (\bibinfo {year}
  {2022})}\BibitemShut {NoStop}%
\bibitem [{\citenamefont {Zhou}\ \emph {et~al.}(2024)\citenamefont {Zhou},
  \citenamefont {Xiao}, \citenamefont {Ma}, \citenamefont {Gofryk},
  \citenamefont {Jiang}, \citenamefont {Manley}, \citenamefont {Hurley},\ and\
  \citenamefont {Marianetti}}]{zhou_phonon_2024}%
  \BibitemOpen
  \bibfield  {author} {\bibinfo {author} {\bibfnamefont {S.}~\bibnamefont
  {Zhou}}, \bibinfo {author} {\bibfnamefont {E.}~\bibnamefont {Xiao}}, \bibinfo
  {author} {\bibfnamefont {H.}~\bibnamefont {Ma}}, \bibinfo {author}
  {\bibfnamefont {K.}~\bibnamefont {Gofryk}}, \bibinfo {author} {\bibfnamefont
  {C.}~\bibnamefont {Jiang}}, \bibinfo {author} {\bibfnamefont {M.~E.}\
  \bibnamefont {Manley}}, \bibinfo {author} {\bibfnamefont {D.~H.}\
  \bibnamefont {Hurley}}, \ and\ \bibinfo {author} {\bibfnamefont {C.~A.}\
  \bibnamefont {Marianetti}},\ }\href {\doibase 10.1103/PhysRevLett.132.106502}
  {\bibfield  {journal} {\bibinfo  {journal} {Physical Review Letters}\
  }\textbf {\bibinfo {volume} {132}},\ \bibinfo {pages} {106502} (\bibinfo
  {year} {2024})},\ \bibinfo {note} {publisher: American Physical
  Society}\BibitemShut {NoStop}%
\bibitem [{\citenamefont {Buckingham}\ and\ \citenamefont
  {Lennard-Jones}(1938)}]{buckingham_classical_1938}%
  \BibitemOpen
  \bibfield  {author} {\bibinfo {author} {\bibfnamefont {R.~A.}\ \bibnamefont
  {Buckingham}}\ and\ \bibinfo {author} {\bibfnamefont {J.~E.}\ \bibnamefont
  {Lennard-Jones}},\ }\href {\doibase 10.1098/rspa.1938.0173} {\bibfield
  {journal} {\bibinfo  {journal} {Proceedings of the Royal Society of London.
  Series A. Mathematical and Physical Sciences}\ }\textbf {\bibinfo {volume}
  {168}},\ \bibinfo {pages} {264} (\bibinfo {year} {1938})}\BibitemShut
  {NoStop}%
\bibitem [{\citenamefont {Morse}(1929)}]{morse_diatomic_1929}%
  \BibitemOpen
  \bibfield  {author} {\bibinfo {author} {\bibfnamefont {P.~M.}\ \bibnamefont
  {Morse}},\ }\href {\doibase 10.1103/PhysRev.34.57} {\bibfield  {journal}
  {\bibinfo  {journal} {Physical Review}\ }\textbf {\bibinfo {volume} {34}},\
  \bibinfo {pages} {57} (\bibinfo {year} {1929})}\BibitemShut {NoStop}%
\bibitem [{\citenamefont {Mitchell}\ and\ \citenamefont
  {Fincham}(1993)}]{mitchell_shell_1993}%
  \BibitemOpen
  \bibfield  {author} {\bibinfo {author} {\bibfnamefont {P.~J.}\ \bibnamefont
  {Mitchell}}\ and\ \bibinfo {author} {\bibfnamefont {D.}~\bibnamefont
  {Fincham}},\ }\href {\doibase 10.1088/0953-8984/5/8/006} {\ \textbf {\bibinfo
  {volume} {5}},\ \bibinfo {pages} {1031} (\bibinfo {year} {1993})}\BibitemShut
  {NoStop}%
\bibitem [{\citenamefont {Gale}(1997)}]{gale_gulp_1997}%
  \BibitemOpen
  \bibfield  {author} {\bibinfo {author} {\bibfnamefont {J.~D.}\ \bibnamefont
  {Gale}},\ }\href {\doibase 10.1039/A606455H} {\bibfield  {journal} {\bibinfo
  {journal} {Journal of the Chemical Society, Faraday Transactions}\ }\textbf
  {\bibinfo {volume} {93}},\ \bibinfo {pages} {629} (\bibinfo {year} {1997})},\
  \bibinfo {note} {publisher: The Royal Society of Chemistry}\BibitemShut
  {NoStop}%
\bibitem [{Sup()}]{Supplemental_Material}%
  \BibitemOpen
  \href@noop {} {}\bibinfo {note} {See Supplemental Material at [link] for
  details of the training and assessment data, the training method for
  empirical potential, the training results for alternative analytical forms,
  the configurations of Frenkel pairs, the analyses and comparisons of second-
  and third-order irreducible derivatives, and the reciprocal coordinates of
  primary q points and their displacement bases. See also
  Ref.~\cite{mommaVESTAThreedimensionalVisualization2011}}\BibitemShut
  {NoStop}%
\bibitem [{\citenamefont {Blöchl}(1994)}]{blochl_projector_1994}%
  \BibitemOpen
  \bibfield  {author} {\bibinfo {author} {\bibfnamefont {P.~E.}\ \bibnamefont
  {Blöchl}},\ }\href {\doibase 10.1103/PhysRevB.50.17953} {\bibfield
  {journal} {\bibinfo  {journal} {Physical Review B}\ }\textbf {\bibinfo
  {volume} {50}},\ \bibinfo {pages} {17953} (\bibinfo {year}
  {1994})}\BibitemShut {NoStop}%
\bibitem [{\citenamefont {Kresse}\ and\ \citenamefont
  {Joubert}(1999)}]{kresse_ultrasoft_1999}%
  \BibitemOpen
  \bibfield  {author} {\bibinfo {author} {\bibfnamefont {G.}~\bibnamefont
  {Kresse}}\ and\ \bibinfo {author} {\bibfnamefont {D.}~\bibnamefont
  {Joubert}},\ }\href {\doibase 10.1103/PhysRevB.59.1758} {\bibfield  {journal}
  {\bibinfo  {journal} {Physical Review B}\ }\textbf {\bibinfo {volume} {59}},\
  \bibinfo {pages} {1758} (\bibinfo {year} {1999})}\BibitemShut {NoStop}%
\bibitem [{\citenamefont {Kresse}\ and\ \citenamefont
  {Hafner}(1993)}]{kresse_ab_1993}%
  \BibitemOpen
  \bibfield  {author} {\bibinfo {author} {\bibfnamefont {G.}~\bibnamefont
  {Kresse}}\ and\ \bibinfo {author} {\bibfnamefont {J.}~\bibnamefont
  {Hafner}},\ }\href {http://link.aps.org/doi/10.1103/PhysRevB.47.558}
  {\bibfield  {journal} {\bibinfo  {journal} {Physical Review B}\ }\textbf
  {\bibinfo {volume} {47}},\ \bibinfo {pages} {558} (\bibinfo {year}
  {1993})}\BibitemShut {NoStop}%
\bibitem [{\citenamefont {Kresse}\ and\ \citenamefont
  {Furthmüller}(1996)}]{kresse_efficient_1996}%
  \BibitemOpen
  \bibfield  {author} {\bibinfo {author} {\bibfnamefont {G.}~\bibnamefont
  {Kresse}}\ and\ \bibinfo {author} {\bibfnamefont {J.}~\bibnamefont
  {Furthmüller}},\ }\href {\doibase 10.1103/PhysRevB.54.11169} {\bibfield
  {journal} {\bibinfo  {journal} {Physical Review B}\ }\textbf {\bibinfo
  {volume} {54}},\ \bibinfo {pages} {11169} (\bibinfo {year}
  {1996})}\BibitemShut {NoStop}%
\bibitem [{\citenamefont {Sun}\ \emph {et~al.}(2015)\citenamefont {Sun},
  \citenamefont {Ruzsinszky},\ and\ \citenamefont
  {Perdew}}]{sunStronglyConstrainedAppropriately2015}%
  \BibitemOpen
  \bibfield  {author} {\bibinfo {author} {\bibfnamefont {J.}~\bibnamefont
  {Sun}}, \bibinfo {author} {\bibfnamefont {A.}~\bibnamefont {Ruzsinszky}}, \
  and\ \bibinfo {author} {\bibfnamefont {J.~P.}\ \bibnamefont {Perdew}},\
  }\href {\doibase 10.1103/PhysRevLett.115.036402} {\bibfield  {journal}
  {\bibinfo  {journal} {Physical Review Letters}\ }\textbf {\bibinfo {volume}
  {115}},\ \bibinfo {pages} {036402} (\bibinfo {year} {2015})}\BibitemShut
  {NoStop}%
\bibitem [{\citenamefont {Perdew}\ \emph {et~al.}(1996)\citenamefont {Perdew},
  \citenamefont {Burke},\ and\ \citenamefont
  {Ernzerhof}}]{perdewGeneralizedGradientApproximation1996a}%
  \BibitemOpen
  \bibfield  {author} {\bibinfo {author} {\bibfnamefont {J.~P.}\ \bibnamefont
  {Perdew}}, \bibinfo {author} {\bibfnamefont {K.}~\bibnamefont {Burke}}, \
  and\ \bibinfo {author} {\bibfnamefont {M.}~\bibnamefont {Ernzerhof}},\ }\href
  {\doibase 10.1103/PhysRevLett.77.3865} {\bibfield  {journal} {\bibinfo
  {journal} {Physical Review Letters}\ }\textbf {\bibinfo {volume} {77}},\
  \bibinfo {pages} {3865} (\bibinfo {year} {1996})}\BibitemShut {NoStop}%
\bibitem [{\citenamefont {Maldonado}\ \emph {et~al.}(2016)\citenamefont
  {Maldonado}, \citenamefont {Paolasini}, \citenamefont {Oppeneer},
  \citenamefont {Forrest}, \citenamefont {Prodi}, \citenamefont {Magnani},
  \citenamefont {Bosak}, \citenamefont {Lander},\ and\ \citenamefont
  {Caciuffo}}]{maldonado2016crystal}%
  \BibitemOpen
  \bibfield  {author} {\bibinfo {author} {\bibfnamefont {P.}~\bibnamefont
  {Maldonado}}, \bibinfo {author} {\bibfnamefont {L.}~\bibnamefont
  {Paolasini}}, \bibinfo {author} {\bibfnamefont {P.}~\bibnamefont {Oppeneer}},
  \bibinfo {author} {\bibfnamefont {T.~R.}\ \bibnamefont {Forrest}}, \bibinfo
  {author} {\bibfnamefont {A.}~\bibnamefont {Prodi}}, \bibinfo {author}
  {\bibfnamefont {N.}~\bibnamefont {Magnani}}, \bibinfo {author} {\bibfnamefont
  {A.}~\bibnamefont {Bosak}}, \bibinfo {author} {\bibfnamefont
  {G.}~\bibnamefont {Lander}}, \ and\ \bibinfo {author} {\bibfnamefont
  {R.}~\bibnamefont {Caciuffo}},\ }\href@noop {} {\bibfield  {journal}
  {\bibinfo  {journal} {Physical Review B}\ }\textbf {\bibinfo {volume} {93}},\
  \bibinfo {pages} {144301} (\bibinfo {year} {2016})}\BibitemShut {NoStop}%
\bibitem [{\citenamefont {Anisimov}\ \emph {et~al.}(1991)\citenamefont
  {Anisimov}, \citenamefont {Zaanen},\ and\ \citenamefont
  {Andersen}}]{anisimov_band_1991}%
  \BibitemOpen
  \bibfield  {author} {\bibinfo {author} {\bibfnamefont {V.~I.}\ \bibnamefont
  {Anisimov}}, \bibinfo {author} {\bibfnamefont {J.}~\bibnamefont {Zaanen}}, \
  and\ \bibinfo {author} {\bibfnamefont {O.~K.}\ \bibnamefont {Andersen}},\
  }\href {\doibase 10.1103/PhysRevB.44.943} {\bibfield  {journal} {\bibinfo
  {journal} {Physical Review B}\ }\textbf {\bibinfo {volume} {44}},\ \bibinfo
  {pages} {943} (\bibinfo {year} {1991})}\BibitemShut {NoStop}%
\bibitem [{\citenamefont {Dudarev}\ \emph {et~al.}(1997)\citenamefont
  {Dudarev}, \citenamefont {Manh},\ and\ \citenamefont
  {Sutton}}]{dudarev_effect_1997}%
  \BibitemOpen
  \bibfield  {author} {\bibinfo {author} {\bibfnamefont {S.~L.}\ \bibnamefont
  {Dudarev}}, \bibinfo {author} {\bibfnamefont {D.~N.}\ \bibnamefont {Manh}}, \
  and\ \bibinfo {author} {\bibfnamefont {A.~P.}\ \bibnamefont {Sutton}},\
  }\href {\doibase 10.1080/13642819708202343} {\bibfield  {journal} {\bibinfo
  {journal} {Philosophical Magazine B}\ }\textbf {\bibinfo {volume} {75}},\
  \bibinfo {pages} {613} (\bibinfo {year} {1997})}\BibitemShut {NoStop}%
\bibitem [{\citenamefont {Monkhorst}\ and\ \citenamefont
  {Pack}(1976)}]{monkhorst_special_1976}%
  \BibitemOpen
  \bibfield  {author} {\bibinfo {author} {\bibfnamefont {H.~J.}\ \bibnamefont
  {Monkhorst}}\ and\ \bibinfo {author} {\bibfnamefont {J.~D.}\ \bibnamefont
  {Pack}},\ }\href {\doibase 10.1103/PhysRevB.13.5188} {\bibfield  {journal}
  {\bibinfo  {journal} {Physical Review B}\ }\textbf {\bibinfo {volume} {13}},\
  \bibinfo {pages} {5188} (\bibinfo {year} {1976})}\BibitemShut {NoStop}%
\bibitem [{\citenamefont {Ewald}(1921)}]{ewald_berechnung_1921}%
  \BibitemOpen
  \bibfield  {author} {\bibinfo {author} {\bibfnamefont {P.~P.}\ \bibnamefont
  {Ewald}},\ }\href {\doibase 10.1002/andp.19213690304} {\bibfield  {journal}
  {\bibinfo  {journal} {Annalen der Physik}\ }\textbf {\bibinfo {volume}
  {369}},\ \bibinfo {pages} {253} (\bibinfo {year} {1921})}\BibitemShut
  {NoStop}%
\bibitem [{\citenamefont {Rushton}()}]{potential-pro-fit}%
  \BibitemOpen
  \bibfield  {author} {\bibinfo {author} {\bibfnamefont {M.~J.}\ \bibnamefont
  {Rushton}},\ }\href@noop {} {\enquote {\bibinfo {title}
  {potential-pro-fit},}\ }\bibinfo {howpublished}
  {\url{https://github.com/mjdrushton/potential-pro-fit}}\BibitemShut {NoStop}%
\bibitem [{\citenamefont {Bryan}\ \emph {et~al.}(2019)\citenamefont {Bryan},
  \citenamefont {Pang}, \citenamefont {Larson}, \citenamefont {Chernatynskiy},
  \citenamefont {Abernathy}, \citenamefont {Gofryk},\ and\ \citenamefont
  {Manley}}]{bryan_impact_2019}%
  \BibitemOpen
  \bibfield  {author} {\bibinfo {author} {\bibfnamefont {M.~S.}\ \bibnamefont
  {Bryan}}, \bibinfo {author} {\bibfnamefont {J.~W.~L.}\ \bibnamefont {Pang}},
  \bibinfo {author} {\bibfnamefont {B.~C.}\ \bibnamefont {Larson}}, \bibinfo
  {author} {\bibfnamefont {A.}~\bibnamefont {Chernatynskiy}}, \bibinfo {author}
  {\bibfnamefont {D.~L.}\ \bibnamefont {Abernathy}}, \bibinfo {author}
  {\bibfnamefont {K.}~\bibnamefont {Gofryk}}, \ and\ \bibinfo {author}
  {\bibfnamefont {M.~E.}\ \bibnamefont {Manley}},\ }\href {\doibase
  10.1103/PhysRevMaterials.3.065405} {\bibfield  {journal} {\bibinfo  {journal}
  {Physical Review Materials}\ }\textbf {\bibinfo {volume} {3}},\ \bibinfo
  {pages} {065405} (\bibinfo {year} {2019})}\BibitemShut {NoStop}%
\bibitem [{\citenamefont {Pang}\ \emph {et~al.}(2013)\citenamefont {Pang},
  \citenamefont {Buyers}, \citenamefont {Chernatynskiy}, \citenamefont
  {Lumsden}, \citenamefont {Larson},\ and\ \citenamefont
  {Phillpot}}]{pang_phonon_2013}%
  \BibitemOpen
  \bibfield  {author} {\bibinfo {author} {\bibfnamefont {J.~W.~L.}\
  \bibnamefont {Pang}}, \bibinfo {author} {\bibfnamefont {W.~J.~L.}\
  \bibnamefont {Buyers}}, \bibinfo {author} {\bibfnamefont {A.}~\bibnamefont
  {Chernatynskiy}}, \bibinfo {author} {\bibfnamefont {M.~D.}\ \bibnamefont
  {Lumsden}}, \bibinfo {author} {\bibfnamefont {B.~C.}\ \bibnamefont {Larson}},
  \ and\ \bibinfo {author} {\bibfnamefont {S.~R.}\ \bibnamefont {Phillpot}},\
  }\href {\doibase 10.1103/PhysRevLett.110.157401} {\bibfield  {journal}
  {\bibinfo  {journal} {Physical Review Letters}\ }\textbf {\bibinfo {volume}
  {110}},\ \bibinfo {pages} {157401} (\bibinfo {year} {2013})}\BibitemShut
  {NoStop}%
\bibitem [{\citenamefont {Macedo}\ \emph {et~al.}(1964)\citenamefont {Macedo},
  \citenamefont {Capps},\ and\ \citenamefont {Wachtman}}]{macedo_elastic_1964}%
  \BibitemOpen
  \bibfield  {author} {\bibinfo {author} {\bibfnamefont {P.~M.}\ \bibnamefont
  {Macedo}}, \bibinfo {author} {\bibfnamefont {W.}~\bibnamefont {Capps}}, \
  and\ \bibinfo {author} {\bibfnamefont {J.~O.}\ \bibnamefont {Wachtman}},\
  }\href {\doibase https://doi.org/10.1111/j.1151-2916.1964.tb13130.x}
  {\bibfield  {journal} {\bibinfo  {journal} {Journal of the American Ceramic
  Society}\ }\textbf {\bibinfo {volume} {47}},\ \bibinfo {pages} {651}
  (\bibinfo {year} {1964})}\BibitemShut {NoStop}%
\bibitem [{\citenamefont {Khanolkar}\ \emph {et~al.}(2023)\citenamefont
  {Khanolkar}, \citenamefont {Wang}, \citenamefont {Dennett}, \citenamefont
  {Hua}, \citenamefont {Mann}, \citenamefont {Khafizov},\ and\ \citenamefont
  {Hurley}}]{khanolkar2023temperature}%
  \BibitemOpen
  \bibfield  {author} {\bibinfo {author} {\bibfnamefont {A.}~\bibnamefont
  {Khanolkar}}, \bibinfo {author} {\bibfnamefont {Y.}~\bibnamefont {Wang}},
  \bibinfo {author} {\bibfnamefont {C.~A.}\ \bibnamefont {Dennett}}, \bibinfo
  {author} {\bibfnamefont {Z.}~\bibnamefont {Hua}}, \bibinfo {author}
  {\bibfnamefont {J.~M.}\ \bibnamefont {Mann}}, \bibinfo {author}
  {\bibfnamefont {M.}~\bibnamefont {Khafizov}}, \ and\ \bibinfo {author}
  {\bibfnamefont {D.~H.}\ \bibnamefont {Hurley}},\ }\href {\doibase
  10.1063/5.0148866} {\bibfield  {journal} {\bibinfo  {journal} {Journal of
  Applied Physics}\ }\textbf {\bibinfo {volume} {133}},\ \bibinfo {pages}
  {195101} (\bibinfo {year} {2023})}\BibitemShut {NoStop}%
\bibitem [{\citenamefont {Brandt}\ and\ \citenamefont
  {Walker}(1967)}]{brandt_temperature_1967}%
  \BibitemOpen
  \bibfield  {author} {\bibinfo {author} {\bibfnamefont {O.~G.}\ \bibnamefont
  {Brandt}}\ and\ \bibinfo {author} {\bibfnamefont {C.~T.}\ \bibnamefont
  {Walker}},\ }\href {\doibase 10.1103/PhysRevLett.18.11} {\bibfield  {journal}
  {\bibinfo  {journal} {Physical Review Letters}\ }\textbf {\bibinfo {volume}
  {18}},\ \bibinfo {pages} {11} (\bibinfo {year} {1967})}\BibitemShut {NoStop}%
\bibitem [{\citenamefont {Muta}\ \emph {et~al.}(2013)\citenamefont {Muta},
  \citenamefont {Murakami}, \citenamefont {Uno}, \citenamefont {Kurosaki},\
  and\ \citenamefont {Yamanaka}}]{muta_thermophysical_2013}%
  \BibitemOpen
  \bibfield  {author} {\bibinfo {author} {\bibfnamefont {H.}~\bibnamefont
  {Muta}}, \bibinfo {author} {\bibfnamefont {Y.}~\bibnamefont {Murakami}},
  \bibinfo {author} {\bibfnamefont {M.}~\bibnamefont {Uno}}, \bibinfo {author}
  {\bibfnamefont {K.}~\bibnamefont {Kurosaki}}, \ and\ \bibinfo {author}
  {\bibfnamefont {S.}~\bibnamefont {Yamanaka}},\ }\href {\doibase
  10.1080/00223131.2013.757468} {\bibfield  {journal} {\bibinfo  {journal}
  {Journal of Nuclear Science and Technology}\ }\textbf {\bibinfo {volume}
  {50}},\ \bibinfo {pages} {181} (\bibinfo {year} {2013})}\BibitemShut
  {NoStop}%
\bibitem [{\citenamefont {Bakker}\ \emph {et~al.}(1997)\citenamefont {Bakker},
  \citenamefont {Cordfunke}, \citenamefont {Konings},\ and\ \citenamefont
  {Schram}}]{bakker_critical_1997}%
  \BibitemOpen
  \bibfield  {author} {\bibinfo {author} {\bibfnamefont {K.}~\bibnamefont
  {Bakker}}, \bibinfo {author} {\bibfnamefont {E.~H.~P.}\ \bibnamefont
  {Cordfunke}}, \bibinfo {author} {\bibfnamefont {R.~J.~M.}\ \bibnamefont
  {Konings}}, \ and\ \bibinfo {author} {\bibfnamefont {R.~P.~C.}\ \bibnamefont
  {Schram}},\ }\href {\doibase 10.1016/S0022-3115(97)00241-9} {\bibfield
  {journal} {\bibinfo  {journal} {Journal of Nuclear Materials}\ }\textbf
  {\bibinfo {volume} {250}},\ \bibinfo {pages} {1} (\bibinfo {year}
  {1997})}\BibitemShut {NoStop}%
\bibitem [{\citenamefont
  {Fink}(2000)}]{finkThermophysicalPropertiesUranium2000}%
  \BibitemOpen
  \bibfield  {author} {\bibinfo {author} {\bibfnamefont {J.~K.}\ \bibnamefont
  {Fink}},\ }\href {\doibase 10.1016/S0022-3115(99)00273-1} {\bibfield
  {journal} {\bibinfo  {journal} {Journal of Nuclear Materials}\ }\textbf
  {\bibinfo {volume} {279}},\ \bibinfo {pages} {1} (\bibinfo {year}
  {2000})}\BibitemShut {NoStop}%
\bibitem [{\citenamefont {Bates}(1965)}]{batesVisibleInfraredAbsorption1965}%
  \BibitemOpen
  \bibfield  {author} {\bibinfo {author} {\bibfnamefont {J.~L.}\ \bibnamefont
  {Bates}},\ }\href {\doibase 10.13182/NSE65-A21011} {\bibfield  {journal}
  {\bibinfo  {journal} {Nuclear Science and Engineering}\ }\textbf {\bibinfo
  {volume} {21}},\ \bibinfo {pages} {26} (\bibinfo {year} {1965})}\BibitemShut
  {NoStop}%
\bibitem [{\citenamefont {Godfrey}\ \emph {et~al.}(1965)\citenamefont
  {Godfrey}, \citenamefont {Fulkerson}, \citenamefont {Kollie}, \citenamefont
  {Moore},\ and\ \citenamefont
  {McELROY}}]{godfreyThermalConductivityUranium1965}%
  \BibitemOpen
  \bibfield  {author} {\bibinfo {author} {\bibfnamefont {T.~G.}\ \bibnamefont
  {Godfrey}}, \bibinfo {author} {\bibfnamefont {W.}~\bibnamefont {Fulkerson}},
  \bibinfo {author} {\bibfnamefont {T.~G.}\ \bibnamefont {Kollie}}, \bibinfo
  {author} {\bibfnamefont {J.~P.}\ \bibnamefont {Moore}}, \ and\ \bibinfo
  {author} {\bibfnamefont {D.~L.}\ \bibnamefont {McELROY}},\ }\href {\doibase
  10.1111/j.1151-2916.1965.tb14745.x} {\bibfield  {journal} {\bibinfo
  {journal} {Journal of the American Ceramic Society}\ }\textbf {\bibinfo
  {volume} {48}},\ \bibinfo {pages} {297} (\bibinfo {year} {1965})}\BibitemShut
  {NoStop}%
\bibitem [{\citenamefont {Ronchi}\ \emph {et~al.}(2004)\citenamefont {Ronchi},
  \citenamefont {Sheindlin}, \citenamefont {Staicu},\ and\ \citenamefont
  {Kinoshita}}]{ronchiEffectBurnupThermal2004}%
  \BibitemOpen
  \bibfield  {author} {\bibinfo {author} {\bibfnamefont {C.}~\bibnamefont
  {Ronchi}}, \bibinfo {author} {\bibfnamefont {M.}~\bibnamefont {Sheindlin}},
  \bibinfo {author} {\bibfnamefont {D.}~\bibnamefont {Staicu}}, \ and\ \bibinfo
  {author} {\bibfnamefont {M.}~\bibnamefont {Kinoshita}},\ }\href {\doibase
  10.1016/j.jnucmat.2004.01.018} {\bibfield  {journal} {\bibinfo  {journal}
  {Journal of Nuclear Materials}\ }\textbf {\bibinfo {volume} {327}},\ \bibinfo
  {pages} {58} (\bibinfo {year} {2004})}\BibitemShut {NoStop}%
\bibitem [{\citenamefont {Hirata}\ \emph {et~al.}(1977)\citenamefont {Hirata},
  \citenamefont {Moriya},\ and\ \citenamefont {Waseda}}]{hirata_high_1977}%
  \BibitemOpen
  \bibfield  {author} {\bibinfo {author} {\bibfnamefont {K.}~\bibnamefont
  {Hirata}}, \bibinfo {author} {\bibfnamefont {K.}~\bibnamefont {Moriya}}, \
  and\ \bibinfo {author} {\bibfnamefont {Y.}~\bibnamefont {Waseda}},\ }\href
  {\doibase 10.1007/BF00548182} {\bibfield  {journal} {\bibinfo  {journal}
  {Journal of Materials Science}\ }\textbf {\bibinfo {volume} {12}},\ \bibinfo
  {pages} {838} (\bibinfo {year} {1977})}\BibitemShut {NoStop}%
\bibitem [{\citenamefont {Momma}\ and\ \citenamefont
  {Izumi}(2011)}]{mommaVESTAThreedimensionalVisualization2011}%
  \BibitemOpen
  \bibfield  {author} {\bibinfo {author} {\bibfnamefont {K.}~\bibnamefont
  {Momma}}\ and\ \bibinfo {author} {\bibfnamefont {F.}~\bibnamefont {Izumi}},\
  }\href {\doibase 10.1107/S0021889811038970} {\bibfield  {journal} {\bibinfo
  {journal} {Journal of Applied Crystallography}\ }\textbf {\bibinfo {volume}
  {44}},\ \bibinfo {pages} {1272} (\bibinfo {year} {2011})}\BibitemShut
  {NoStop}%
\end{thebibliography}%


\begin{thebibliography}{10}%
\makeatletter
\providecommand \@ifxundefined [1]{%
 \@ifx{#1\undefined}
}%
\providecommand \@ifnum [1]{%
 \ifnum #1\expandafter \@firstoftwo
 \else \expandafter \@secondoftwo
 \fi
}%
\providecommand \@ifx [1]{%
 \ifx #1\expandafter \@firstoftwo
 \else \expandafter \@secondoftwo
 \fi
}%
\providecommand \natexlab [1]{#1}%
\providecommand \enquote  [1]{``#1''}%
\providecommand \bibnamefont  [1]{#1}%
\providecommand \bibfnamefont [1]{#1}%
\providecommand \citenamefont [1]{#1}%
\providecommand \href@noop [0]{\@secondoftwo}%
\providecommand \href [0]{\begingroup \@sanitize@url \@href}%
\providecommand \@href[1]{\@@startlink{#1}\@@href}%
\providecommand \@@href[1]{\endgroup#1\@@endlink}%
\providecommand \@sanitize@url [0]{\catcode `\\12\catcode `\$12\catcode
  `\&12\catcode `\#12\catcode `\^12\catcode `\_12\catcode `\%12\relax}%
\providecommand \@@startlink[1]{}%
\providecommand \@@endlink[0]{}%
\providecommand \url  [0]{\begingroup\@sanitize@url \@url }%
\providecommand \@url [1]{\endgroup\@href {#1}{\urlprefix }}%
\providecommand \urlprefix  [0]{URL }%
\providecommand \Eprint [0]{\href }%
\providecommand \doibase [0]{http://dx.doi.org/}%
\providecommand \selectlanguage [0]{\@gobble}%
\providecommand \bibinfo  [0]{\@secondoftwo}%
\providecommand \bibfield  [0]{\@secondoftwo}%
\providecommand \translation [1]{[#1]}%
\providecommand \BibitemOpen [0]{}%
\providecommand \bibitemStop [0]{}%
\providecommand \bibitemNoStop [0]{.\EOS\space}%
\providecommand \EOS [0]{\spacefactor3000\relax}%
\providecommand \BibitemShut  [1]{\csname bibitem#1\endcsname}%
\let\auto@bib@innerbib\@empty
\bibitem [{\citenamefont {Momma}\ and\ \citenamefont
  {Izumi}(2011)}]{mommaVESTAThreedimensionalVisualization2011}%
  \BibitemOpen
  \bibfield  {author} {\bibinfo {author} {\bibfnamefont {K.}~\bibnamefont
  {Momma}}\ and\ \bibinfo {author} {\bibfnamefont {F.}~\bibnamefont {Izumi}},\
  }\href {\doibase 10.1107/S0021889811038970} {\bibfield  {journal} {\bibinfo
  {journal} {Journal of Applied Crystallography}\ }\textbf {\bibinfo {volume}
  {44}},\ \bibinfo {pages} {1272} (\bibinfo {year} {2011})}\BibitemShut
  {NoStop}%
\bibitem [{\citenamefont {Muta}\ \emph {et~al.}(2013)\citenamefont {Muta},
  \citenamefont {Murakami}, \citenamefont {Uno}, \citenamefont {Kurosaki},\
  and\ \citenamefont {Yamanaka}}]{muta_thermophysical_2013}%
  \BibitemOpen
  \bibfield  {author} {\bibinfo {author} {\bibfnamefont {H.}~\bibnamefont
  {Muta}}, \bibinfo {author} {\bibfnamefont {Y.}~\bibnamefont {Murakami}},
  \bibinfo {author} {\bibfnamefont {M.}~\bibnamefont {Uno}}, \bibinfo {author}
  {\bibfnamefont {K.}~\bibnamefont {Kurosaki}}, \ and\ \bibinfo {author}
  {\bibfnamefont {S.}~\bibnamefont {Yamanaka}},\ }\href {\doibase
  10.1080/00223131.2013.757468} {\bibfield  {journal} {\bibinfo  {journal}
  {Journal of Nuclear Science and Technology}\ }\textbf {\bibinfo {volume}
  {50}},\ \bibinfo {pages} {181} (\bibinfo {year} {2013})}\BibitemShut
  {NoStop}%
\bibitem [{\citenamefont {Bakker}\ \emph {et~al.}(1997)\citenamefont {Bakker},
  \citenamefont {Cordfunke}, \citenamefont {Konings},\ and\ \citenamefont
  {Schram}}]{bakker_critical_1997}%
  \BibitemOpen
  \bibfield  {author} {\bibinfo {author} {\bibfnamefont {K.}~\bibnamefont
  {Bakker}}, \bibinfo {author} {\bibfnamefont {E.~H.~P.}\ \bibnamefont
  {Cordfunke}}, \bibinfo {author} {\bibfnamefont {R.~J.~M.}\ \bibnamefont
  {Konings}}, \ and\ \bibinfo {author} {\bibfnamefont {R.~P.~C.}\ \bibnamefont
  {Schram}},\ }\href {\doibase 10.1016/S0022-3115(97)00241-9} {\bibfield
  {journal} {\bibinfo  {journal} {Journal of Nuclear Materials}\ }\textbf
  {\bibinfo {volume} {250}},\ \bibinfo {pages} {1} (\bibinfo {year}
  {1997})}\BibitemShut {NoStop}%
\bibitem [{\citenamefont
  {Fink}(2000)}]{finkThermophysicalPropertiesUranium2000}%
  \BibitemOpen
  \bibfield  {author} {\bibinfo {author} {\bibfnamefont {J.~K.}\ \bibnamefont
  {Fink}},\ }\href {\doibase 10.1016/S0022-3115(99)00273-1} {\bibfield
  {journal} {\bibinfo  {journal} {Journal of Nuclear Materials}\ }\textbf
  {\bibinfo {volume} {279}},\ \bibinfo {pages} {1} (\bibinfo {year}
  {2000})}\BibitemShut {NoStop}%
\bibitem [{\citenamefont {Bates}(1965)}]{batesVisibleInfraredAbsorption1965}%
  \BibitemOpen
  \bibfield  {author} {\bibinfo {author} {\bibfnamefont {J.~L.}\ \bibnamefont
  {Bates}},\ }\href {\doibase 10.13182/NSE65-A21011} {\bibfield  {journal}
  {\bibinfo  {journal} {Nuclear Science and Engineering}\ }\textbf {\bibinfo
  {volume} {21}},\ \bibinfo {pages} {26} (\bibinfo {year} {1965})}\BibitemShut
  {NoStop}%
\bibitem [{\citenamefont {Godfrey}\ \emph {et~al.}(1965)\citenamefont
  {Godfrey}, \citenamefont {Fulkerson}, \citenamefont {Kollie}, \citenamefont
  {Moore},\ and\ \citenamefont
  {McELROY}}]{godfreyThermalConductivityUranium1965}%
  \BibitemOpen
  \bibfield  {author} {\bibinfo {author} {\bibfnamefont {T.~G.}\ \bibnamefont
  {Godfrey}}, \bibinfo {author} {\bibfnamefont {W.}~\bibnamefont {Fulkerson}},
  \bibinfo {author} {\bibfnamefont {T.~G.}\ \bibnamefont {Kollie}}, \bibinfo
  {author} {\bibfnamefont {J.~P.}\ \bibnamefont {Moore}}, \ and\ \bibinfo
  {author} {\bibfnamefont {D.~L.}\ \bibnamefont {McELROY}},\ }\href {\doibase
  10.1111/j.1151-2916.1965.tb14745.x} {\bibfield  {journal} {\bibinfo
  {journal} {Journal of the American Ceramic Society}\ }\textbf {\bibinfo
  {volume} {48}},\ \bibinfo {pages} {297} (\bibinfo {year} {1965})}\BibitemShut
  {NoStop}%
\bibitem [{\citenamefont {Ronchi}\ \emph {et~al.}(2004)\citenamefont {Ronchi},
  \citenamefont {Sheindlin}, \citenamefont {Staicu},\ and\ \citenamefont
  {Kinoshita}}]{ronchiEffectBurnupThermal2004}%
  \BibitemOpen
  \bibfield  {author} {\bibinfo {author} {\bibfnamefont {C.}~\bibnamefont
  {Ronchi}}, \bibinfo {author} {\bibfnamefont {M.}~\bibnamefont {Sheindlin}},
  \bibinfo {author} {\bibfnamefont {D.}~\bibnamefont {Staicu}}, \ and\ \bibinfo
  {author} {\bibfnamefont {M.}~\bibnamefont {Kinoshita}},\ }\href {\doibase
  10.1016/j.jnucmat.2004.01.018} {\bibfield  {journal} {\bibinfo  {journal}
  {Journal of Nuclear Materials}\ }\textbf {\bibinfo {volume} {327}},\ \bibinfo
  {pages} {58} (\bibinfo {year} {2004})}\BibitemShut {NoStop}%
\bibitem [{\citenamefont {Macedo}\ \emph {et~al.}(1964)\citenamefont {Macedo},
  \citenamefont {Capps},\ and\ \citenamefont {Wachtman}}]{macedo_elastic_1964}%
  \BibitemOpen
  \bibfield  {author} {\bibinfo {author} {\bibfnamefont {P.~M.}\ \bibnamefont
  {Macedo}}, \bibinfo {author} {\bibfnamefont {W.}~\bibnamefont {Capps}}, \
  and\ \bibinfo {author} {\bibfnamefont {J.~O.}\ \bibnamefont {Wachtman}},\
  }\href {\doibase https://doi.org/10.1111/j.1151-2916.1964.tb13130.x}
  {\bibfield  {journal} {\bibinfo  {journal} {Journal of the American Ceramic
  Society}\ }\textbf {\bibinfo {volume} {47}},\ \bibinfo {pages} {651}
  (\bibinfo {year} {1964})}\BibitemShut {NoStop}%
\bibitem [{\citenamefont {Khanolkar}\ \emph {et~al.}(2023)\citenamefont
  {Khanolkar}, \citenamefont {Wang}, \citenamefont {Dennett}, \citenamefont
  {Hua}, \citenamefont {Mann}, \citenamefont {Khafizov},\ and\ \citenamefont
  {Hurley}}]{khanolkar2023temperature}%
  \BibitemOpen
  \bibfield  {author} {\bibinfo {author} {\bibfnamefont {A.}~\bibnamefont
  {Khanolkar}}, \bibinfo {author} {\bibfnamefont {Y.}~\bibnamefont {Wang}},
  \bibinfo {author} {\bibfnamefont {C.~A.}\ \bibnamefont {Dennett}}, \bibinfo
  {author} {\bibfnamefont {Z.}~\bibnamefont {Hua}}, \bibinfo {author}
  {\bibfnamefont {J.~M.}\ \bibnamefont {Mann}}, \bibinfo {author}
  {\bibfnamefont {M.}~\bibnamefont {Khafizov}}, \ and\ \bibinfo {author}
  {\bibfnamefont {D.~H.}\ \bibnamefont {Hurley}},\ }\href {\doibase
  10.1063/5.0148866} {\bibfield  {journal} {\bibinfo  {journal} {Journal of
  Applied Physics}\ }\textbf {\bibinfo {volume} {133}},\ \bibinfo {pages}
  {195101} (\bibinfo {year} {2023})}\BibitemShut {NoStop}%
\bibitem [{\citenamefont {Brandt}\ and\ \citenamefont
  {Walker}(1967)}]{brandt_temperature_1967}%
  \BibitemOpen
  \bibfield  {author} {\bibinfo {author} {\bibfnamefont {O.~G.}\ \bibnamefont
  {Brandt}}\ and\ \bibinfo {author} {\bibfnamefont {C.~T.}\ \bibnamefont
  {Walker}},\ }\href {\doibase 10.1103/PhysRevLett.18.11} {\bibfield  {journal}
  {\bibinfo  {journal} {Physical Review Letters}\ }\textbf {\bibinfo {volume}
  {18}},\ \bibinfo {pages} {11} (\bibinfo {year} {1967})}\BibitemShut {NoStop}%
\end{thebibliography}%

\end{document}


\renewcommand{\thefigure}{S\arabic{figure}}
\renewcommand{\thetable}{S\arabic{table}}
\renewcommand\cellalign{cl}

\raggedright{\textbf{\Large Supplemental materials to}}
\title{Parameterizing empirical interatomic potentials for predicting thermophysical properties via an irreducible derivative approach: the case of ThO$_2$ and UO$_2$}
\author{Shuxiang Zhou$^1$, Chao Jiang$^1$, Enda Xiao$^2$, Sasaank Bandi$^3$, Michael W.D. Cooper$^4$, Miaomiao Jin$^5$, David H Hurley$^1$, Marat Khafizov$^6$, and Chris A. Marianetti$^3$}
\affiliation{$^1$Idaho National Laboratory, Idaho Falls, ID 83415, USA\\$^2$Department of Chemistry, Columbia University, New York, NY 10027, USA\\$^3$Department of Applied Physics and Applied Mathematics, Columbia University, New York, NY 10027, USA\\$^4$Los Alamos National Laboratory, Los Alamos, NM 87545, USA\\$^5$Department of Nuclear Engineering, The Pennsylvania State University, University Park, PA 16802, USA\\$^6$Department of Mechanical and Aerospace Engineering, The Ohio State University, Columbus, OH 43210, USA}

\maketitle

\section{\label{sec:propertySummary}Properties in the training procedure and model assessment}
\justifying
\begin{table*}[h]
\caption{\label{tab:SMpropertySummary}%
Summary of DFT-calculated properties in the training and assessment.
}
\begin{ruledtabular}
\begin{tabular}{@{}llclll
}
\multirow{ 2}{*}{Property} &
\multicolumn{3}{c}{Training data}&
\multicolumn{2}{c}{Assessment data}\\
\cmidrule(lr){2-4}\cmidrule(lr){5-6}
& {Data} & {Weight} & {See in} & {Data} & {See in}\\ [0.5em]
\hline\hline \\[-0.8em]
Lattice parameter  & \multirow{ 2}{*}{All data} & \multirow{ 2}{*}{3700} & \multirow{ 2}{*}{Table III} & & \\ 
at $T=0$, $a(0)$ & & & & &  \\[0.5em]

Normalized Born  & \multirow{ 2}{*}{All data} & \multirow{ 2}{*}{3700} & \multirow{ 2}{*}{Table III} & & \\ 
effective charge $Z^*_\alpha$  & & & & &  \\[0.5em]

Frenkel pair (FP)   & a Th or U FP & \multirow{ 2}{*}{20} & \multirow{ 2}{*}{Table III} &  a Th or U FP and two  & \multirow{ 2}{*}{Table IV} \\ 
formation energy $E_F$  & in the supercell $\mathbf{\hat{S}}_\mathrm{C}$ & & & O FPs in the supercell $2\mathbf{\hat{S}}_\mathrm{C}$ &  \\[0.5em]

Strain IDs & \multirow{ 2}{*}{All data} & \multirow{ 2}{*}{55} & \multirow{ 2}{*}{Table III} & & \\ 
$ d_{\alpha_{1}}{}_{\alpha_{2}^{}} $  & & & & &  \\[0.5em]  

Second-order IDs & \multirow{ 2}{*}{within the supercell $2\mathbf{\hat{S}}_\mathrm{I}$}  & \multirow{ 2}{*}{75} & Table III and &  within the supercell $4\mathbf{\hat{S}}_\mathrm{I}$, &  \multirow{ 2}{*}{Figure S2 S3}  \\
$ d^{\alpha_{1}^{}\alpha_{2}^{}}_{q_1 q_2 } $ & & & Figure S2 S3 & excluding the training data & \\[0.5em]

Third-order IDs &  \multirow{ 2}{*}{within the supercell $\mathbf{\hat{S}}_\mathrm{C}$}  & \multirow{ 2}{*}{37}  & Table S2 and  &  within the supercell $\mathbf{\hat{S}}_\mathrm{O}$,     &  \multirow{ 2}{*}{Figure S2 S3}  \\
$ d^{\alpha_{1}^{}\alpha_{2}^{}\alpha_{3}^{}}_{q_1 q_2 q_3} $ & & & Figure S2 S3 & excluding the training data & \\[0.5em]


Phonon dispersion & & & & All data & Figure 1 \\[0.5em]

Thermal conductivity & & & & All data & Figure 2 \\[0.5em]


\end{tabular}
\end{ruledtabular}

\end{table*}

ThO$_2$ and UO$_2$ have a face-centered cubic lattice; the lattice vectors chosen are given in a row-stacked matrix:
\begin{equation}
\hat{\mathbf{a}}=\frac{a}{2}\left[\begin{array}{lll}
0 & 1 & 1 \\
1 & 0 & 1 \\
1 & 1 & 0
\end{array}\right],
\end{equation}
where $a$ is the lattice parameter of a conventional cubic cell. There are three classes of supercells used to define the discretization of the phonon interactions:
$n\hat{\mathbf{S}}_{I}=n\hat{\mathbf1}$ (i.e., uniform supercells),  $n\hat{\mathbf{S}}_{C}=n(\hat{\mathbf{J}}-2\hat{\mathbf{1}})$,
and  $n\hat{\mathbf{S}}_{O}=n(4\hat{\mathbf{1}}-\hat{\mathbf{J}})$, where $n$
is a positive integer, $\hat{\mathbf{1}}$ is the $3\times3$ identity matrix,
and $\hat{\mathbf{J}}$ is a $3\times3$ matrix with each element being 1. 
The crystal structures used for FPs are given in Section~\ref{sec:FPs} of SM.

\clearpage

\section{\label{sec:propertySummary}Method for training the empirical potential }
\justifying

\begin{figure}[h]
    \centering
    \includegraphics[width=0.7\columnwidth]{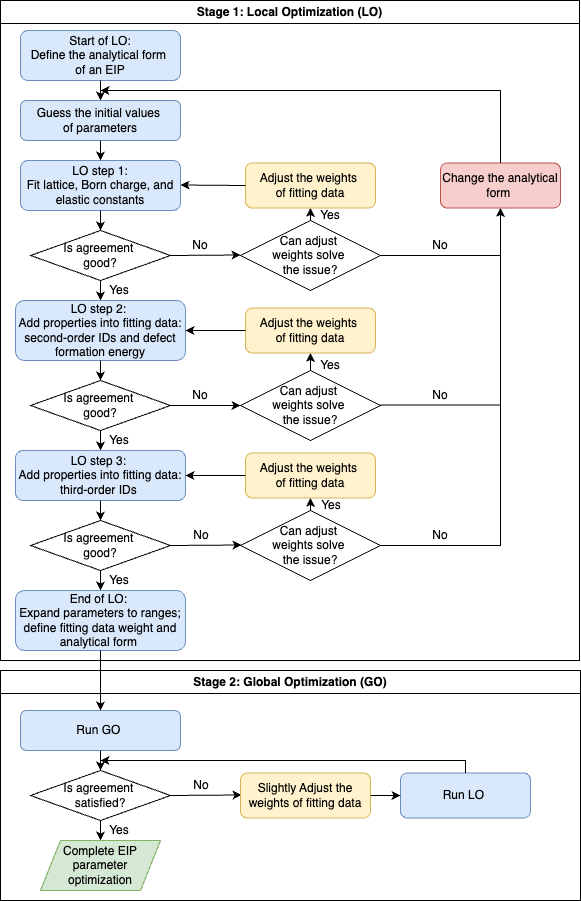}
    \caption{Flowchart of parameter optimization for EIP training in this work.}
    \label{fig:SM_method}
\end{figure}

Here we provide the method for the parameter optimization of EIP training in this work, where a two-stage approach is developed (see the flowchart in Figure~\ref{fig:SM_method}). The challenge of parameter optimization of EIP results from all the variables, including the analytical form of EIP, the range of parameters, and the training data set and its weight. If all this information is already known, one can use a given global optimization (GO) algorithm, e.g., the particle swarm algorithm as applied in this work, and complete the parameter optimization process in one shot.
Therefore, in our two-stage approach, the main target of Stage 1 is to obtain the aforementioned information for variables by applying local optimization (LO) algorithms, e.g., the Nelder–Mead method as applied in this work, then run GO in Stage 2. Stage 1 starts by defining the analytical form of an EIP and the initial values of parameters, and the existing EIPs for the same material or similar systems can be valuable references. Then we add the training data into LO step by step, generally in the order from basic properties to complex properties. To improve the predictions of phonon properties, we designed three steps: first, the training data containing the lattice, elastic constants, and the effective Born charge; secondly, the second-order IDs and defect formation energy are added to the training data; then finally, the third-order IDs are added. In each step, the weights of training data are adjusted, so that a reasonable agreement is obtained. If some issues cannot be fixed by adjusting the weights of training data, these issues are likely due to the analytical form of the EIP. For example, the phonon energies of the highest two optical branches are always approximately identical without the core-shell model for ThO$_2$, thus we add the core-shell model for improvement.

After Stage 1, the range of each parameter can be estimated by expanding its value, e.g., by 2 times or 10 times, depending on the computational resources available for GO. With the analytical form and the training data set and its weight defined in Stage 1, one can start Stage 2 of GO. After GO, if small improvements are required, one can always slightly adjust the weights and re-run LO. Finally, the EIP parameter optimization process is completed.

\clearpage

\section{\label{sec:compareIDs}Comparison of displacement irreducible derivatives}
\justifying

\begin{figure}[h]
    \centering
    \includegraphics[width=0.5\columnwidth]{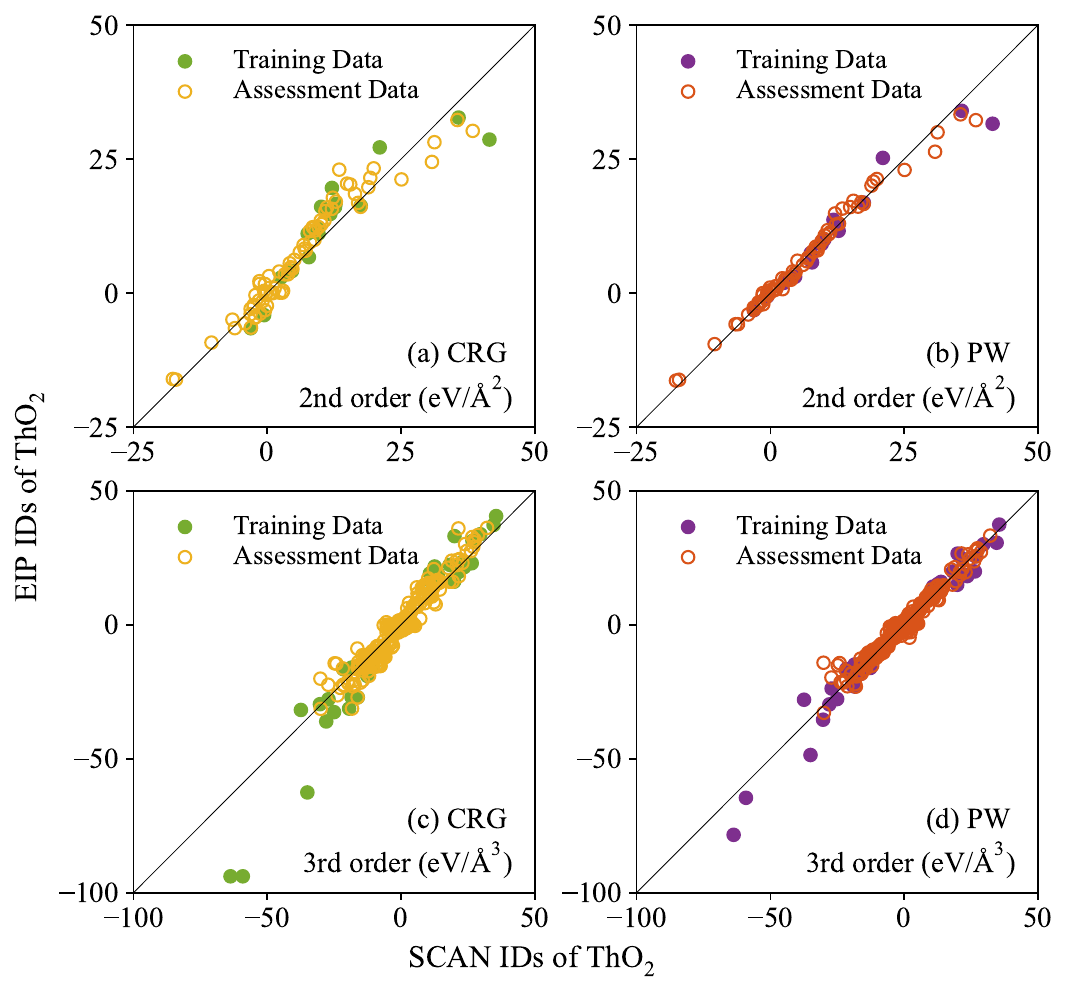}
    \caption{Comparison of computed second- and third-order displacement irreducible derivatives between EIPs (the CRG and the PW) and SCAN for ThO$_2$. Only solid dots were used in the potential training process.}
    \label{fig:SM_ids_tho2}
\end{figure}

\begin{figure}[h]
    \centering
    \includegraphics[width=0.5\columnwidth]{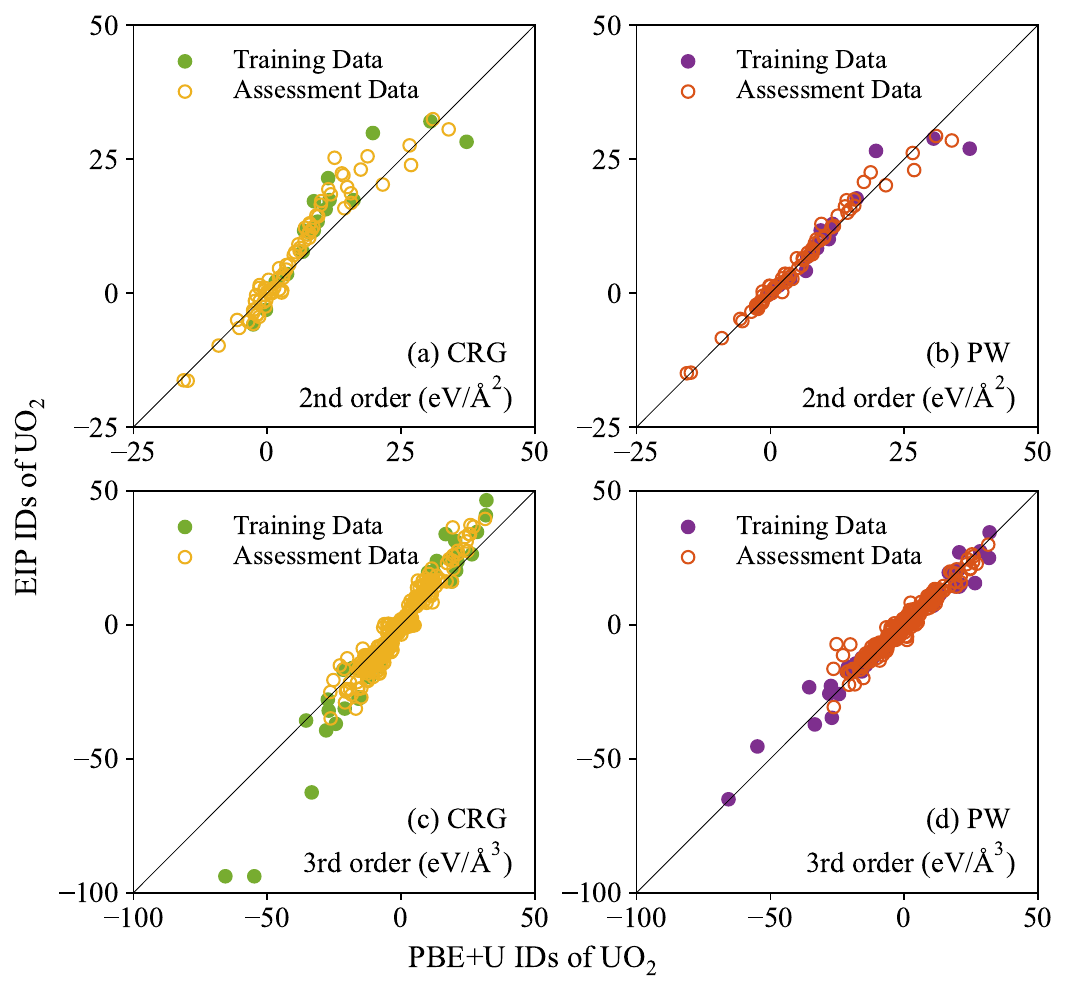}
    \caption{Comparison of computed second- and third-order displacement irreducible derivatives between EIPs (the CRG and the PW) and PBE+$U$ for UO$_2$. Only solid dots were used in the potential training process.}
    \label{fig:SM_ids_uo2}
\end{figure}

We compare the second- and third-order displacement irreducible derivatives computed from EIPs (the CRG and the PW) and DFT, for both ThO$_2$ (see Figure~\ref{fig:SM_ids_tho2}) and UO$_2$ (see Figure~\ref{fig:SM_ids_uo2}). Please note that the training and assessment data are according to the PW, as the CRG was not trained based on DFT data. The second- and third-order displacement irreducible derivatives are computed using the CRG for the sake of comparison. 

\clearpage

\section{\label{sec:FPs}The configuration of Frenkel Pairs}
\justifying

The training set contains two data points: Th (or U) FP in the conventional cubic supercell ($\mathbf{\hat{S}}_\mathrm{C}$).
\begin{itemize}
    \item Th (or U) FP: the Th (or U) atom at $(0, 0, 0)$ is moved to the octahedral interstitial site at $(0.5, 0.5, 0.5)$.
\end{itemize}
\begin{figure}[h]
    \centering
    \includegraphics[width=0.3\columnwidth]{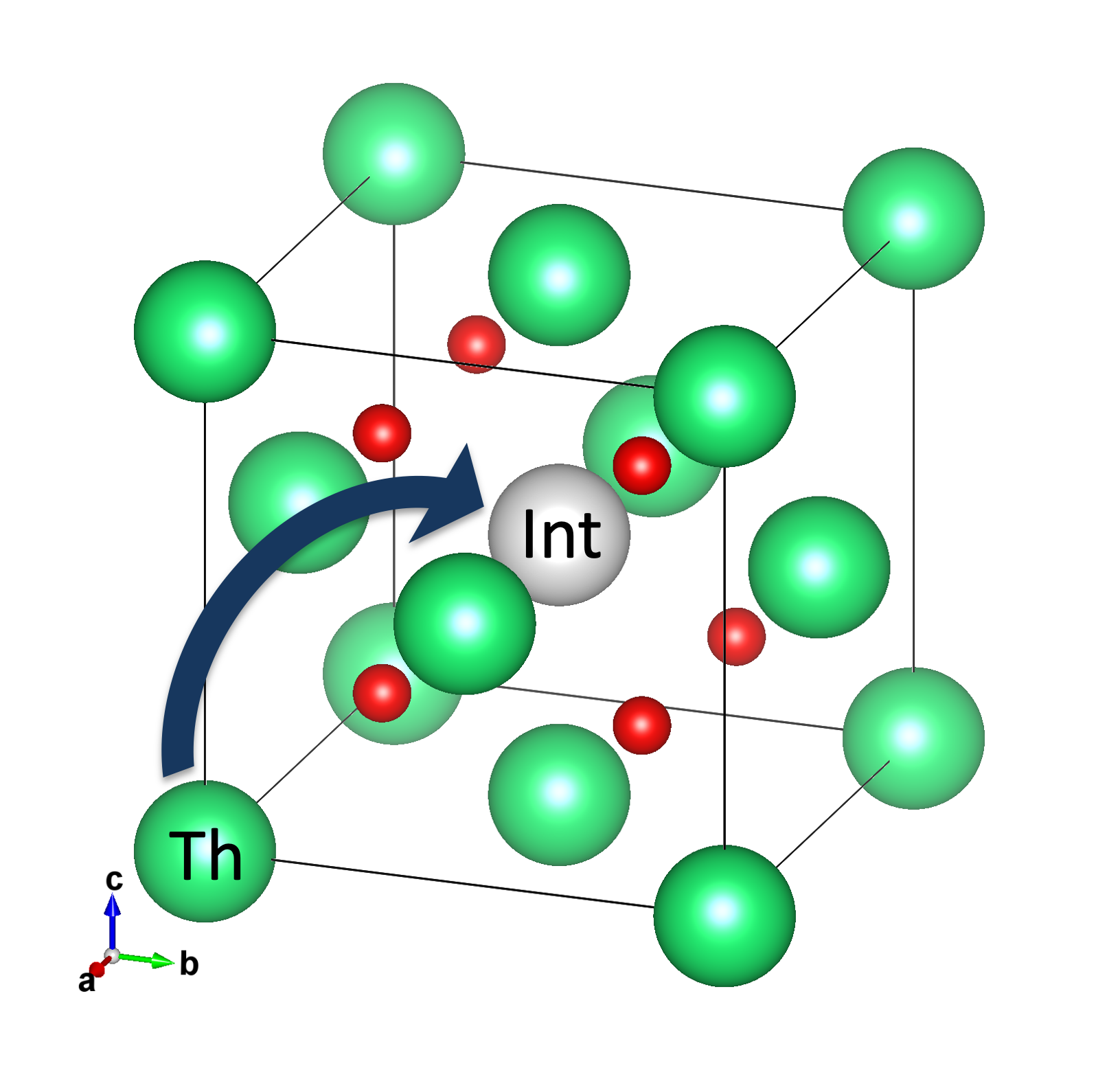}
\end{figure}

The assessment set contains six data points: one Th (or U) FP and two O FPs in the $2\times2\times2$ conventional cubic supercell ($2\mathbf{\hat{S}}_\mathrm{C}$) .
\begin{itemize}
    \item Th (or U) FP: the Th (or U) atom at $(0, 0, 0)$ is moved to the octahedral interstitial site at $(0.25, 0.25, 0.25)$.
    \item O FP1: the O atom at $(0.675, 0.675, 0.125)$ is moved to the octahedral interstitial site at $(0.25, 0.25, 0.25)$.
    \item O FP2: the O atom at $(0.675, 0.675, 0.675)$ is moved to the octahedral interstitial site at $(0.25, 0.25, 0.25)$.
\end{itemize}
\begin{figure}[h]
    \centering
    \includegraphics[width=0.6\columnwidth]{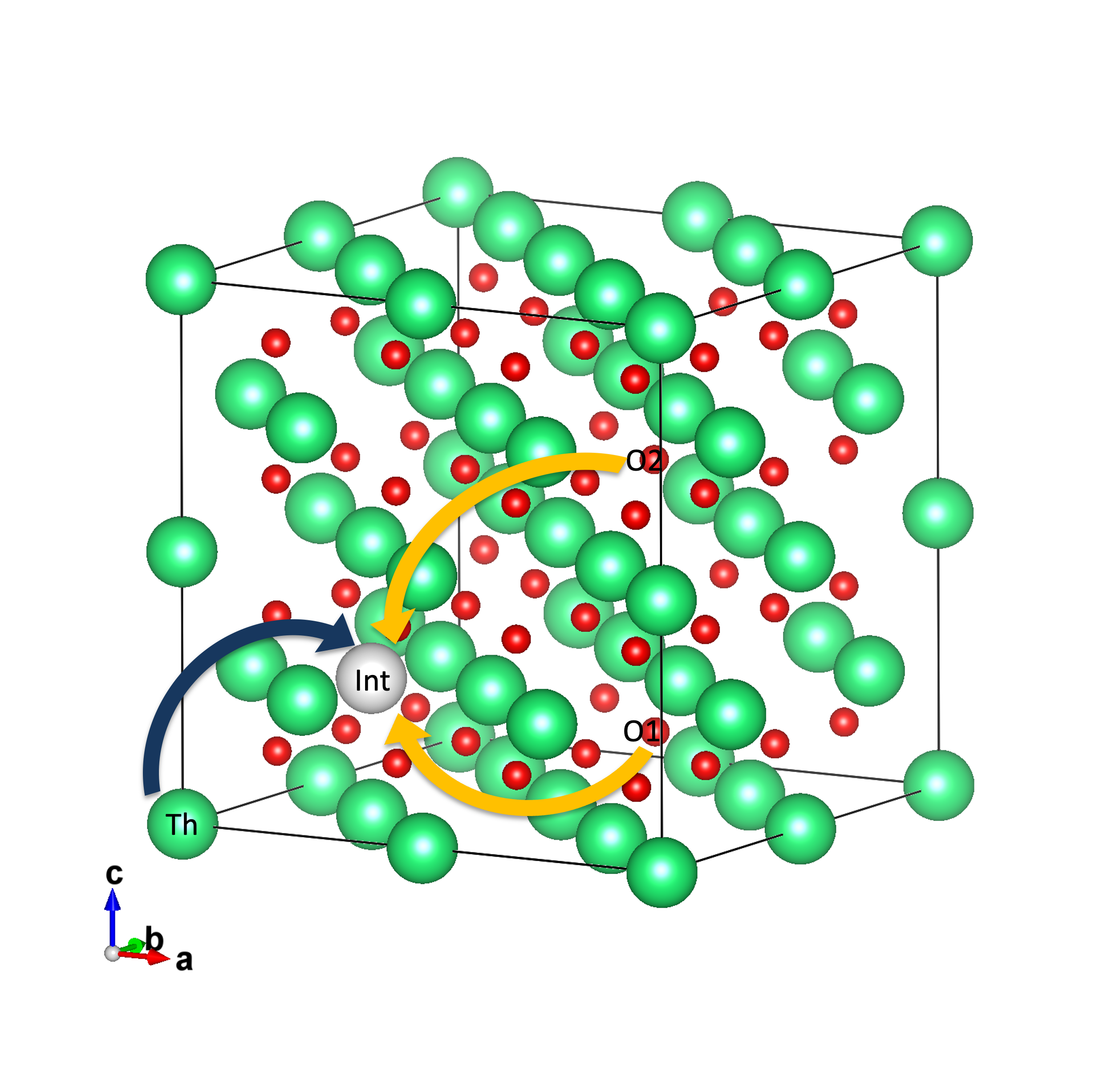}
\end{figure}

The crystal structure is visualized by the VESTA package \cite{mommaVESTAThreedimensionalVisualization2011}.


\section{\label{sec:thirdorder}Training data of third-order irreducible derivatives}
\justifying

\begin{table}
\caption{\label{tab:SMtrainingo3}%
The third-order displacement IDs in the training data computed by DFT, and compared with the CRG and the PW.
}
\begin{ruledtabular}
\begin{tabular}{@{}c
S [table-format=4.2]
S [table-format=4.2]
S [table-format=4.2]
S [table-format=4.2]
S [table-format=4.2]
S [table-format=4.2]
}
\multirow{ 2}{*}{Property} &
\multicolumn{3}{c}{ThO$_2$}&
\multicolumn{3}{c}{UO$_2$}\\
\cmidrule(lr){2-4}\cmidrule(lr){5-7}
& {SCAN} & {CRG} & {PW} & {PBE+$U$} & {CRG} & {PW}\\ [0.2em]
\hline\hline \\[-0.8em]
\multicolumn{7}{c}{Third-order irreducible derivatives within the conventional cubic supercell (eV/\AA$^3$)}\\[0.2em]
\colrule\\[-0.8em]
 $ d^{T_{2g}^{}T_{2g}^{}T_{2g}^{}}_{\Gamma \Gamma \Gamma } $	& -63.74 & -93.90  & -78.41 	& -65.66 & -93.88  & -65.15 \\[0.2em]
 $ d^{T_{1u}^{}T_{1u}^{}T_{2g}^{}}_{\Gamma \Gamma \Gamma } $	& -59.14 & -93.90  & -64.60 	& -54.86 & -93.89  & -45.44 \\[0.2em]
 $ d^{T_{2g}^{}A_{1g}^{}E_{g}^{}}_{\Gamma X X } $ 	& -18.55 & -16.06  & -14.97 	& -18.12 & -16.13  & -14.48 \\[0.2em]
 $ d^{T_{1u}^{}A_{1g}^{}\tensor*[^{2}]{\hspace{-0.2em}A}{}_{2u}^{}}_{\Gamma X X } $ 	& 35.55 & 40.55  & 37.38 	& 31.93 & 46.45  & 34.46 \\[0.2em]
 $ d^{T_{1u}^{}A_{1g}^{}E_{u}^{}}_{\Gamma X X } $ 	& -27.12 & -27.82  & -23.81 	& -27.37 & -27.93  & -22.86 \\[0.2em]
 $ d^{T_{1u}^{}A_{1g}^{}\tensor*[^{1}]{\hspace{-0.2em}E}{}_{u}^{}}_{\Gamma X X } $ 	& 29.66 & 33.78  & 30.02 	& 28.39 & 34.64  & 27.40 \\[0.2em]
 $ d^{T_{2g}^{}E_{g}^{}E_{g}^{}}_{\Gamma X X } $ 	& 26.69 & 31.30  & 25.50 	& 20.26 & 31.30  & 20.53 \\[0.2em]
 $ d^{T_{1u}^{}E_{g}^{}\tensor*[^{2}]{\hspace{-0.2em}A}{}_{2u}^{}}_{\Gamma X X } $ 	& 34.58 & 37.22  & 30.60 	& 31.72 & 40.84  & 24.93 \\[0.2em]
 $ d^{T_{1u}^{}B_{1u}^{}E_{g}^{}}_{\Gamma X X } $ 	& -18.40 & -27.11  & -19.78 	& -15.09 & -27.10  & -14.48 \\[0.2em]
 $ d^{T_{1u}^{}E_{g}^{}E_{u}^{}}_{\Gamma X X } $ 	& -16.17 & -27.11  & -18.33 	& -15.83 & -27.10  & -12.41 \\[0.2em]
 $ d^{T_{1u}^{}E_{g}^{}\tensor*[^{1}]{\hspace{-0.2em}E}{}_{u}^{}}_{\Gamma X X } $ 	& 19.98 & 33.07  & 26.54 	& 16.67 & 33.82  & 19.43 \\[0.2em]
 $ d^{T_{2g}^{}B_{1u}^{}\tensor*[^{2}]{\hspace{-0.2em}A}{}_{2u}^{}}_{\Gamma X X } $ 	& 26.42 & 22.94  & 19.88 	& 26.47 & 26.29  & 15.53 \\[0.2em]
 $ d^{T_{2g}^{}\tensor*[^{2}]{\hspace{-0.2em}A}{}_{2u}^{}E_{u}^{}}_{\Gamma X X } $ 	& 12.57 & 21.73  & 15.18 	& 13.36 & 23.78  & 11.35 \\[0.2em]
 $ d^{T_{2g}^{}\tensor*[^{2}]{\hspace{-0.2em}A}{}_{2u}^{}\tensor*[^{1}]{\hspace{-0.2em}E}{}_{u}^{}}_{\Gamma X X } $ 	& -27.96 & -36.11  & -29.74 	& -27.99 & -39.44  & -25.74 \\[0.2em]
 $ d^{T_{2g}^{}B_{1u}^{}E_{u}^{}}_{\Gamma X X } $ 	& -11.66 & -15.65  & -11.91 	& -10.42 & -15.65  & -9.13 \\[0.2em]
 $ d^{T_{2g}^{}B_{1u}^{}\tensor*[^{1}]{\hspace{-0.2em}E}{}_{u}^{}}_{\Gamma X X } $ 	& 13.78 & 19.09  & 15.99 	& 12.29 & 19.52  & 12.34 \\[0.2em]
 $ d^{T_{2g}^{}E_{u}^{}E_{u}^{}}_{\Gamma X X } $ 	& -19.42 & -31.30  & -22.45 	& -21.05 & -31.29  & -15.90 \\[0.2em]
 $ d^{T_{2g}^{}E_{u}^{}\tensor*[^{1}]{\hspace{-0.2em}E}{}_{u}^{}}_{\Gamma X X } $ 	& 10.90 & 19.09  & 11.98 	& 10.84 & 19.52  & 7.07 \\[0.2em]
 $ d^{T_{2g}^{}\tensor*[^{1}]{\hspace{-0.2em}E}{}_{u}^{}\tensor*[^{1}]{\hspace{-0.2em}E}{}_{u}^{}}_{\Gamma X X } $ 	& -35.01 & -62.60  & -48.62 	& -33.40 & -62.59  & -37.23 \\[0.2em]
 $ d^{A_{1g}^{}A_{1g}^{}A_{1g}^{}}_{X_{2} X_{1} X } $ 	& -21.61 & -16.56  & -16.57 	& -21.57 & -16.81  & -17.44 \\[0.2em]
 $ d^{A_{1g}^{}E_{g}^{}E_{g}^{}}_{X_{2} X_{1} X } $ 	& 19.83 & 16.06  & 14.90 	& 18.94 & 16.13  & 14.35 \\[0.2em]
 $ d^{A_{1g}^{}\tensor*[^{2}]{\hspace{-0.2em}A}{}_{2u}^{}E_{u}^{}}_{X_{2} X_{1} X } $ 	& 18.20 & 22.34  & 19.88 	& 18.64 & 24.49  & 19.38 \\[0.2em]
 $ d^{A_{1g}^{}\tensor*[^{2}]{\hspace{-0.2em}A}{}_{2u}^{}\tensor*[^{1}]{\hspace{-0.2em}E}{}_{u}^{}}_{X_{2} X_{1} X } $ 	& -25.04 & -32.63  & -27.74 	& -24.38 & -36.98  & -25.87 \\[0.2em]
 $ d^{A_{1g}^{}B_{1u}^{}B_{1u}^{}}_{X_{2} X_{1} X } $ 	& -18.46 & -16.06  & -14.88 	& -16.92 & -16.13  & -14.32 \\[0.2em]
 $ d^{A_{1g}^{}E_{u}^{}E_{u}^{}}_{X_{2} X_{1} X } $ 	& -13.73 & -16.06  & -14.60 	& -14.28 & -16.13  & -15.08 \\[0.2em]
 $ d^{A_{1g}^{}E_{u}^{}\tensor*[^{1}]{\hspace{-0.2em}E}{}_{u}^{}}_{X_{2} X_{1} X } $ 	& 21.57 & 19.67  & 24.72 	& 20.60 & 20.20  & 27.01 \\[0.2em]
 $ d^{A_{1g}^{}\tensor*[^{1}]{\hspace{-0.2em}E}{}_{u}^{}\tensor*[^{1}]{\hspace{-0.2em}E}{}_{u}^{}}_{X_{2} X_{1} X } $ 	& -30.31 & -29.67  & -35.45 	& -27.04 & -32.06  & -34.74 \\[0.2em]
 $ d^{E_{g}^{}E_{g}^{}E_{g}^{}}_{X_{2} X_{1} X } $ 	& -11.65 & -15.65  & -12.58 	& -7.96 & -15.65  & -9.95 \\[0.2em]
 $ d^{E_{g}^{}\tensor*[^{2}]{\hspace{-0.2em}A}{}_{2u}^{}B_{1u}^{}}_{X_{2} X_{1} X } $ 	& 23.59 & 21.47  & 18.14 	& 21.10 & 23.56  & 14.99 \\[0.2em]
 $ d^{E_{g}^{}B_{1u}^{}E_{u}^{}}_{X_{2} X_{1} X } $ 	& 11.48 & 15.65  & 11.57 	& 9.99 & 15.65  & 8.56 \\[0.2em]
 $ d^{E_{g}^{}B_{1u}^{}\tensor*[^{1}]{\hspace{-0.2em}E}{}_{u}^{}}_{X_{2} X_{1} X } $ 	& -12.64 & -19.09  & -16.12 	& -11.73 & -19.52  & -12.29 \\[0.2em]
 $ d^{E_{g}^{}E_{u}^{}\tensor*[^{2}]{\hspace{-0.2em}A}{}_{2u}^{}}_{X_{2} X_{1} X } $ 	& 22.21 & 22.97  & 18.67 	& 20.60 & 26.31  & 14.26 \\[0.2em]
 $ d^{E_{g}^{}E_{u}^{}E_{u}^{}}_{X_{2} X_{1} X } $ 	& -10.36 & -15.65  & -10.85 	& -10.17 & -15.65  & -7.43 \\[0.2em]
 $ d^{E_{g}^{}E_{u}^{}\tensor*[^{1}]{\hspace{-0.2em}E}{}_{u}^{}}_{X_{2} X_{1} X } $ 	& 10.80 & 19.09  & 14.21 	& 10.10 & 19.52  & 9.85 \\[0.2em]
 $ ^{1}d^{E_{g}^{}E_{u}^{}\tensor*[^{1}]{\hspace{-0.2em}E}{}_{u}^{}}_{X_{2} X_{1} X } $ 	& 12.28 & 19.09  & 14.21 	& 11.51 & 19.52  & 9.85 \\[0.2em]
 $ d^{E_{g}^{}\tensor*[^{1}]{\hspace{-0.2em}E}{}_{u}^{}\tensor*[^{2}]{\hspace{-0.2em}A}{}_{2u}^{}}_{X_{2} X_{1} X } $ 	& -37.43 & -31.81  & -28.00 	& -35.49 & -35.75  & -23.34 \\[0.2em]
 $ d^{E_{g}^{}\tensor*[^{1}]{\hspace{-0.2em}E}{}_{u}^{}\tensor*[^{1}]{\hspace{-0.2em}E}{}_{u}^{}}_{X_{2} X_{1} X } $ 	& -18.46 & -27.00  & -23.08 	& -15.83 & -27.61  & -17.45 \\[0.2em]

\end{tabular}
\end{ruledtabular}

\end{table}
\clearpage

\section{\label{sec:thirdorder}The spectral and cumulative thermal conductivity}

\begin{figure}[h]
\includegraphics[width=0.40\columnwidth]{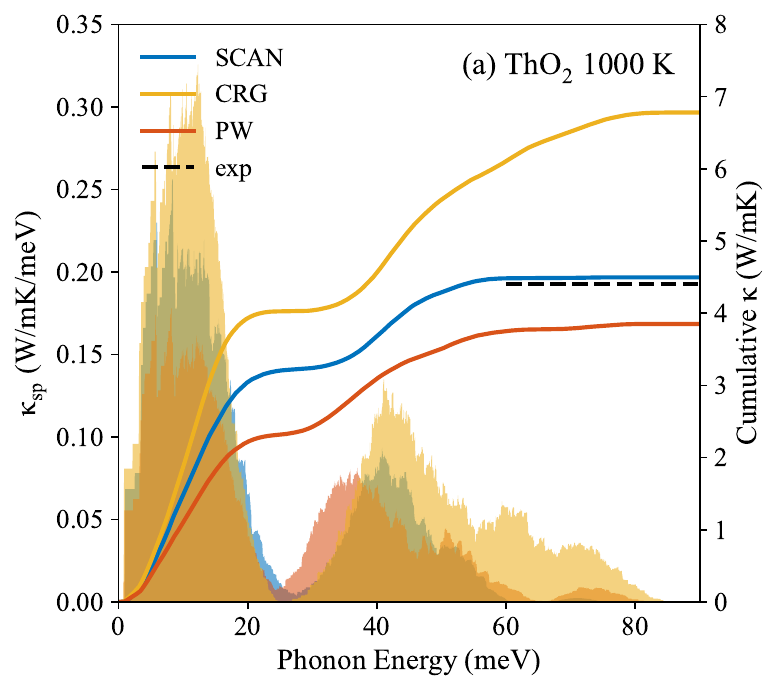}
\includegraphics[width=0.40\columnwidth]{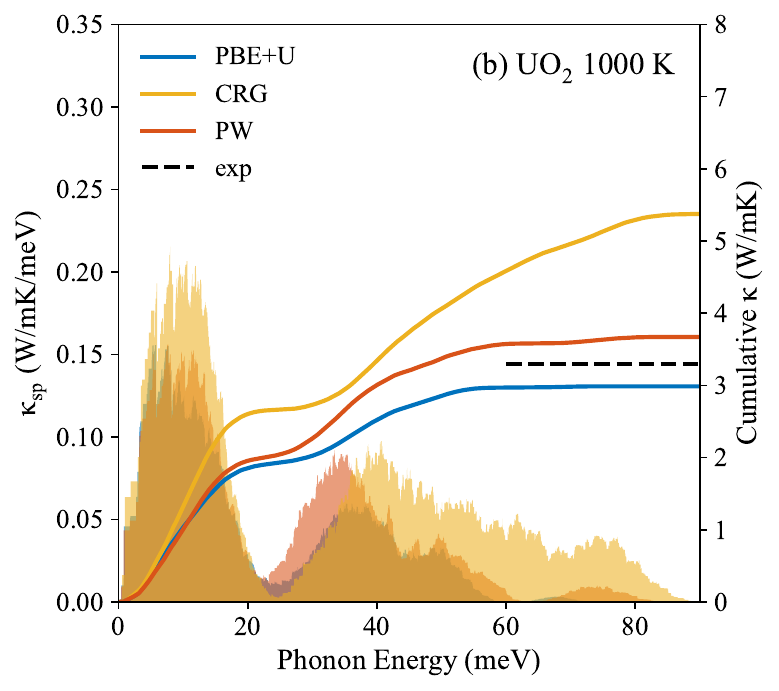}
\caption{\label{fig:mode_cond} Calculated spectral and cumulative thermal conductivity as functions of phonon energy at $T= 1000$ K obtained from the BTE within the RTA for (a) ThO$_2$ and (b) UO$_2$, using DFT, the CRG, and the PW. The black dashed lines, denoted as $exp$, represent the mean values of linearly interpolated experimental data for ThO$_2$ \cite{muta_thermophysical_2013,bakker_critical_1997} and UO$_2$ \cite{finkThermophysicalPropertiesUranium2000, batesVisibleInfraredAbsorption1965, godfreyThermalConductivityUranium1965, ronchiEffectBurnupThermal2004} at $T=1000$ K, respectively. }
\end{figure}

In this section, we report the calculated spectral and cumulative thermal conductivity as
functions of phonon energy at $T=1000$ K obtained from BTE-RTA calculations, shown in Fig.
~\ref{fig:mode_cond}. The black dashed lines, denoted as $exp$, represent the mean values of linearly interpolated experimental data for ThO$_2$ \cite{muta_thermophysical_2013,bakker_critical_1997} and UO$_2$ \cite{finkThermophysicalPropertiesUranium2000, batesVisibleInfraredAbsorption1965, godfreyThermalConductivityUranium1965, ronchiEffectBurnupThermal2004} at $T=1000$ K, respectively. For both the DFT and the PW results, the highest optical phonon mode, where the phonon energy $> 65$ meV and $62$ meV for ThO$_2$ and UO$_2$, respectively, has a thermal conductivity contribution $< 3\%$.

\clearpage

\section{\label{sec:alterForms}Alternative analytical forms}
\justifying

The analytical functional form of our EIP in the main text contains three portions (pairwise potentials, many-body potentials, and the core-shell model), here we benchmark two alternative analytical forms: (1) only pairwise potentials,  denoted as PW\_pair; (2) pairwise potentials plus many-body potentials of EAM, denoted as PW\_eam.  PW\_eam is in the same form as the CRG potential. 
PW\_pair and PW\_eam were trained using the method provided in Section II of SM, and their parameter values and the computed properties are presented in this section. 

Note that, in an EAM potential without the core-shell model, for each element, the normalized Born effective charge is equal to the charge ($Z^*_{\alpha} = q_{\alpha}$). As $q_{Th} = q_{U} = -2 q_{O}$, we have a constraint of $Z^*_{Th} = Z^*_{U}$. However, $Z^*_{Th}$ and $Z^*_{U}$ are different when computed within DFT (as $Z^*_{Th} = 2.448 |e|$, $Z^*_{U} = 2.325 |e|$). Therefore, the values of $Z^*_{Th}$ and $Z^*_{U}$ in EIP should be the mean value of $Z^*_{Th}$ and $Z^*_{U}$ from DFT;
we obtained $q_{Th} = q_{U} = 2.3865 |e|$ and  $q_{O} = -1.19325 |e|$, for both PW\_pair and PW\_eam.

Both PW\_pair and PW\_eam models are simpler than the PW, and thus inevitably, they have larger overall errors than the PW, as well as some other limitations. For PW\_pair, we observed two major errors: the highest two optical branches are nearly degenerate at the $L$ point, and  $C_{12} \approx C_{44}$. While the many-body potential differentiates $C_{12} $ and $ C_{44}$ in PW\_eam, the highest two optical branches are still approximately degenerate at the $L$ point, as only a small gap is observed. The core-shell model, as shown in the PW, can overcome this limitation of nearly degenerate optical modes at the $L$ point, while improving the thermal expansion prediction.  

\begin{table*}[h]
\caption{
Parameters of PW\_pair for the short-range pairwise potentials. 
}
\begin{ruledtabular}
\begin{tabular}{@{}c
S [table-format=5.2]
S [table-format=1.4]
S [table-format=2.4]
S [table-format=2.4]
S [table-format=1.4]
S [table-format=1.4]}
\textrm{Interaction}&
\multicolumn{3}{c}{$\phi_B(r_{ij})$}&
\multicolumn{3}{c}{$\phi_M(r_{ij})$}\\ \cmidrule(lr){1-1}\cmidrule(lr){2-4}\cmidrule(lr){5-7} 
{$\alpha$-$\beta$} &
{$A_{\alpha\beta}(eV)$} &
{$\rho_{\alpha\beta}(\AA)$} &
{$C_{\alpha\beta}(eV\AA^6)$} &
{$D_{\alpha\beta}(eV)$} & 
{$\gamma_{\alpha\beta}(\AA^{-1})$}&
{$r_0(\AA)$}\\[0.2em]
\colrule\\[-0.8em]
Th-Th & 33435.38 & 0.3189 & 2.0174 & 10.9021 & 6.7716 & 2.5580 \\ [0.2em]
U-U   & 32188.29 & 0.3142 & 2.0017 & 11.0094 & 6.7674 & 2.5421  \\ [0.2em]
Th-O  & 184.81 & 0.4286 & 10.2301 & 0.4770 & 2.2433 & 2.3957 \\ [0.2em]
U-O   & 190.91 & 0.4278 & 9.8838 & 0.4678 & 2.1416 & 2.3639 \\ [0.2em]
O-O   & 813.34 & 0.3264 & 0.4932 & {-} & {-} & {-}  
\end{tabular}
\end{ruledtabular}
\end{table*}
\normalsize

\begin{table*}[h]
\caption{
Parameters of PW\_eam for the short-range pairwise potentials.  
}
\begin{ruledtabular}
\begin{tabular}{@{}c
S [table-format=5.2]
S [table-format=1.4]
S [table-format=2.4]
S [table-format=2.4]
S [table-format=1.4]
S [table-format=1.4]}
\textrm{Interaction}&
\multicolumn{3}{c}{$\phi_B(r_{ij})$}&
\multicolumn{3}{c}{$\phi_M(r_{ij})$}\\ \cmidrule(lr){1-1}\cmidrule(lr){2-4}\cmidrule(lr){5-7} 
{$\alpha$-$\beta$} &
{$A_{\alpha\beta}(eV)$} &
{$\rho_{\alpha\beta}(\AA)$} &
{$C_{\alpha\beta}(eV\AA^6)$} &
{$D_{\alpha\beta}(eV)$} & 
{$\gamma_{\alpha\beta}(\AA^{-1})$}&
{$r_0(\AA)$}\\[0.2em]
\colrule\\[-0.8em]
Th-Th & 41277.44 & 0.3107 & 4.2405 & 11.2342 & 7.6147 & 2.5705 \\ [0.2em]
U-U   & 39471.38 & 0.3087 & 4.6978 & 10.8305 & 7.7293 & 2.5609  \\ [0.2em]
Th-O  & 305.57 & 0.4160 & 6.8704 & 0.4342 & 2.2179 & 2.4731 \\ [0.2em]
U-O   & 305.72 & 0.4194 & 5.5307 & 0.4496 & 2.2057 & 2.4315 \\ [0.2em]
O-O   & 630.18 & 0.3461 & 0.8693 & {-} & {-} & {-}  
\end{tabular}
\end{ruledtabular}
\end{table*}
\normalsize

\begin{table}[h]
\caption{
Parameters of PW\_eam for many-body interactions and the Coulomb potential. The charge values ($q$) of PW\_eam are as same as PW\_pair.  
}
\begin{ruledtabular}
\begin{tabular}{@{}c
S [table-format=1.4]
S [table-format=4.2]
S [table-format=3.6]
}
\textrm{Species}&
{$G_\alpha(eV\AA^{1.5})$}&
{$n_\alpha(\AA^5)$}&
{$q_\alpha(|e|)$}\\[0.2em]
\colrule\\[-0.8em]
\textrm{Th} & 2.4472 & 1173.44 & 2.38650 \\[0.2em]
\textrm{U}  & 2.6547 & 1151.67 & 2.38650 \\[0.2em]
\textrm{O}  & 0.4619 & 252.07 & -1.19325 \\
\end{tabular}
\end{ruledtabular}
\end{table}
\normalsize

\begin{figure*}[h]
    \centering
    \includegraphics[width=0.3\columnwidth]{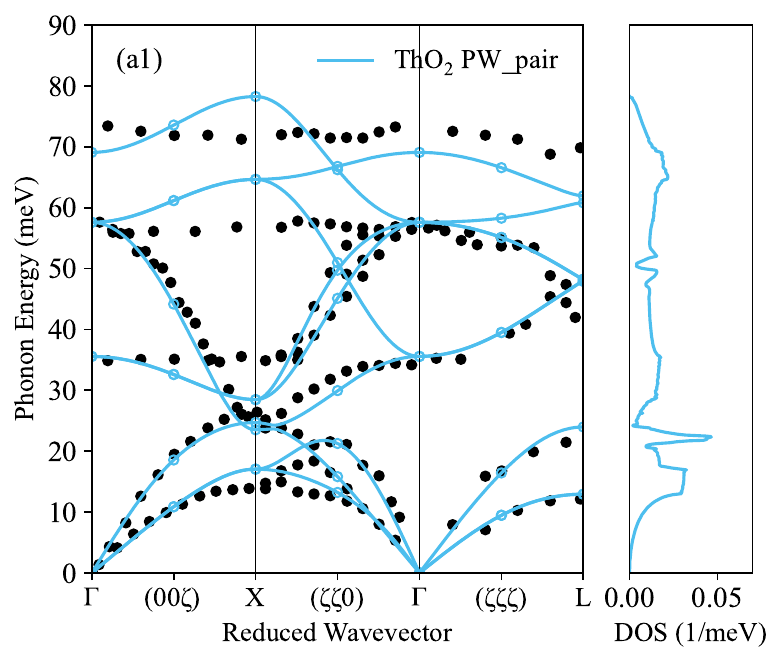}
    \includegraphics[width=0.3\columnwidth]{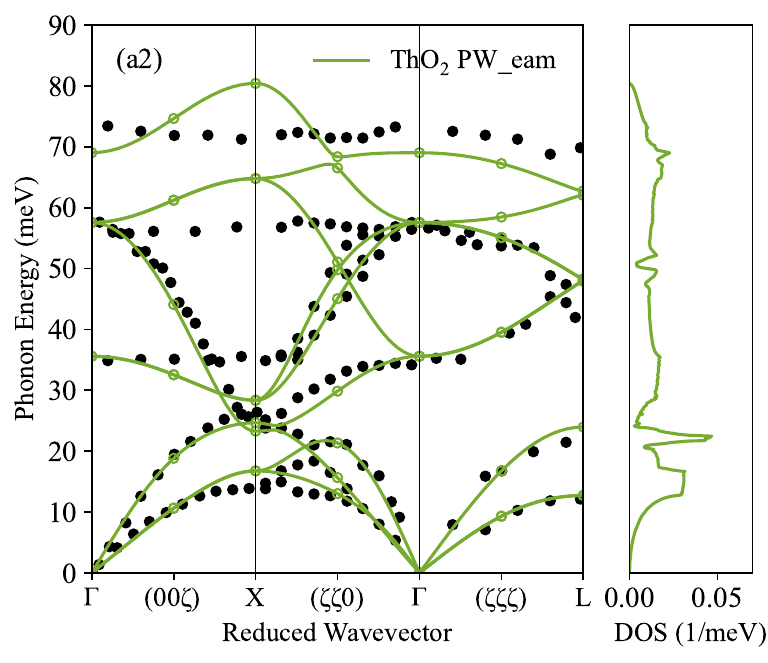}

    \includegraphics[width=0.3\columnwidth]{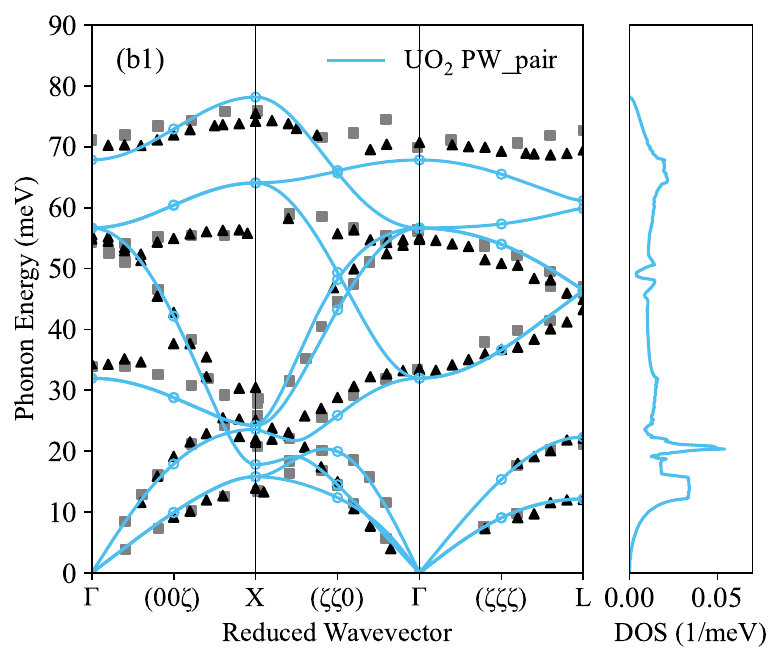}
    \includegraphics[width=0.3\columnwidth]{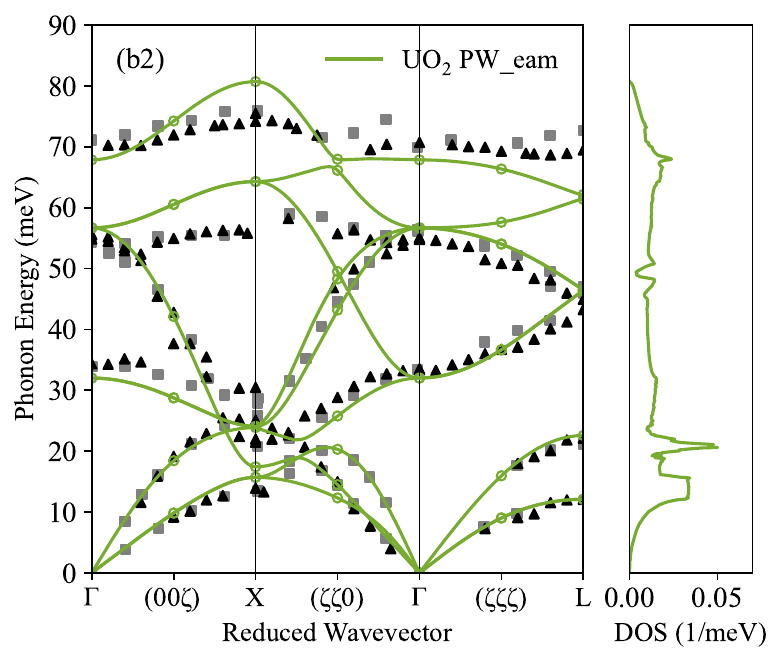}
    \caption{Calculated phonon dispersion and density of states for (a) ThO$_2$ and (b) UO$_2$, using (1) PW\_pair and (2) PW\_eam, compared to experimental data.}
    \label{fig:SMphononSpectra}
\end{figure*}

\begin{table*}[h]
\caption{\label{tab:SMelasticConstant}%
Elastic constants and defect formation energy $E_F$ for FPs calculated using DFT, the PW\_pair, and the PW\_eam in comparison to experimental data.
}
\begin{ruledtabular}
\begin{tabular}{@{}c
cccccccc
}
\multirow{ 2}{*}{Property} &
\multicolumn{4}{c}{ThO$_2$}&
\multicolumn{4}{c}{UO$_2$}\\
\cmidrule(lr){2-5}\cmidrule(lr){6-9}
& {SCAN} & {PW\_pair} & {PW\_eam} & {exp} & {PBE+$U$} & {PW\_pair} & {PW\_eam} & {exp}\\ [0.2em]
\hline\hline \\[-0.8em]
\multicolumn{9}{c}{Lattice constant at $T=0$ K (\AA)}\\[0.2em]
\colrule\\[-0.8em]
$ a(0) $ 
&  5.594 & 5.590  & 5.594  &    &  5.546  & 5.546 & 5.546 &   \\[0.8em]

\multicolumn{9}{c}{Elastic constants (GPa)}\\[0.2em]
\colrule\\[-0.8em]
$ C_{11} $ 
& 376 & 350 & 375 &  367\textsuperscript{a}, 366\textsuperscript{b}  &  380  &  343  & 380  & 400\textsuperscript{c} \\[0.2em]
$ C_{12} $ 
& 117 & 96 & 127 &  106\textsuperscript{a}, 114\textsuperscript{b}  &  120  & 82  & 119 & 125\textsuperscript{c}  \\[0.2em]
$ C_{44} $ 
&  81 &  95 &  90 &  80\textsuperscript{a}, 81\textsuperscript{b}  &  63  & 82 & 80 & 59\textsuperscript{c}  \\[0.8em]

\multicolumn{9}{c}{Defect formation energy $E_F$ (eV)}\\[0.2em]
\colrule\\[-0.8em]
O FP1 & 4.55 & 5.27 & 5.32 &  & 4.02 & 4.94 & 5.05 & \\ [0.2em]
O FP2 & 4.63 & 5.60 & 5.66 &  & 4.06 & 5.25 & 5.38 & \\ [0.2em]
Th or U FP & 13.15 & 12.49 & 14.02 &  & 10.26 & 8.33 & 12.27 & 

\end{tabular}
\end{ruledtabular}
{\raggedright
\textsuperscript{a} Ref.~\cite{macedo_elastic_1964}, \textsuperscript{b} Ref.~\cite{khanolkar2023temperature}, \textsuperscript{c} Ref.~\cite{brandt_temperature_1967}.
\par}

\end{table*}
\normalsize

\begin{figure}[h]
    \centering
    \includegraphics[width=0.4\columnwidth]{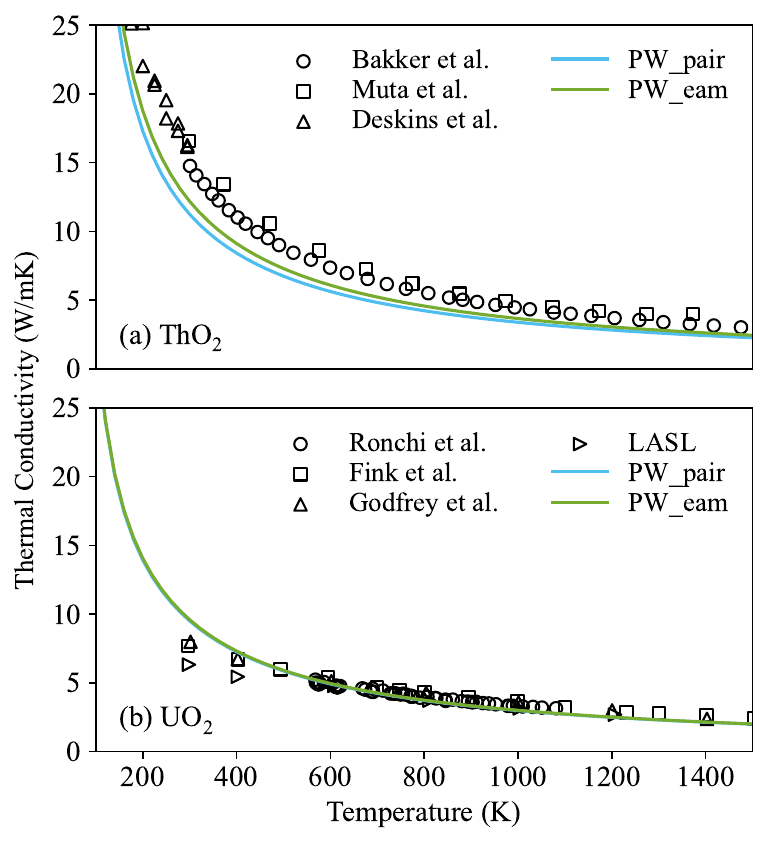}
    \caption{Calculated phonon thermal conductivity of (a) ThO$_2$ and (b) UO$_2$, using the PW\_pair and the PW\_eam at $T=100-1500$ K, compared with the experimental data. }
    \label{fig:SMthermalConductivity}
\end{figure}

\begin{figure}[h]
    \centering
    \includegraphics[width=0.40\columnwidth]{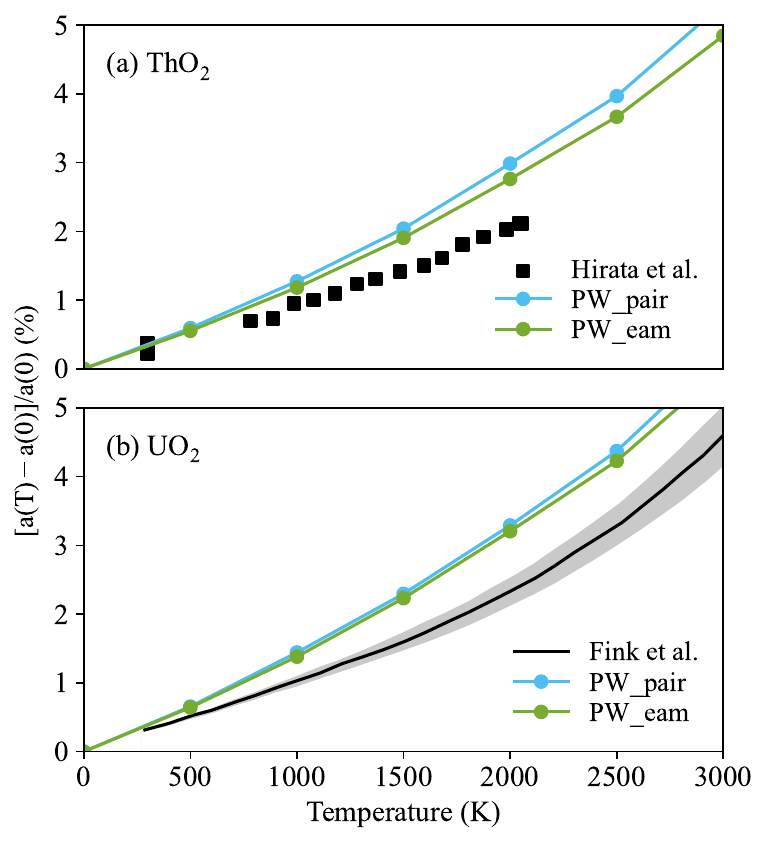}
    \caption{Calculated percentage change of the lattice parameter as a function of temperature at $T=0-3000$ K for (a) ThO$_2$ and (b) UO$_2$, compared with the experiments.}
    \label{fig:SMthermalExpansion}
\end{figure}

\begin{figure}[h]
    \centering
    \includegraphics[width=0.5\columnwidth]{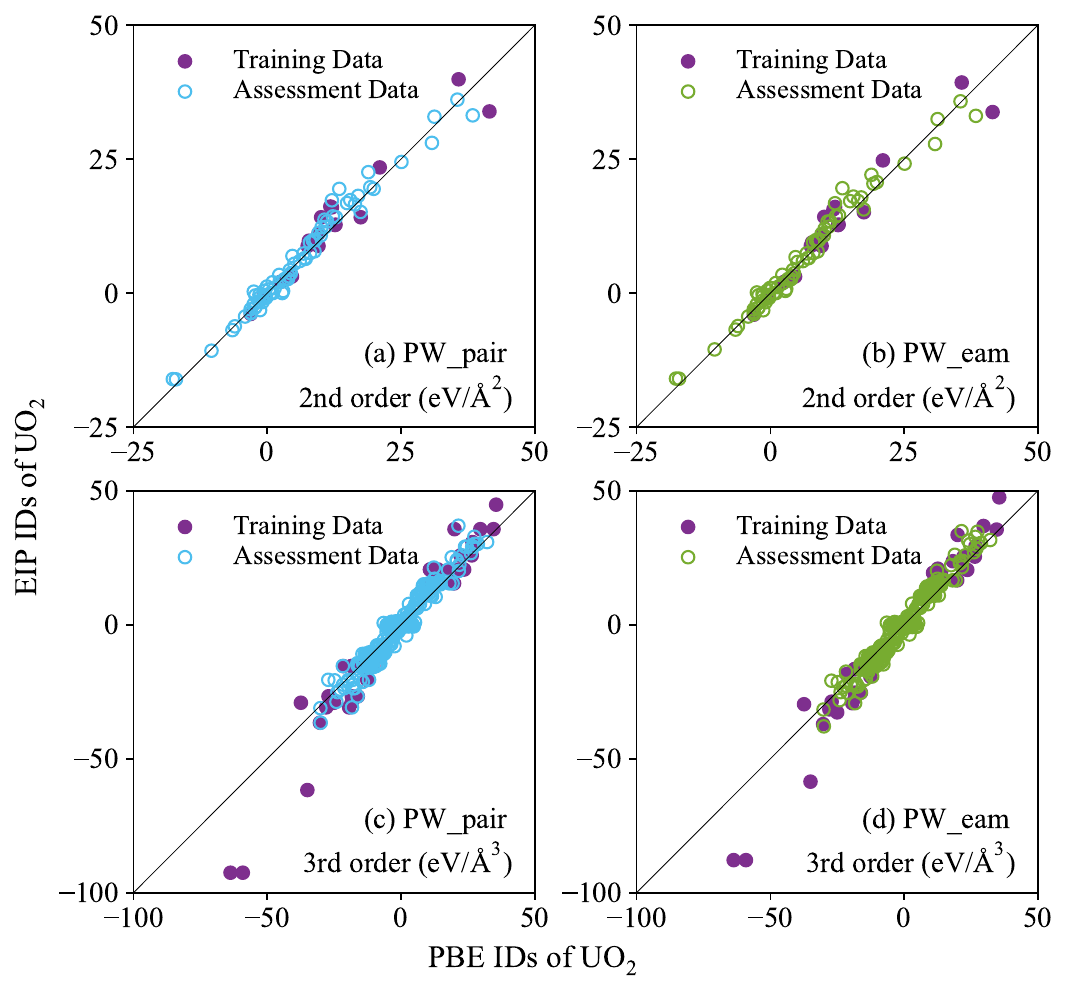}
    \caption{Comparison of computed second- and third-order displacement irreducible derivatives between EIPs (the PW\_pair and the PW\_eam) and SCAN for ThO$_2$. Only solid dots were used in the potential training process.}
\end{figure}

\begin{figure}[h]
    \centering
    \includegraphics[width=0.5\columnwidth]{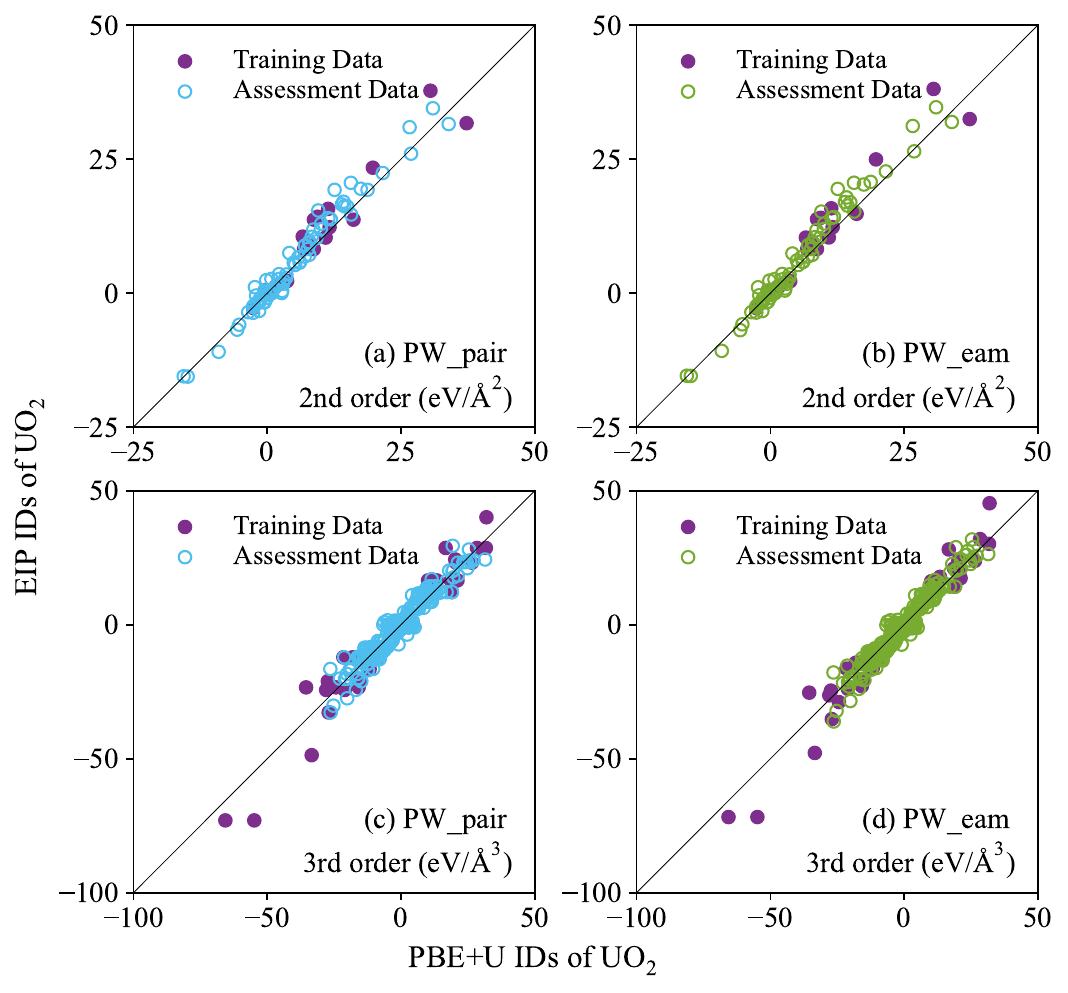}
    \caption{Comparison of computed second- and third-order displacement irreducible derivatives between EIPs (the PW\_pair and the PW\_eam) and PBE+$U$ for UO$_2$. Only solid dots were used in the potential training process.}
\end{figure}

\clearpage

\section{Primary $\mathbf{q}$ points and symmetrized atomic displacement basis}
\justifying
The primary $\mathbf{q}$ points for ThO$_2$ or UO$_2$ are listed in Eq. (\ref{eq:qtable}), and the symmetrized atomic displacement basis of the primary $\mathbf{q}$ points are tabulated in Table~\ref{tab:SMqbasis}.

\begin{equation}
\label{eq:qtable}
\begin{aligned}
 &{\Gamma} = {\left( 0 0 0\right) } \\
 &{L} = {\left( \frac{1}{2} 0 0\right) } & 
 &{L_{1}} = {\left( \frac{1}{2} \frac{1}{2} \frac{1}{2}\right) } & 
 &{L_{2}} = {\left( 0 \frac{1}{2} 0\right) } & 
 &{L_{3}} = {\left( 0 0 \frac{1}{2}\right) } \\
 &{X} = {\left( \frac{1}{2} \frac{1}{2} 0\right) } & 
 &{X_{1}} = {\left( 0 \frac{1}{2} \frac{1}{2}\right) } & 
 &{X_{2}} = {\left( \frac{1}{2} 0 \frac{1}{2}\right) } \\
 &{\Delta} = {\left( \frac{1}{4} \frac{1}{4} 0\right) } & 
 &{\bar{\Delta}} = {\left( \frac{3}{4} \frac{3}{4} 0\right) } & 
 &{\Delta_{1}} = {\left( 0 \frac{1}{4} \frac{1}{4}\right) } & 
 &{\bar{\Delta}_{1}} = {\left( 0 \frac{3}{4} \frac{3}{4}\right) } & 
 &{\Delta_{2}} = {\left( \frac{3}{4} 0 \frac{3}{4}\right) } & 
 &{\bar{\Delta}_{2}} = {\left( \frac{1}{4} 0 \frac{1}{4}\right) }\\
 &{W} = {\left( \frac{1}{4} \frac{3}{4} \frac{1}{2}\right) } & 
 &{\bar{W}} = {\left( \frac{3}{4} \frac{1}{4} \frac{1}{2}\right) } & 
 &{W_{1}} = {\left( \frac{1}{2} \frac{1}{4} \frac{3}{4}\right) } & 
 &{\bar{W}_{1}} = {\left( \frac{1}{2} \frac{3}{4} \frac{1}{4}\right) } & 
 &{W_{2}} = {\left( \frac{1}{4} \frac{1}{2} \frac{3}{4}\right) } & 
 &{\bar{W}_{2}} = {\left( \frac{3}{4} \frac{1}{2} \frac{1}{4}\right) }\\
 &{A} = {\left( \frac{1}{4} \frac{3}{4} 0\right) } & 
 &{\bar{A}} = {\left( \frac{3}{4} \frac{1}{4} 0\right) } & 
 &{A_{1}} = {\left( \frac{1}{2} \frac{3}{4} \frac{3}{4}\right) } & 
 &{\bar{A}_{1}} = {\left( \frac{1}{2} \frac{1}{4} \frac{1}{4}\right) } & 
 &{A_{2}} = {\left( \frac{1}{4} 0 \frac{3}{4}\right) } & 
 &{\bar{A}_{2}} = {\left( \frac{3}{4} 0 \frac{1}{4}\right) }\\[0.4em] 
 &{A_{3}} = {\left( 0 \frac{3}{4} \frac{1}{4}\right) } & 
 &{\bar{A}_{3}} = {\left( 0 \frac{1}{4} \frac{3}{4}\right) } & 
 &{A_{4}} = {\left( \frac{3}{4} \frac{1}{2} \frac{3}{4}\right) } & 
 &{\bar{A}_{4}} = {\left( \frac{1}{4} \frac{1}{2} \frac{1}{4}\right) } & 
 &{A_{5}} = {\left( \frac{1}{4} \frac{1}{4} \frac{1}{2}\right) } & 
 &{\bar{A}_{5}} = {\left( \frac{3}{4} \frac{3}{4} \frac{1}{2}\right) }\\
 &{\Lambda} = {\left( \frac{1}{4} 0 0\right) } & 
 &{\bar{\Lambda}} = {\left( \frac{3}{4} 0 0\right) } & 
 &{\Lambda_{1}} = {\left( 0 \frac{3}{4} 0\right) } & 
 &{\bar{\Lambda}_{1}} = {\left( 0 \frac{1}{4} 0\right) } & 
 &{\Lambda_{2}} = {\left( 0 0 \frac{3}{4}\right) } & 
 &{\bar{\Lambda}_{2}} = {\left( 0 0 \frac{1}{4}\right) }\\[0.4em] 
 &{\Lambda_{3}} = {\left( \frac{1}{4} \frac{1}{4} \frac{1}{4}\right) } & 
 &{\bar{\Lambda}_{3}} = {\left( \frac{3}{4} \frac{3}{4} \frac{3}{4}\right) } & 
\\
 &{B} = {\left( \frac{1}{4} \frac{1}{2} 0\right) } & 
 &{\bar{B}} = {\left( \frac{3}{4} \frac{1}{2} 0\right) } & 
 &{B_{1}} = {\left( \frac{1}{2} \frac{3}{4} \frac{1}{2}\right) } & 
 &{\bar{B}_{1}} = {\left( \frac{1}{2} \frac{1}{4} \frac{1}{2}\right) } & 
 &{B_{2}} = {\left( \frac{3}{4} \frac{3}{4} \frac{1}{4}\right) } & 
 &{\bar{B}_{2}} = {\left( \frac{1}{4} \frac{1}{4} \frac{3}{4}\right) }\\[0.4em] 
 &{B_{3}} = {\left( \frac{1}{4} \frac{3}{4} \frac{3}{4}\right) } & 
 &{\bar{B}_{3}} = {\left( \frac{3}{4} \frac{1}{4} \frac{1}{4}\right) } & 
 &{B_{4}} = {\left( \frac{1}{2} \frac{1}{2} \frac{1}{4}\right) } & 
 &{\bar{B}_{4}} = {\left( \frac{1}{2} \frac{1}{2} \frac{3}{4}\right) } & 
 &{B_{5}} = {\left( \frac{3}{4} \frac{1}{4} \frac{3}{4}\right) } & 
 &{\bar{B}_{5}} = {\left( \frac{1}{4} \frac{3}{4} \frac{1}{4}\right) }\\
 &{B_{6}} = {\left( \frac{1}{4} \frac{1}{2} \frac{1}{2}\right) } & 
 &{\bar{B}_{6}} = {\left( \frac{3}{4} \frac{1}{2} \frac{1}{2}\right) } & 
 &{B_{7}} = {\left( \frac{1}{2} \frac{3}{4} 0\right) } & 
 &{\bar{B}_{7}} = {\left( \frac{1}{2} \frac{1}{4} 0\right) } & 
 &{B_{8}} = {\left( \frac{1}{4} 0 \frac{1}{2}\right) } & 
 &{\bar{B}_{8}} = {\left( \frac{3}{4} 0 \frac{1}{2}\right) }\\
 &{B_{9}} = {\left( \frac{1}{2} 0 \frac{1}{4}\right) } & 
 &{\bar{B}_{9}} = {\left( \frac{1}{2} 0 \frac{3}{4}\right) } & 
 &{B_{10}} = {\left( 0 \frac{1}{4} \frac{1}{2}\right) } & 
 &{\bar{B}_{10}} = {\left( 0 \frac{3}{4} \frac{1}{2}\right) } & 
 &{B_{11}} = {\left( 0 \frac{1}{2} \frac{1}{4}\right) } & 
 &{\bar{B}_{11}} = {\left( 0 \frac{1}{2} \frac{3}{4}\right) }\\
\end{aligned}
\end{equation}

\clearpage
\begin{longtable}{l|rrrrrrrrr}
\caption{\label{tab:SMqbasis}
Atomic displacement basis for the primary $\mathbf{q}$ points. 
}
\\\hline\hline \\[\dimexpr-\normalbaselineskip+0.4em]
 & Th$_x  $  & Th$_y  $ &  Th$_z $ 
 & O$^1_x $  & O$^1_y $ &  O$^1_z $ 
 & O$^2_x $  & O$^2_y $ &  O$^2_z $ \\[0.4em]
\hline
\\[\dimexpr-\normalbaselineskip+0.4em]
\multirow{3}{*}{${u_{\Gamma}^{T_{2g}^{}}}$} & $  0 $ & $  0 $ & $  0 $ & $  0 $ & $  0 $ & $  \frac{ \sqrt{2}}{2} $ & $  0 $ & $  0 $ & $ -\frac{ \sqrt{2}}{2} $ \\[0.4em]
                                            & $  0 $ & $  0 $ & $  0 $ & $  \frac{ \sqrt{2}}{2} $ & $  0 $ & $  0 $ & $ -\frac{ \sqrt{2}}{2} $ & $  0 $ & $  0 $ \\[0.4em]
                                            & $  0 $ & $  0 $ & $  0 $ & $  0 $ & $  \frac{ \sqrt{2}}{2} $ & $  0 $ & $  0 $ & $ -\frac{ \sqrt{2}}{2} $ & $  0 $ \\[0.4em]
\multirow{3}{*}{${u_{\Gamma}^{T_{1u}^{}}}$} & $  \frac{ \sqrt{6}}{3} $ & $  0 $ & $  0 $ & $ -\frac{ \sqrt{6}}{6} $ & $  0 $ & $  0 $ & $ -\frac{ \sqrt{6}}{6} $ & $  0 $ & $  0 $ \\
\hline\hline
\end{longtable}
\bibliographystyle{apsrev4-1} 
\bibliography{tho2,extra}